%% file: main.tex
\documentclass[11pt]{article}
\usepackage[margin=0.75in]{geometry}
\usepackage{graphicx}
\usepackage{caption}
\graphicspath{{figures/}}
\usepackage{amsmath,amssymb}
\usepackage{booktabs}
\usepackage{algorithm}
\usepackage[noend]{algpseudocode}
\algblockdefx[Phase]{Phase}{EndPhase}[1]{\textbf{#1}}{}
\algnotext{EndPhase}
\usepackage[natbibapa]{apacite}
\makeatletter\NAT@longnamesfalse\makeatother
\AtBeginDocument{}
\usepackage{textcomp}
\usepackage[hidelinks]{hyperref}
\hypersetup{linktoc=all, pdftitle={Geometric and algebraic inference in humans and neural networks}, pdfauthor={Jacob Russin, Joonhwa Kim, Ellie Pavlick, Michael J. Frank}}
\usepackage{etoc}
\makeatletter
\let\paper@origparagraph\paragraph
\renewcommand\paragraph{\@ifstar{\paper@origparagraph*}%
  {\@ifnextchar[{\paper@origparagraph}{\paper@paragraph}}}
\def\paper@paragraph#1{\paper@splitdot#1\paper@end.\paper@end\paper@stop{#1}}
\def\paper@splitdot#1.\paper@end#2\paper@stop#3{%
  \if\relax\detokenize{#2}\relax
    \expandafter\@firstoftwo
  \else
    \expandafter\@secondoftwo
  \fi
  {\paper@origparagraph{#3}}%
  {\paper@origparagraph[{#1}]{#3}}}
\newdimen\paper@numwd \newdimen\paper@hang \newdimen\paper@pagewd
\newcommand\paper@tocline[7]{%
  \par\begingroup
  \normalfont\normalsize
  \paper@numwd=\dimexpr#2\relax
  \paper@hang=\dimexpr#3\relax
  \paper@pagewd=1.8em
  \leftskip=\dimexpr#1+\paper@numwd+\paper@hang\relax
  \rightskip=2.4em plus 1fil
  \parfillskip=-\rightskip
  \parindent=\z@ \interlinepenalty=\@M
  \leavevmode#4\hskip-\dimexpr\paper@numwd+\paper@hang\relax
  \makebox[\paper@numwd][l]{#5}#6\nobreak
  {\normalfont\normalsize\color{black!35}\nobreak
   \leaders\hbox{$\m@th\mkern4.5mu\hbox{.}\mkern4.5mu$}\hfill}%
  \nobreak\hb@xt@\paper@pagewd{\hss#7}\par
  \endgroup}
\etocsetstyle{section}{}
  {\addvspace{12pt}}
  {\paper@tocline{\z@}{1.4em}{\z@}{\bfseries}{\etocnumber}{\etocname}{\etocpage}%
   \nobreak\vskip2pt}
  {}
\etocsetstyle{subsection}{}
  {\addvspace{3pt}}
  {\paper@tocline{1.4em}{2.0em}{\z@}{}{\etocnumber}{\etocname}{\etocpage}\nobreak}
  {}
\newcommand\paper@paraindent{3.4em}
\etocsetstyle{subsubsection}
  {\gdef\paper@paraindent{6.2em}}
  {\addvspace{2pt}}
  {\paper@tocline{3.4em}{2.8em}{\z@}{}{\etocnumber}{\etocname}{\etocpage}\nobreak}
  {\gdef\paper@paraindent{3.4em}}
\etocsetstyle{paragraph}{}
  {}
  {\paper@tocline{\paper@paraindent}{\z@}{1em}{\small}{}{\etocname}{\etocpage}}
  {}
\makeatother
\usepackage[section]{placeins}
\usepackage{enumitem}
\usepackage{xcolor}

\usepackage{fvextra}
\DefineVerbatimEnvironment{tracetext}{Verbatim}{fontsize=\footnotesize,
  formatcom=\fontencoding{T1}\fontfamily{lmtt}\selectfont,
  breaklines, frame=lines, framesep=4pt, rulecolor=\color{black!40},
  labelposition=topline}
\newcounter{trace}
\renewcommand{\thetrace}{S\arabic{trace}}

\newcommand{\tracehead}[2]{\par\medskip\noindent\refstepcounter{trace}\label{#1}%
  \textbf{Trace~\thetrace.}~#2\par\nopagebreak\smallskip}

\newcommand{\rankm}{\pi^{(m)}}
\newcommand{\dist}{d}

\title{Geometric and algebraic inference in humans and neural networks}

\author{
  Jacob Russin\textsuperscript{1,2,$\ast$} \and
  Joonhwa Kim\textsuperscript{3,$\ast$} \and
  Ellie Pavlick\textsuperscript{1,$\dagger$} \and
  Michael J. Frank\textsuperscript{2,4,$\dagger$}
}
\date{\vspace{1ex}\small
  \textsuperscript{1}Department of Computer Science, Brown University\\
  \textsuperscript{2}Department of Cognitive and Psychological Sciences, Brown University\\
  \textsuperscript{3}Department of Neuroscience, Brown University\\
  \textsuperscript{4}Carney Institute for Brain Science, Brown University\\[1ex]
  \textsuperscript{$\ast$}Equal contribution.\quad
  \textsuperscript{$\dagger$}Joint senior authors.
}

\begin{document}
\etocdepthtag.toc{main}
\maketitle
{\NoHyper\renewcommand{\thefootnote}{}\footnotetext{Correspondence: jake\_russin@brown.edu}\endNoHyper}

\begin{abstract}
Humans possess a remarkable ability to draw inferences from limited data.  
While some inferences are made intuitively from past experience, others require online reasoning based on novel information given in context.   
Here, we formalize these strategies, which we term \emph{geometric} and \emph{algebraic} inference, and study their computational tradeoffs in humans and neural networks.
The geometric strategy gradually shapes representational geometry to reflect underlying relational structure, making subsequent inference cheap.
The algebraic strategy flexibly attends to and reasons over novel relations given in context, but costs additional computation at inference time. 
We identify opposing signatures of these two strategies in transitive inference, a domain widely studied across species, and characterize the conditions under which each is observed.
Our framework predicts that the geometric strategy should be dominant when relations are familiar, whereas the algebraic should emerge to support \emph{de novo} inferences drawn in context.
Multiple experiments demonstrate these opposing signatures in humans, neural networks trained from scratch, and large language models (LLMs).
Causal interventions in the neural networks reveal that the geometric and algebraic strategies depend on representational geometry and selective attention mechanisms, respectively, and reproduce the effects of analogous neurobiological disruptions in humans.
Our framework unifies these effects with disparate phenomena, including content effects observed in human and LLM reasoning and data-distribution effects observed in machine learning.
Our findings reconcile competing views of cognitive flexibility that emphasize either representational geometry or algebraic manipulation by showing that each can have unique strengths, and can characterize different aspects of both human and neural network behavior.

\end{abstract}

\input{sections/introduction}
\input{sections/framework}
\input{sections/results}
\input{sections/discussion}

\input{sections/methods}

\section*{Acknowledgments}
We thank Roman Feiman, Jason Ritt, Jae-Young Son, Daniel Carstensen, and the members of the Language Understanding and Representation Lab, the Laboratory of Neural Computation and Cognition, and the Brown Language and Thought Lab for helpful discussions.

\paragraph*{Funding:}
J.R. and J.K. were supported by the National Institute of Mental Health of the National Institutes of Health (NIH) under Award Numbers T32MH126388 and T32MH115895, respectively.
The content is solely the responsibility of the authors and does not necessarily represent the official views of the NIH.
This work used computational resources supported by the NIH S10 instrumentation grant S10OD036341 (High-Performance Compute Cluster for Brain Science) via the Center for Computation and Visualization at Brown University.
M.J.F. was supported by an Office of Naval Research MURI Award N00014-23-1-2792.
E.P. was supported in part by Schmidt Sciences Grant \#GR5300958, and the NSF AI Research Institute on Interaction for AI Assistants Grant \#GR5300593.
The content of this article does not necessarily reflect the views of the US Government, and no official endorsement should be inferred.

\paragraph*{Competing interests:}
E.P. is a paid consultant for Google DeepMind.
The content of this article does not necessarily reflect the views of Google, and no official endorsement should be inferred.
The other authors declare no competing interests.

\paragraph*{Data and materials availability:}
All code is publicly available at the following repositories:
\begin{itemize}[leftmargin=*]
    \item Transformer experiments: \url{https://github.com/jlrussin/geometric-algebraic-ti-transformers}
    \item Human experiments: \url{https://github.com/jlrussin/geometric-algebraic-ti-humans}
    \item LLM experiments: \url{https://github.com/joonhwakim/geometric-algebraic-ti-llms}
\end{itemize}
The human data cannot be shared publicly under the terms of participants' consent.

\paragraph*{Use of AI tools:}
Under the authors' direction, AI tools (Anthropic Claude Opus and Fable 4.7--5.5 and OpenAI GPT 5.5--6.1) were used to help write code for experiments, analysis, and visualization.
AI tools were also used for assisting with literature search, editing text for grammar, spelling, and clarity, and for generating initial drafts of specific methods paragraphs.
The authors reviewed and verified all AI-assisted content and take full responsibility for all aspects of the work.

\bibliographystyle{apacite}
\bibliography{refs}

\clearpage
\appendix
\etocdepthtag.toc{appendix}
\setcounter{figure}{0}
\renewcommand{\thefigure}{S\arabic{figure}}
\renewcommand{\theHfigure}{appendix.S\arabic{figure}}
\setcounter{table}{0}
\renewcommand{\thetable}{S\arabic{table}}
\renewcommand{\theHtable}{appendix.S\arabic{table}}
\etocsetnexttocdepth{paragraph}
\etocsettagdepth{main}{none}
\etocsettagdepth{appendix}{subsubsection}
\etocsettagdepth{extendedresults}{paragraph}
\etocsettocstyle{}{}
\pdfbookmark[1]{Appendix contents}{appendix-contents}
{\noindent\Large\bfseries Appendix\par}
\tableofcontents
\clearpage
\input{appendix/formalization}
\input{appendix/extended_methods}
\etocdepthtag.toc{extendedresults}
\input{appendix/extended_results}

\end{document}

%% file: sections/introduction.tex
\section{Introduction}
Windmills tend to be larger than toasters. 
You have probably never entertained that specific comparison, yet you know without much thought that it is true.
If, on the other hand, you were told that Leo is taller than Max, Owen is taller than Zoe, and Max is taller than Owen, you might have to reason step by step to deduce that Leo is taller than Zoe.

These examples seem to rely on different underlying processes, but both can be understood as instances of \emph{transitive inference} (TI), the derivation of an unseen relation ($A \succ C$) from learned ones ($A \succ B$, $B \succ C$) via the transitive property.
Such inferences are a textbook form of logical deduction, long considered a hallmark of rational thought~\citep{InhelderPiaget58}---yet they are exhibited in a remarkably wide range of nonhuman animals, including pigeons~\citep{vonFersenWynneDeliusEtAl91}, corvids~\citep{Paz-y-MinoCBondKamilEtAl04}, rats~\citep{Davis92,DusekEichenbaum97}, and monkeys~\citep{DAmatoColombo90,GazesCheeHampton12,JensenAlkanMunozEtAl17}.
These behaviors may not require the kind of explicit logical deduction typically associated with human syllogistic reasoning, and may instead be explained by simpler processes, such as ones that rely on incrementally building up a continuum of ordered representations~\citep{Wynne95,FrankRudyOReilly03,FrankRudyLevyEtAl05,JensenMunozAlkanEtAl15,Vasconcelos08}.
However, no one doubts that humans \emph{can}, at least under certain circumstances, perform logical deductions over completely novel relations~\citep{DehaeneAlRoumiLakretzEtAl22,Johnson-Laird10,Rips94}.
Here, rather than asking which strategy underlies TI in general, we show that humans solve TI problems in two fundamentally distinct ways---one that relies on ordered internal representations and one that relies on step-by-step logical deduction---and provide a normative account of when each strategy should be used.

These debates are not specific to TI, but reflect a contrast between two longstanding perspectives on human cognitive flexibility~\citep{FodorPylyshyn88,Smolensky88,Sloman96,McClellandBotvinickNoelleEtAl10}.
The \emph{geometric} view emphasizes that generalization is possible when experience is distilled into representations whose geometry mirrors the structure of a domain, allowing novel relations to be read directly from that geometry.
This view is exemplified in work on neural geometry~\citep{BernardiBennaRigottiEtAl20,ChungAbbott21,SorscherGanguliSompolinsky22,NelliBraunDumbalskaEtAl23,KayBidermanKhajehEtAl24}, cognitive maps~\citep{Tolman48,BehrensMullerWhittingtonEtAl18,ConstantinescuOReillyBehrens16,ParkMillerNiliEtAl20,WhittingtonMullerMarkEtAl20,ParkZolfagharRussinEtAl24}, distributional semantics~\citep{LandauerDumais97,RogersMcClelland04,MikolovYihZweig13,Boleda20,GrandBlankPereiraEtAl22}, and representation learning in neural networks more generally~\citep{RumelhartMcClellandPDPResearchGroup86a,BengioCourvilleVincent13,LeCunBengioHinton15}.
The \emph{algebraic} view emphasizes that generalization is possible when knowledge is encoded as abstract, content-independent rules, allowing novel relations to be derived by systematically applying those rules step by step.
This view has been emphasized in formal approaches to language~\citep{Chomsky57, PinkerPrince88}, rule-based or hypothesis-driven learning~\citep{MarcusVijayanBandiRaoEtAl99,Marcus01,GoodmanTenenbaumFeldmanEtAl08,TenenbaumGriffiths01,KempTenenbaum08,PiantadosiTenenbaumGoodman12,TenenbaumKempGriffithsEtAl11}, compositionality~\citep{FodorPylyshyn88,RussinMcGrathWilliamsEtAl24}, and in symbolic models of cognition more generally~\citep{Fodor75,NewellSimon72,Rips94,AndersonBothellByrneEtAl04,LairdNewellRosenbloom87,DehaeneAlRoumiLakretzEtAl22}.
Although these perspectives emerged from distinct research traditions and have historically been framed as rivals~\citep{PinkerPrince88,Smolensky88,McClellandBotvinickNoelleEtAl10,LakeUllmanTenenbaumEtAl17}, we demonstrate that humans and neural networks can flexibly generalize in both ways on TI tasks, and characterize their unique computational advantages and when each is expressed.

Computational models of TI have generally realized one of these generalization strategies or the other, either building up ordered representations~\citep{vonFersenWynneDeliusEtAl91,Wynne95,FrankRudyOReilly03,JensenMunozAlkanEtAl15,CirankaLinde-DomingoPadezhkiEtAl22,KayBidermanKhajehEtAl24,MiconiKay25} or executing a step-by-step reasoning process~\citep{SternbergWeil80,HummelHolyoak01,KumaranMcClelland12b}.
Studying how the two strategies relate and interact, however, requires a single system capable of both.
Although neural networks have typically been associated with the geometric strategy~\citep{FrankRudyOReilly03,LipplKayJensenEtAl24,RussinZolfagharParkEtAl21,NoroozizadehNagarajanRosenfeldEtAl26}, recent models, including large language models (LLMs), exhibit both \emph{in-weight learning} (IWL), in which weights change incrementally with experience, and \emph{in-context learning} (ICL), an emergent ability to make inferences about novel premises provided in context, by re-routing activation dynamics rather than weights~\citep{BrownMannRyderEtAl20,OlssonElhageNandaEtAl22}.
These two forms of learning have been shown to generalize differently~\citep{ChanDasguptaKimEtAl22,LampinenChaudhryChanEtAl25,AnandLeporiMerulloEtAl25,GeertsLiuChanEtAl25,LeporiLinzenYuanEtAl26}.
Furthermore, modern neural networks can increasingly \emph{reason} in context as well, producing chains of thought (CoTs) akin to step-by-step algebraic reasoning~\citep{NyeAndreassenGur-AriEtAl21,WeiWangSchuurmansEtAl22,KojimaGuReidEtAl22,OpenAI24,GuoYangZhangEtAl25}.
The distinction between IWL and ICL has parallels in human learning~\citep{RussinPavlickFrank25,PesnotLerousseauSummerfield26,CollinsFrank12,BotvinickRitterWangEtAl19}: IWL is analogous to incremental synaptic plasticity supporting statistical learning in basal ganglia and hippocampus, whereas ICL is analogous to activation dynamics supporting working memory gating and hierarchical generalization in prefrontal cortex 
\citep{OReillyFrank06a, WangKurth-NelsonKumaranEtAl18, RougierNoelleBraverEtAl05c, CollinsFrank13, RussinOReillyBengio20, OReillyRanganathRussin22, SoniTraylorMerulloEtAl26}.
Thus, modern neural networks may present an opportunity to study geometric and algebraic generalization in a system other than humans.

In this work we develop a theoretical framework that formalizes the distinction between geometric and algebraic generalization, and show that it makes testable predictions about learning, inference, and their underlying mechanisms in humans and neural networks.
We derive our predictions formally, validate them in networks trained from scratch, and confirm them empirically in humans and LLMs evaluated on the same problems.
Our framework unifies disparate phenomena from cognitive science, neuroscience, and machine learning, including symbolic distance effects~\citep{MoyerLandauer67,Potts74}, content effects~\citep{EvansBarstonPollard83,LampinenDasguptaChanEtAl24}, and data-distribution effects~\citep{ChanSantoroLampinenEtAl22,Reddy24}, showing that they can be grounded in a single set of principles.
We also use mechanistic interventions to demonstrate that the two strategies rely on dissociable mechanisms in our neural networks, reproducing findings from analogous neural perturbations across species \citep{FrankOReillyCurran06,WaltzKnowltonHolyoakEtAl99,DusekEichenbaum97}.

%% file: sections/framework.tex
\section{Geometric and algebraic generalization in transitive inference}

\begin{figure}
  \centering
  \includegraphics[width=0.5069\textwidth]{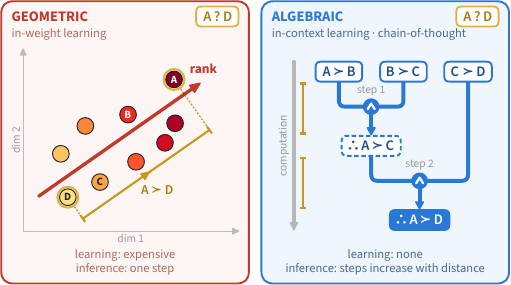}
  \caption{\textbf{Two strategies for transitive inference.}
  The geometric strategy pre-compiles item ranks into ordered representations.
  The algebraic strategy reasons step by step over relations given in context.
  }
  \label{fig:framework}
\end{figure}

In this section we formalize the distinction between the geometric and algebraic strategies (Fig.~\ref{fig:framework}; Appendix~\ref{app:formalization}).
Consider a TI problem in which the given premises are adjacent ($A \succ B$, $B \succ C$, $C \succ D$, \ldots), and the learner must answer which of two items ranks higher (e.g., $B \ ? \ D$). 
In this context, the \textbf{geometric} strategy can be realized by an algorithm that \emph{pre-compiles} (via prior learning) the premises into ordered representations (see Alg.~\ref{alg:geo}), and the \textbf{algebraic} strategy can be realized by an algorithm that \emph{traverses} the premises, applying the transitivity rule step by step from one queried item to the other (see Alg.~\ref{alg:alg}).

This formalization clarifies the distinct computational advantages of each strategy (Fig.~\ref{fig:framework}), yielding behavioral predictions that can be tested by manipulating the query's \emph{symbolic distance} (premises separating the two items; Figs.~\ref{fig:sde}D,E and \ref{fig:content}B,C).
The geometric strategy can only answer queries on familiar items whose relative ranks have been learned (e.g., windmills are larger than toasters), but can do so in a single step by comparing pre-compiled ranks. 
The algebraic strategy can answer \emph{any} query given premises provided in context even after a single shot (e.g., Leo is taller than Max), but requires more computational steps as symbolic distance increases.
A single learner capable of both strategies should therefore deploy the geometric strategy on items familiar from past experience (i.e., learned in weights), and the algebraic strategy when novel relations are given in context.

Because the two strategies use different representations and computations, they generate distinct behavioral predictions.
Under simple noise assumptions, inference under the two strategies exhibits opposite symbolic distance effects (Fig.~\ref{fig:sde}D,E): with greater distance, the geometric strategy's ordered representations are more discriminable (assuming Gaussian noise in learned ranks; see Appendix~\ref{app:formalization}), whereas the algebraic strategy requires a greater number of reasoning steps (and thus opportunities for error; see Appendix~\ref{app:formalization}).
When items are familiar and context is uninformative, the learner should rely on the geometric strategy, resulting in better accuracy with larger distance---the classic \emph{symbolic distance effect} (SDE) that has been observed in many prior TI studies across species~\citep{MoyerLandauer67,Potts74,TrabassoRiley75,Banks77,DAmatoColombo90,vonFersenWynneDeliusEtAl91,FrankRudyLevyEtAl05,JensenAlkanMunozEtAl17,LipplKayJensenEtAl24}.  
Conversely, when items are novel and context is informative, our framework predicts the learner should use the algebraic strategy, requiring longer response times and worse accuracy with larger distance---a \emph{reverse SDE}.
While it would be possible for a learner in this case to use the geometric strategy to pre-compile in-context relations on the fly (in-context representation learning; \citealp{ParkLeeLubanaEtAl25}), our framework predicts that the algebraic strategy should be preferred in this case due to its lower cost (Appendix~\ref{app:formalization}).
In the following, we test these predictions in neural networks trained from scratch, humans, and LLMs, all evaluated on the same TI problems.

%% file: sections/results.tex
\section{Results}

\subsection{Symbolic distance effects}

\begin{figure}[!htbp]
  \centering
  \includegraphics[width=\textwidth]{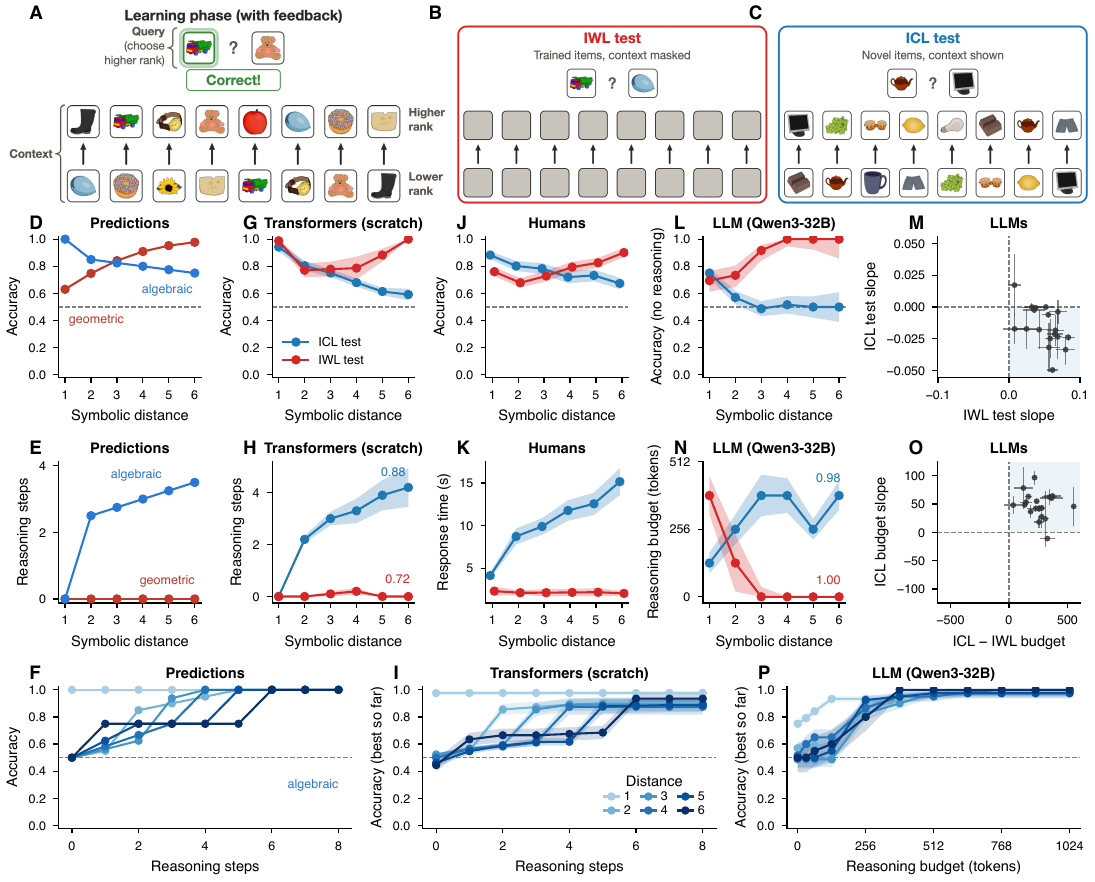}
  \caption{\textbf{Symbolic distance effects (SDE).}
  (\textbf{A})~Learning phase: participants were shown a query above a row of relations in context, and were given feedback.
  (\textbf{B})~IWL test (red): to test whether participants committed ranks of items to memory, we tested trained items with an uninformative context.
  (\textbf{C})~ICL test (blue): to test whether participants could make TIs in context, we tested them on novel items.
  (\textbf{D})~Under simple noise assumptions, the two strategies produce opposite distance effects on accuracy. 
  (\textbf{E})~Expected reasoning steps grow with distance in the algebraic but not the geometric strategy.
  (\textbf{F})~Under the algebraic strategy, accuracy rises with reasoning budget, and longer distances require more steps.
  (\textbf{G})~Transformers trained from scratch exhibited an SDE on the IWL test and a reverse SDE on the ICL test.
  (\textbf{H})~The number of reasoning steps required to reach 90\% of maximum accuracy increased with distance on the ICL, but not the IWL test. 
  (\textbf{I})~ICL test accuracy increased with the number of reasoning steps, and longer distances required more steps.
  (\textbf{J})~Humans exhibited an SDE on the IWL test and a reverse SDE on the ICL test. 
  (\textbf{K})~Human response times increased with distance on the ICL, but not the IWL test. 
  (\textbf{L})~LLMs (Qwen3-32B shown) exhibited an SDE when discriminating familiar relations (country population), and a reverse SDE when tested on novel relations in context.
  (\textbf{M})~Across LLMs, the accuracy increased with symbolic distance on the IWL test and decreased on the ICL test for 15 of 18 models (lower right quadrant).
  (\textbf{N})~When evaluated with chain of thought (CoT), LLMs (Qwen3-32B shown) used more reasoning tokens for longer distances on the ICL test, and used fewer reasoning tokens on the IWL test (country population).
  (\textbf{O})~Across LLMs, the reasoning budget needed was larger on the ICL test than on the IWL test and increased with distance on the ICL test for 18 of 19 models (upper right quadrant).
  (\textbf{P})~The same held on the LLM ICL test as a function of the reasoning budget in tokens (Qwen3-32B shown).
  Panels F, I, and P show one line per symbolic distance (darker for larger distances); I and P show the best accuracy reached at each budget or below.
  Numbers show accuracy at the largest reasoning budget (eight steps; 1,024 tokens) over distances 2--6.
  }
  \label{fig:sde}
\end{figure}

Because the two strategies predict opposite symbolic distance effects (Fig.~\ref{fig:sde}D,E), and strategy choice should depend on \emph{familiarity}, a learner capable of both should show an SDE when familiar orderings are queried and a reverse SDE when inferences must be made on novel premises.
We tested this prediction in neural networks trained from scratch, humans, and LLMs in a setting where correct relations could be derived from relations given in the context (Fig.~\ref{fig:sde}A).
Transformers and humans were trained with feedback on a single ordering of seven items.
Inspired by prior work~\citep{PesnotLerousseauSummerfield26,ChanSantoroLampinenEtAl22,ChanDasguptaKimEtAl22}, we designed tests to isolate each strategy: an \textbf{IWL test} querying pairs of familiar (trained) items without an informative context, and an \textbf{ICL test} querying novel items with the context shown (Fig.~\ref{fig:sde}B,C).

To demonstrate the principles of our framework in a controlled setting where they could be isolated, we first tested these predictions in neural networks trained from scratch, using the transformer architecture \citep{VaswaniShazeerParmarEtAl17}.
To emulate the prior step-by-step reasoning experience of human adults, we pretrained these networks in a setting where they learned (or ``metalearned''; \citealp{WangKurth-NelsonKumaranEtAl18,BinzDasguptaJagadishEtAl24}) how to solve TI problems by reasoning in context.
Models were trained on a distribution of TI problems to generate intermediate reasoning steps before responding (see Alg.~\ref{alg:model_cot}).
After pretraining, the networks were finetuned on a single ordering of seven items, analogous to humans learning a single ordering during a TI experiment.
As predicted, the trained networks exhibited an SDE on the IWL test~\citep{LipplKayJensenEtAl24,Singh26,KayBidermanKhajehEtAl24,GeertsLiuChanEtAl25} but a reverse SDE on the ICL test (Fig.~\ref{fig:sde}G), and required more reasoning steps on more distant queries on the ICL, but not the IWL test (Fig.~\ref{fig:sde}H,I).
Pair-level results and the complete reasoning-step sweeps are reported in Fig.~\ref{fig:app-basic-tests}, and results with other reasoning algorithms in Fig.~\ref{fig:app-cot-algorithms}.

We ran the same experiment with human participants (Experiment 1; $N = 40$): each learned a single ordering with feedback, then completed the IWL test and the ICL test in the same session (order counterbalanced).
As expected, accuracy increased with distance on the IWL test (an SDE; $p < .001$) but decreased with distance on the ICL test (a reverse SDE; $p < .001$), producing a distance-by-test interaction ($p < .001$; Fig.~\ref{fig:sde}J; Tables~\ref{tab:exp1-iwl-accuracy}--\ref{tab:exp1-accuracy-by-test}).
Response times (RTs) also showed the predicted dissociation (distance-by-test interaction: $p < .001$; Fig.~\ref{fig:sde}K; Tables~\ref{tab:exp1-iwl-rt}--\ref{tab:exp1-rt-by-test}): on the ICL test RT increased steeply with distance ($p < .001$), whereas on the IWL test it was shorter and did not depend significantly on distance ($p = .442$).
Learning-phase and pair-level results are reported in Figs.~\ref{fig:app-human-learning} and~\ref{fig:app-human-pairs}.
Overall, these results are consistent with the hypothesis that humans use a geometric strategy in which relations are read from pre-compiled ordered representations when tested on familiar items, and use an algebraic strategy that makes step-by-step inferences when tested on novel items.

The above simulations explicitly trained transformers to solve TI tasks. 
To determine whether the same principles arise spontaneously and at scale, we tested a variety of LLMs on the same task (Methods \ref{sec:methods_llms}). 
Previous studies have reported a classic SDE in LLMs given TI problems in context~\citep{WuDeng25,RaneKirkmanToddEtAl25}, and accuracy that declines with distance on other relational problems~\citep{LiuPriorBalasubramaniamEtAl25}, but to our knowledge, none has investigated whether LLMs produce a reverse symbolic distance effect when reasoning on TI problems.
Our IWL test queried five real-world dimensions (e.g., countries ordered by population) with no relations shown; the ICL test queried novel arbitrary relations (names of people ordered by height) with adjacent relations shown in context. 
As predicted, LLMs consistently showed a classic SDE on the IWL test due to the semantic knowledge acquired in their weights during pretraining \citep{ShahMarupudiKoenenEtAl23,El-ShangitiHiraokaAlQuabehEtAl25,HeinzerlingInui24,ShakiLaMalfaWooldridgeEtAl25}, and a reverse SDE on the ICL test (15 of 18 models; Fig.~\ref{fig:sde}L,M).
These models performed better on the ICL test when they reasoned step by step before producing a response (with CoT), and the reasoning budget required (i.e., tokens required to reach 90\% of maximum accuracy) was generally higher on the ICL test compared to the IWL test, and increased with distance on the ICL test (18 of 19 models; Fig.~\ref{fig:sde}N--P). 
Appendix \ref{app:extended-results-llm} reports variation observed across models and dimensions.
Overall, these results are consistent with predictions, showing that even though they were not explicitly trained to do so, LLMs in general exhibit the same signatures of the geometric and algebraic generalization strategies we observed in humans. 

Taken together, these results confirm our prediction in humans and in both custom and large-scale neural networks that the classic SDE, a signature of the geometric strategy, emerges when relying on IWL, and can reverse if instead an algebraic strategy is flexibly deployed to handle novel premises in context.

\subsection{Content effects}

\begin{figure}[!htbp]
  \centering
  \includegraphics[width=0.99\textwidth]{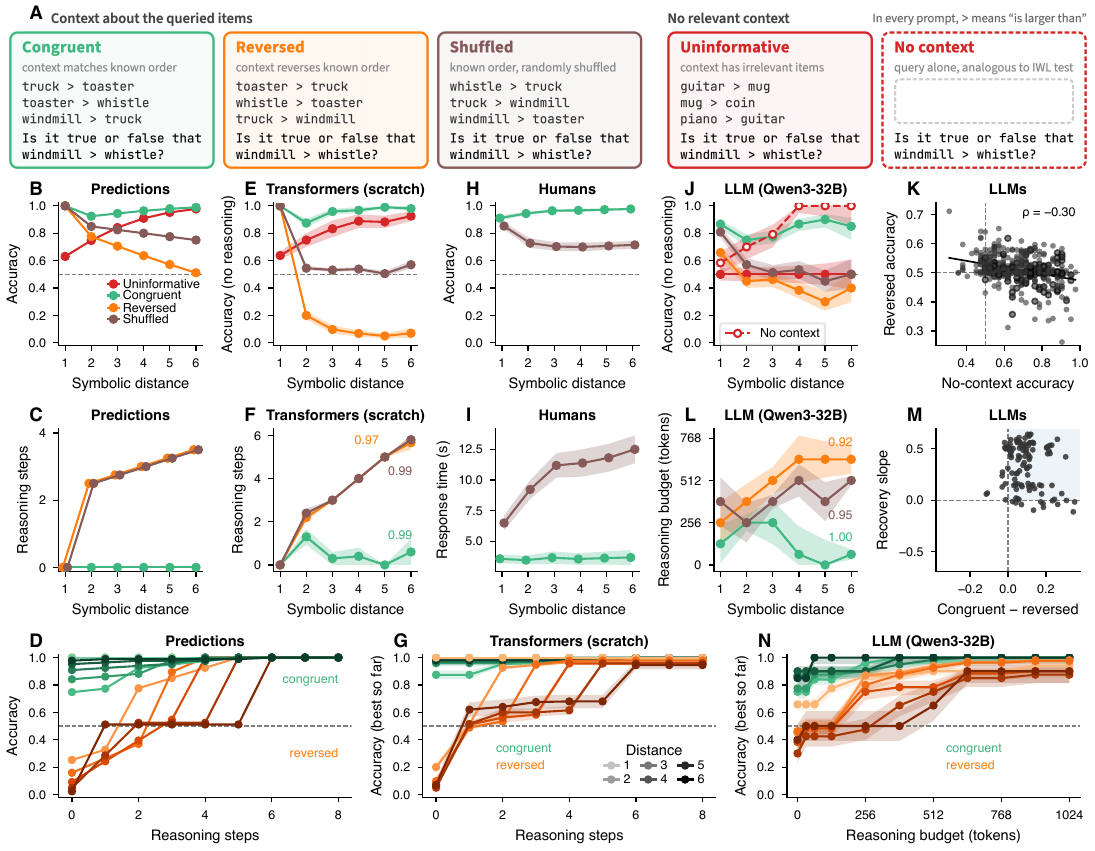}
  \caption{\textbf{Content effects in transitive inference.}
  (\textbf{A})~Context conditions. The ordering shown in the context was either congruent, reversed, or shuffled relative to prior knowledge, or was uninformative.
  (\textbf{B})~Content effects emerge in a system capable of both strategies.
  When context is congruent, the system performs well regardless of distance, but when it is reversed or shuffled, accuracy falls with distance.
  When context is uninformative (IWL test), the system shows an SDE.
  (\textbf{C})~Expected reasoning steps grow with distance on reversed and shuffled trials (identical; offset for visibility) but not on congruent trials.
  (\textbf{D})~On reversed trials, accuracy rises with reasoning steps; on congruent trials it stays near ceiling.
  (\textbf{E})~Transformer accuracy by symbolic distance with no reasoning steps.
  In the IWL test, an SDE was observed.
  Accuracy was near ceiling when the context was congruent but fell below chance when it was reversed and near chance when it was shuffled.
  (\textbf{F})~The number of reasoning steps required to reach 90\% of maximum accuracy increased with distance on reversed and shuffled trials, but not on congruent.
  (\textbf{G})~Transformer accuracy on reversed trials recovered with reasoning steps, with longer distances requiring more steps; congruent accuracy remained high.
  (\textbf{H})~Humans exhibited an SDE in the congruent condition and a reverse SDE in the shuffled condition (uninformative and reversed contexts were not tested).
  (\textbf{I})~Human response times increased with distance in the shuffled, but not the congruent condition.
  (\textbf{J})~LLMs showed the predicted content effects without reasoning (Qwen3-32B, US state population).
  (\textbf{K})~Across LLMs and sets of items (one point per model $\times$ set), accuracy in the reversed condition decreased as accuracy in the no-context condition increased, suggesting a stronger prior increases interference.
  (\textbf{L})~With reasoning, the budget required to reach 90\% of maximum accuracy increased with distance under a reversed and shuffled context (Qwen3-32B, US state population).
  (\textbf{M})~Across LLMs and real-world dimensions (one point per model $\times$ dimension), models showed content effects (higher accuracy on congruent than reversed), and reasoning effects (accuracy on reversed trials increased with reasoning; upper right quadrant).
  (\textbf{N})~LLM accuracy on reversed trials recovered with reasoning; longer distances required more reasoning tokens (Qwen3-32B, US state population).
  Panels G and N show the best accuracy reached at or below each budget.
  Numbers show accuracy at the largest reasoning budget (eight steps; 1,024 tokens) over distances 2--6.}
  \label{fig:content}
\end{figure}

The tests above were designed to isolate each of the two strategies.
However, when familiar items appear alongside informative contextual information, both strategies are available, and conflicts can arise.
The two strategies reinforce one another when the information given in context is \emph{congruent} with prior knowledge (e.g., ``windmills are larger than trucks''),
but can compete when it is \emph{incongruent}---when relations are reversed or shuffled (e.g., ``trucks are larger than windmills'').
The resulting \emph{content effects} have been observed across many domains in human reasoning~\citep{EvansBarstonPollard83,WasonShapiro71,ChengHolyoak85,Byrne89, Wilkins28, OaksfordChater94,Andrews10} and have recently been shown to emerge in LLMs~\citep{LampinenDasguptaChanEtAl24,BertolazziPezzelleBernardi26,LiuPriorBalasubramaniamEtAl25}.
We hypothesized that these content effects arise from competition between the two strategies, and further predicted that prior knowledge should dominate direct responses, whereas responses should become context-informed only after sufficient reasoning, with larger distances requiring more reasoning (Fig.~\ref{fig:content}B--D).

We first tested this prediction in transformers trained from scratch, using a pretraining distribution in which one ordering was very frequent, supporting the formation of a content prior (analogous to real-world cases where a consistent ``ordering,'' such as the relative sizes of common objects, is frequently observed).
To determine how this content prior interacted with contextual evidence, we tested the network in the presence of contextual information that was either \emph{congruent},  \emph{reversed}, or \emph{shuffled} relative to the prior, as well as in the usual IWL test, where context was completely uninformative (Fig.~\ref{fig:content}A).

As predicted, the model's direct responses (i.e., without reasoning) were strongly shaped by prior knowledge (Figs.~\ref{fig:content}E and~\ref{fig:app-content-grid}).
On the IWL test, the model exhibited a classic SDE, with accuracy increasing with symbolic distance.
A congruent context reinforced prior knowledge, increasing accuracy relative to the IWL test at every symbolic distance.
When the ordering given in context was reversed and the model answered without reasoning, prior knowledge still drove direct prepotent responses, yielding below-chance agreement with the context across non-adjacent distances.
With a shuffled context, in which the queried relation agreed with the prior on half of the trials, direct accuracy was near chance at every distance above one.
However, when the network was allowed to reason through the relations given in context, it could progressively overcome this interference, shifting responses toward the context-appropriate inference (Figs.~\ref{fig:content}G and~\ref{fig:app-content-grid}B).
Furthermore, as predicted, the number of reasoning steps required to overcome the prior increased with symbolic distance under both reversed and shuffled contexts, whereas congruent contexts required almost none (Fig.~\ref{fig:content}F,G).
Thus, the model could leverage the algebraic strategy to overcome content interference, and in doing so, exhibited its characteristic reverse SDE on incongruent trials.
But in the absence of such interference, the geometric strategy was sufficient and so no such effect was observed.

Next, we tested the same content effects in humans with an analogous task (Experiment 2; Methods~\ref{sec:methods_exp2}).
Each question showed eight statements ordering nine animals by weight (e.g., ``The horse is heavier than the dog.''), listed in a scrambled order, and asked which of two animals was heavier according to the statements.
Between subjects, the statements were either all congruent with their real-world ordering ($n = 49$) or followed a shuffled ordering ($n = 50$).
As expected, accuracy was lower in the shuffled than in the congruent condition ($p < .001$; Table~\ref{tab:exp2-accuracy}), particularly when the queried relation conflicted with the real-world ordering (Fig.~\ref{fig:app-human2-agree-conflict}). 
Notably, symbolic distance had opposite effects across conditions (condition-by-distance interaction: $p < .001$; Fig.~\ref{fig:content}H): accuracy increased with distance in the congruent condition (an SDE; $p < .001$) but decreased with distance in the shuffled condition (a reverse SDE; $p < .001$).
RTs showed the same dissociation (condition-by-distance interaction: $p < .001$; Fig.~\ref{fig:content}I; Table~\ref{tab:exp2-rt}): in the shuffled condition RT increased with distance ($p < .001$), whereas in the congruent condition it was shorter and did not increase with distance ($p = .050$).
These results are consistent with our hypothesis that participants would leverage the geometric strategy when the context was congruent with their prior knowledge, and the algebraic strategy when it was incongruent with it.

Finally, we tested LLMs on TI problems involving multiple real-world dimensions, varying whether the relations shown in context were congruent, reversed, or randomly shuffled with respect to their real-world order (Fig.~\ref{fig:content}A).
As predicted, direct answers were again shaped by the content prior (Fig.~\ref{fig:content}J): in 81 of 110 model--dimension combinations, congruent accuracy exceeded reversed (Appendix \ref{app:extended-results-llm}).
Furthermore, poorer performance on reversed trials was greater in settings in which the model had a stronger prior (Fig.~\ref{fig:content}K), consistent with the idea that poor performance was due to interference from the prior. 
Importantly, however, as in the transformers trained from scratch, LLM performance on reversed-context problems recovered with step-by-step reasoning (Fig.~\ref{fig:content}L--N): in 98 of 110 model--dimension combinations, reversed-context accuracy increased with reasoning budget (Fig.~\ref{fig:content}M; Figs.~\ref{fig:app-llm-recovery-qwen3}--\ref{fig:app-llm-recovery-llama-olmo}).

Together, these results show that our framework can also account for the content effects observed in humans~\citep{EvansBarstonPollard83,EvansCurtis-Holmes05,SchubertFerreiraMataEtAl21} and in LLMs~\citep{LampinenDasguptaChanEtAl24}.
When a system has encoded knowledge in its representational geometry but can also execute an algebraic strategy via in-context reasoning, a natural competition between the two emerges: interference hurts performance on incongruent trials, but can be overcome via step-by-step reasoning.

\subsection{Independent mechanisms underlie the geometric and algebraic strategies}

\begin{figure}[!t]
  \centering
  \includegraphics[width=\textwidth]{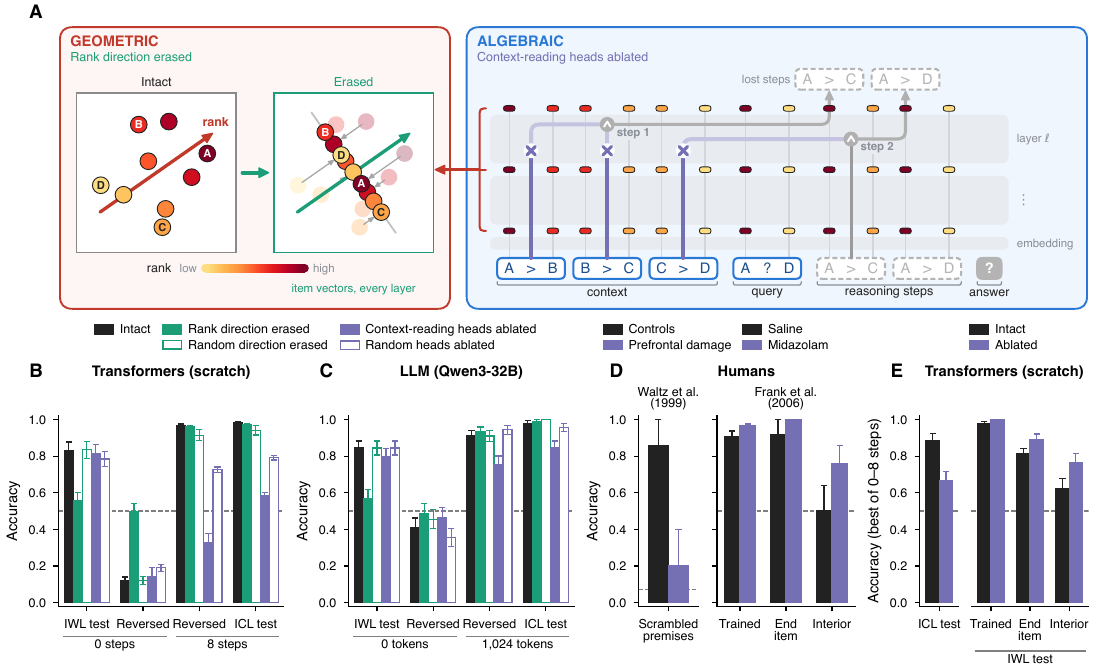}
  \caption{\textbf{Mechanisms underlying the geometric and algebraic strategies.}
  (\textbf{A})~Interventions. The geometric strategy was targeted by erasing the rank direction in the model's representations; the algebraic strategy was targeted by ablating context-reading heads.
  (\textbf{B})~In our transformers, erasing the rank direction impaired direct IWL responses and removed the bias on reversed trials, but left accuracy after reasoning intact. Ablating two context-reading heads impaired accuracy after reasoning but not direct responses. 
  (\textbf{C})~The same interventions in an LLM (Qwen3-32B, US state population; 32 context-reading heads).
  Open bars show controls; accuracy is pooled over distances 2--6.
  (\textbf{D})~Patients with prefrontal damage could not order names from scrambled premises~\citep{WaltzKnowltonHolyoakEtAl99}. Humans given midazolam were more accurate on novel pairs~\citep{FrankOReillyCurran06}.
  (\textbf{E})~Transformers that reasoned during finetuning, with two context-reading heads per network intact or ablated during finetuning and at test.
  Error bars: $\pm$1~SE; dashed lines indicate chance.}
  \label{fig:mechanisms}
\end{figure}

The above experiments show that human and neural network behavior can be consistent with the geometric and algebraic strategies, but do not directly investigate their underlying mechanisms.
We hypothesized that the geometric strategy would be reflected in network weights encoding the ordering, but that algebraic reasoning would depend on step-by-step attentional computations guiding activations toward the solution consistent with the context \citep{JinCaoYuanEtAl24, OrtuJinDoimoEtAl24a, YuMerulloPavlick23}. 
To test this, we intervened on each strategy (Fig.~\ref{fig:mechanisms}A) in the transformers and in an LLM (Qwen3-32B).
To target the geometric strategy, we identified and erased the \emph{rank direction}, the direction in an item's representation that encodes the familiar ordering, at every layer~\citep{BelroseSchneider-JosephRavfogelEtAl23,GurneeTegmark24,HeinzerlingInui24,El-ShangitiHiraokaAlQuabehEtAl25}.
To target the algebraic strategy, we ablated \emph{context-reading heads}, the attention heads that attended most to the relevant relations in context while the model reasoned.

In our transformers, erasing the rank direction substantially reduced IWL accuracy and removed the content-prior bias on reversed trials, but left accuracy after reasoning largely intact (Fig.~\ref{fig:mechanisms}B).
These results are similar to ones from neuroscience showing that lesions to hippocampus can impair transitive inferences about learned relations in rodents~\citep{DusekEichenbaum97}.
Conversely, ablating two context-reading heads per network had the opposite effect: accuracy after reasoning fell sharply on reversed trials and the ICL test, whereas IWL test accuracy was unchanged.
The LLM showed the same pattern (Fig.~\ref{fig:mechanisms}C): erasing the rank direction reduced IWL accuracy without impairing accuracy after reasoning, and ablating 32 context-reading heads reduced accuracy after reasoning while having little effect on IWL accuracy.
These results resemble ones from patients with damage to prefrontal cortex, who could not complete inferences on premises given in scrambled order (Fig.~\ref{fig:mechanisms}D; \citealp{WaltzKnowltonHolyoakEtAl99}).
Additional analyses are reported in Figs.~\ref{fig:app-representation} and~\ref{fig:app-llm-representation}.
Together, these results suggest that the mechanisms underlying the two strategies are largely independent: while the geometric strategy relies on rank information encoded in representational geometry, the algebraic strategy relies on attention heads that read contextual information during reasoning and focus network activity on that information to overcome the prior.

Even though the mechanisms underlying the two strategies appear largely independent, they may still compete during learning.
\citet{FrankOReillyCurran06} found that when humans took midazolam, a drug that transiently impairs explicit memory while leaving implicit memory intact, before learning a TI task, their performance on novel pairs on a subsequent test \emph{improved} ($p=.03$, $n=11$ per drug; Fig.~\ref{fig:mechanisms}D).
This suggests that in humans, reliance on reasoning over explicit memories during learning may reduce implicit learning that would otherwise occur.
To test the analogous hypothesis in our networks, we ablated context-reading heads while the networks reasoned during finetuning on a fixed ordering (simulating midazolam's inhibition of the explicit strategy during learning).
Similar to midazolam's effect in humans, this ablation \emph{increased} performance on the subsequent IWL test on non-adjacent pairs, while degrading performance on the ICL test (Fig.~\ref{fig:mechanisms}E).
This suggests that although their mechanisms may be largely independent, the two strategies compete during learning to reduce errors~\citep{RussinPavlickFrank25}.

\subsection{Benefits and costs of each strategy depend on data distribution}

\begin{figure}[!htbp]
  \centering
  \includegraphics[width=\textwidth]{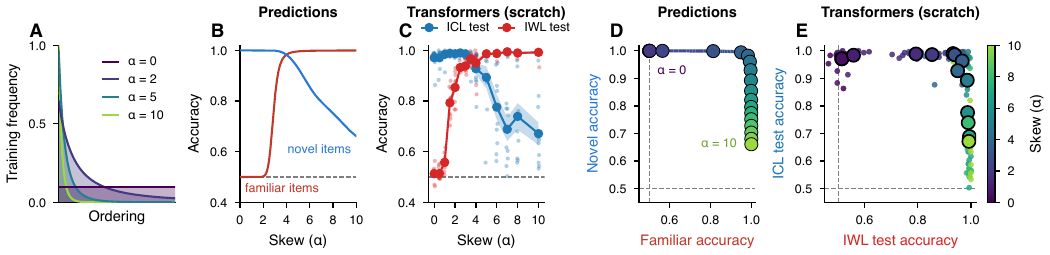}
  \caption{\textbf{Data-distribution effects.}
  (\textbf{A})~Schematic of the training frequency of orderings under Zipf distributions with different skew ($\alpha$).
  (\textbf{B})~Learners should shift from geometric to algebraic as the data distribution becomes more diverse (smaller $\alpha$).
  (\textbf{C})~Final test accuracy versus skew. As data diversity decreases (i.e., $\alpha$ increases), final ICL test accuracy decreases, and final IWL test accuracy increases. 
  (\textbf{D})~Predicted accuracy on novel versus familiar items as skew ($\alpha$) varies.
  (\textbf{E})~ICL versus IWL test accuracy per seed. As skew ($\alpha$) increases, ICL accuracy decreases and IWL accuracy increases, with a large range where models perform well on both tests.}
  \label{fig:diversity}
\end{figure}

In the simulations above, we showed that a single network can acquire both strategies through pretraining.
Prior work has shown that the diversity of the training distribution determines whether transformers and humans acquire an IWL or ICL strategy~\citep{ChanSantoroLampinenEtAl22,Reddy24,RaventosPaulChenEtAl23,ChanChenGyorgyEtAl25,PesnotLerousseauSummerfield26,KuGriffithsChan26}.
Here we show that these results extend to the benefits arising from step-by-step reasoning (CoT), and, moreover, that this is predicted by a simple model of each strategy's \emph{learning and inference costs} (Fig.~\ref{fig:diversity}B,D and Appendix~\ref{app:formalization}).
Pre-compiling an ordering is worthwhile only if it will be experienced consistently (e.g., the sizes of common objects), so a learner should prefer the geometric strategy when training is concentrated on a few such orderings (i.e., when training data is highly skewed).
The algebraic strategy, on the other hand, is costly at inference but applies to any ordering (e.g., arbitrary newly learned heights), so should be preferred when many conflicting orderings are seen during training (e.g., when learning algebra in math class it is not a good idea to try to memorize what ``x'' is, as it can change on every new problem).

Because it would be difficult to perform experiments manipulating the pretraining distribution of LLMs, or the developmental environments of humans, we tested these predictions only in transformers trained from scratch.
Networks were pretrained on 100 orderings of the same items, each indicated by a distinct cue, where their frequencies followed a Zipf distribution with skew ranging from uniform ($\alpha = 0$) to highly skewed ($\alpha = 10$; Fig.~\ref{fig:diversity}A).
After pretraining, we applied the same IWL and ICL tests (Fig.~\ref{fig:sde}B,C).

More diverse training distributions drove networks to become less accurate on the IWL test but more accurate on the ICL test (Figs.~\ref{fig:diversity}C,E and \ref{fig:app-skew-learning}).
Across a broad intermediate range ($\alpha \in [2,4]$), however, the same networks performed well on both tests, memorizing frequent orderings while also generalizing in context to novel ones (Fig.~\ref{fig:diversity}E).
To isolate the effect of conflict, we finetuned networks on 1 to 10 orderings comprising either conflicting orderings of the same items or separate orderings of disjoint items.
IWL test accuracy was more sensitive to conflict than to the total number of orderings alone: accuracy declined more steeply across conflicting orderings than across disjoint orderings (Fig.~\ref{fig:app-orderings}B).

Together, these results support the predicted cost tradeoff: networks acquired the geometric strategy when a few orderings were frequent, the algebraic when orderings were diverse, and acquired both across an intermediate regime.
Thus, repeated exposure incentivizes use of pre-compiled representational geometry, while varied experience encourages development of a more flexible algebraic strategy.

\FloatBarrier

%% file: sections/discussion.tex
\section{Discussion}

We have shown that both humans and neural networks, including LLMs, are capable of two fundamentally different forms of inference. 
Geometric inference works via incremental weight changes that build internal representations whose geometry reflects a domain's relational structure, allowing generalization to unseen relations through direct representational comparison. 
Algebraic inference uses selective-attention mechanisms acquired through diverse experience to derive new relations step by step from information available in context, akin to working-memory-based inferences thought to unfold via activation dynamics in prefrontal cortex~\citep{OReillyFrank06a,RougierNoelleBraverEtAl05c,KrieteNoelleCohenEtAl13a,WaltzKnowltonHolyoakEtAl99}.
These two forms of generalization complement each other: the geometric strategy crystallizes knowledge and makes subsequent inference cheap, while the algebraic strategy enables fluid adaptation even when no previous knowledge is available, at the cost of additional computation.
Our framework demonstrates how these two strategies can coexist and interact within a single system, providing a unified account of three disparate sets of phenomena: symbolic distance effects, content effects, and data-distribution effects. 

The distinct representational and computational requirements of the two strategies explain the symbolic distance effect that has been observed across species~\citep{DAmatoColombo90,JensenAlkanMunozEtAl17,Vasconcelos08}, but also predict that this effect can be \textit{reversed} when making \emph{de novo} inferences about novel premises ~\citep{SedekvonHecker04,Barrouillet96,vonHeckerMullerKirianDillEtAl23}.
Our results from both humans and LLMs provide direct evidence of this reversal (Fig.~\ref{fig:sde}), thus demonstrating the circumstances under which TI resembles classical logical deduction~\citep{InhelderPiaget58,Wright12}.
When relying on prior knowledge in the IWL test, humans and neural networks exhibited the classic SDE. 
Mechanistic interventions revealed in the neural networks that this prior knowledge was encoded in representational geometry such that larger distances were easier to discriminate.
But when forced to solve \emph{de novo} TI problems on the ICL test, both humans and networks exhibited the reverse SDE, even though in principle it was possible to use the geometric strategy (by pre-compiling ranks in context before answering the query, which would produce a classic SDE).
This finding was predicted by our framework (Appendix~\ref{app:formalization}), but may seem especially surprising for the LLMs, which are forced to process every relation given in context in their initial forward pass (before CoT), and have been shown to learn sophisticated representations in context \citep{ParkLeeLubanaEtAl25}.
We speculate that LLMs nevertheless show a reverse SDE because their fixed depth limits the serial computation possible in a single forward pass~\citep{MerrillSabharwal24,LiLiuZhouEtAl24}, or because a single-pass ICL strategy that accommodates larger symbolic distances is difficult to (meta)learn~\citep{WenZhangLinEtAl25,AbbeBengioLotfiEtAl24,WiesLevineShashua23a}, as we observed in networks trained from scratch (Fig.~\ref{fig:app-cot-proportion}). 
Interventions showed that step-by-step reasoning on these problems required selective-attention mechanisms to guide activation dynamics toward contextual information.
Thus, despite their different architectures and learning histories, both humans and neural networks appear to converge on the same two strategies~\citep{SripadaLewis26,WebbHolyoakLu23,McGrathRussin24,RussinPavlickFrank25,LanHuntSummerfield25}: learned representational geometry supports inference about familiar relations, while step-by-step reasoning enables inference about novel ones.

These findings bear on a longstanding debate about whether TI in humans and other animals involves explicit logical deduction or more primitive incremental learning~\citep{DeSotoLondonHandel65,BryantTrabasso71,vonFersenWynneDeliusEtAl91,Vasconcelos08,LazarevaGazesElkinsEtAl20,Trabasso77,AcunaSanesDonoghue02,JensenTerraceFerrera19}.
In our framework, the reverse SDE arises from the algebraic strategy, and the classic SDE from the geometric strategy.
To our knowledge, no nonhuman animal has shown a reverse SDE, so future work could test whether the algebraic strategy is unique to humans~\citep{DehaeneAlRoumiLakretzEtAl22}, or is shared with other species such as nonhuman primates~\citep{CantlonPiantadosi24}.
The geometric strategy, by contrast, can be realized either by a simple associative mechanism~\citep{Wynne95,CirankaLinde-DomingoPadezhkiEtAl22} or by richer relational coding~\citep{BehrensMullerWhittingtonEtAl18,NelliBraunDumbalskaEtAl23,KayBidermanKhajehEtAl24}.
Which of these supports TI in nonhuman animals has been debated~\citep{vonFersenWynneDeliusEtAl91,Wynne95,GazesCheeHampton12,LazarevaGazesElkinsEtAl20,DusekEichenbaum97}.
In our networks the geometric strategy is likely closer to relational coding~\citep{NoroozizadehNagarajanRosenfeldEtAl26}, but a conclusive test may require TI paradigms with two or more dimensions~\citep{ParkMillerNiliEtAl20,ParkZolfagharRussinEtAl24}.

Our findings also demonstrate how the two strategies can interact with one another in a system capable of both, cooperating when information in context is congruent with prior knowledge, and competing when it is incongruent.
This framework is consistent with a dual-process account of TI~\citep{Wright01,FrankRudyLevyEtAl05,FrankOReillyCurran06,WendelkenBunge10}, which posits two separate routes for making transitive inferences---for example, an ``analytic'' route that relies heavily on working memory and prefrontal cortex and is especially engaged in novel situations, and an ``intuitive'' route that relies heavily on long-term memory and hippocampus and is especially engaged when information has been memorized~\citep{Wright12}.
Our framework extends this dual-process proposal by specifying an algorithmic account of how both processes can be implemented in a single neural network (Fig.~\ref{fig:mechanisms}), with the intuitive route accomplished by representational geometry learned via IWL, and the analytic route accomplished by a step-by-step reasoning (CoT) procedure operating over information available in context.

Our framework also makes specific predictions about what should happen when the two strategies come into conflict, thereby explaining the content effects observed in both human and LLM reasoning~\citep{EvansBarstonPollard83,EvansCurtis-Holmes05,Lupyan13,StuddifordLupyan26,HuAngstadtStorksEtAl26, LampinenDasguptaChanEtAl24, KimValentinoFreitas25, HuLeporiFranke25}.
In TI, our transformers, human participants, and the LLMs all exhibited content effects, performing better on congruent trials where context agreed with prior knowledge than on incongruent trials where these came into conflict (Fig.~\ref{fig:content}).
Because the two strategies rely on relatively independent mechanisms, we predicted that step-by-step reasoning should overcome the prior encoded in representational geometry, with larger symbolic distances requiring more steps.
This prediction was confirmed behaviorally in all three systems (Fig.~\ref{fig:content}), and mechanistically in the neural networks (Fig.~\ref{fig:mechanisms}).
These results relate to prior work identifying separate mechanisms for memorized versus in-context answers in LLMs~\citep{YuMerulloPavlick23,JinCaoYuanEtAl24,OrtuJinDoimoEtAl24a,WuWangXiaoEtAl25,KahardiprajaAchtibatWiegandEtAl25}, and to neural network models of cognitive control \citep{CohenDunbarMcClelland90,OReillyFrank06a,OReillyNairRussinEtAl20,CollinsFrank13}.
Our framework also explains why imposing a time limit on participants increases belief bias in syllogistic reasoning~\citep{EvansCurtis-Holmes05}.
However, belief bias has been observed even when participants take their time and reason aloud from the premises~\citep{EvansBarstonPollard83}, suggesting that content can also interfere with the reasoning process itself.
This may occur because the reasoning process has been metalearned, and is therefore sensitive to whether problems are far outside its training distribution (Fig.~\ref{fig:app-content-grid}).
If true, then experience with such conflicts should improve performance, as has indeed been observed in humans~\citep{Ball13}.
Overall, our framework offers a mechanistic account of why content effects arise, and why reasoning can sometimes overcome them.
 
Because the geometric strategy is implemented via IWL and the algebraic strategy proceeds via a form of ICL (CoT), our framework also connects to and extends existing work from machine learning to predict the conditions under which each strategy can be acquired.
In particular, transformers have been shown to develop ICL abilities when their training data is diverse, but to rely on IWL otherwise~\citep{ChanSantoroLampinenEtAl22,SinghChanMoskovitzEtAl23,Reddy24,RaventosPaulChenEtAl23,ChanChenGyorgyEtAl25,WurgaftLubanaParkEtAl25b}.
This pattern has also been observed in humans~\citep{PesnotLerousseauSummerfield26,RavivLupyanGreen22}.
We observed the same dependence on training diversity (Fig.~\ref{fig:diversity}), extending prior work to a setting where the ICL strategy required CoT reasoning steps rather than being accomplished in a single forward pass.
Furthermore, our framework grounds this dependence in the same learning and inference costs that predict the symbolic distance effects (Fig.~\ref{fig:framework}), bridging research from machine learning and cognitive science.
Because ICL is itself (meta)learned, our framework is also consistent with the proposal that sophisticated human reasoning abilities are metalearned over the course of development rather than innately specified~\citep{WangKurth-NelsonKumaranEtAl18,BotvinickRitterWangEtAl19,BinzDasguptaJagadishEtAl24,LakeBaroni23}, a hypothesis that, while difficult to test in practice, may make specific empirical predictions~\citep{RussinEtAl2024,RussinMcGrathWilliamsEtAl24}. 
Our framework thus not only characterizes how each strategy operates, but also explains why certain kinds of experiences give rise to each.

We assume that a learner capable of both strategies will select between the two according to their relative learning and inference costs (Fig.~\ref{fig:diversity}B,D, Appendix~\ref{app:formalization}). 
Similar tradeoffs have been proposed in many areas of cognitive science, including in the context of model-based versus model-free reinforcement learning \citep{DawNivDayan05,KoolGershmanCushman17a}, the automatization of skills \citep{Logan88}, amortized inference \citep{GershmanGoodman14,DasguptaGershman21}, and resource-rational analysis \citep{LiederGriffiths20}.
In particular, the successor representation, which caches multi-step relations among states so that they do not need to be computed at inference time, is closely related to the geometric strategy we discuss here, and has similar tradeoffs \citep{Dayan93a,StachenfeldBotvinickGershman17d,MomennejadRussekCheongEtAl17b,RussekMomennejadBotvinickEtAl17,Momennejad20a,BrunecMomennejad21,SonBhandariFeldmanHall23,SonCarstensenFavila26,LehnertLittmanFrank20}.
Because our transformers were never explicitly trained to trade off the cost of reasoning against accuracy, we did not probe their ability to make this tradeoff for themselves, and instead simply characterized the relationship between accuracy and reasoning budget (Fig.~\ref{fig:sde}H,I).
However, humans and LLMs explicitly trained to reason are capable of making such tradeoffs, and appear to do so in similar ways \citep{deVardaDEliaKeanEtAl25,LiuGengWuEtAl25}.
An important direction for future work is to determine the extent to which shared costs shape these tradeoffs in both humans and LLMs.

Theories of human cognitive flexibility and generalization have historically divided into those emphasizing structured distributed representations~\citep{McClellandBotvinickNoelleEtAl10,RogersMcClelland04,Sloman93,BehrensMullerWhittingtonEtAl18,OReillyRanganathRussin22} and those emphasizing explicit reasoning with abstract or symbolic rules~\citep{NewellSimon72,NewellSimon76,Braine78,LairdNewellRosenbloom87,AndersonBothellByrneEtAl04,FodorPylyshyn88,Marcus01,SackurDehaene09,FeimanModyCarey22c,MandelbaumDunhamFeimanEtAl22a,DehaeneAlRoumiLakretzEtAl22,Quilty-DunnPorotMandelbaum23}.
Our framework is consistent with the idea that these two views need not be seen as alternatives \citep{Smolensky88,Sloman96,RussinOReillyBengio20,SmolenskyMcCoyFernandezEtAl22c,McGrathRussinPavlickEtAl23,McGrathRussinPavlickEtAl24,GriffithsLakeMcCoyEtAl26}, and can instead be understood as characterizing two generalization strategies with different advantages.
Furthermore, our results are consistent with recent evidence that symbolic or rule-like mechanisms can emerge in both artificial and biologically inspired neural networks, even when these networks are not explicitly trained to develop them \citep{RougierNoelleBraverEtAl05c,KrieteNoelleCohenEtAl13a,CollinsFrank13,RussinJoOReillyEtAl,WebbSinhaCohen21a,WebbFranklandAltabaaEtAl24,RussinMcGrathWilliamsEtAl24,YangCampbellHuangEtAl25,ToddBrinkmannGandikotaEtAl26}.

%% file: sections/methods.tex
\section{Methods}
\label{sec:methods}

\subsection{Transformers trained from scratch}

\paragraph{Task and training data.}
Each sequence contained a cue token that indexed the ordering, a query, a shuffled and randomly oriented set of adjacent relations specifying that ordering, and an answer.
Sequences contained nine ordered items, but the two endpoints were never queried, yielding seven queryable ranks and symbolic distances $d=1,\ldots,6$.
On CoT trials, the model was trained to produce the output from a predefined reasoning algorithm before answering. 
On non-CoT (direct) trials, it was trained to give the answer immediately.

\paragraph{Architecture and optimization.}
We trained decoder-only causal transformers from scratch~\citep{VaswaniShazeerParmarEtAl17}.
Each transformer had four layers, eight attention heads, a residual width of 256, a feedforward width of 1,024, rotary position embeddings~\citep{SuAhmedLuEtAl24}, pre-layer normalization~\citep{XiongYangHeEtAl20}, tied input and output embeddings~\citep{PressWolf17}, and no dropout.
Models were trained with Adam~\citep{KingmaBa15} for 100,000 pretraining steps (batch size 64, learning rate 0.001, no weight decay) and evaluated using greedy decoding with zero to eight generated reasoning steps.
Every experiment ran 10 independently initialized networks.

\paragraph{Symbolic distance effects experiments.}
During pretraining, each of 100 training cue tokens deterministically selected an ordering over items sampled from a pool of 300, while 16 further reserved cues were never trained; every pretraining trial carried a teacher-forced reasoning chain, and symbolic distances were sampled uniformly from one to six.
We then finetuned each network for 1,600 gradient steps at a learning rate of 0.0001 on adjacent queries from one fixed seven-item ordering, supervising on answers only.
To avoid catastrophic forgetting~\citep{McCloskeyCohen89}, 10\% of finetuning trials were replaced by fresh pretraining trials (Fig.~\ref{fig:app-finetune-replay}), and 1 in 10 of the remaining trials showed an uninformative context of other items in place of the fixed ordering, matching the masked-context share of the human learning phase.
The IWL test queried items from the finetuned ordering under an uninformative context; the ICL test queried novel items, outside the finetuned ordering and its distractors, whose ordering was shown in context under a reserved cue.
In every test condition, a response was scored as correct when it agreed with the condition's reference ordering: the finetuned ordering when the context was uninformative, and the ordering shown in context otherwise.
Each test condition comprised all 42 ordered pairs of the seven queryable ranks, each presented with 10 independently sampled contexts.
ICL accuracy was averaged over one to five generated reasoning steps while IWL accuracy was measured with no intermediate steps (Fig.~\ref{fig:sde}G).
To measure the number of reasoning steps required at each symbolic distance, we took the smallest number of steps reaching 90\% of each seed's maximum accuracy over the zero-to-eight-step sweep.
Accuracy as a function of reasoning steps (Figs.~\ref{fig:sde}I and~\ref{fig:content}G) shows, at each budget, the best accuracy reached at that budget or below within each seed, averaged over seeds.

\paragraph{Content effects experiments.}
Ten models were pretrained for 100,000 gradient steps, without finetuning, on 100 cue-defined orderings in which a frequent cue $c^*$ selected the same ordering on 20\% of trials, and reasoning chains were included on 90\% of trials.
Additional frequent-cue and chain-of-thought proportions are reported in Fig.~\ref{fig:app-content-grid}.

\paragraph{Mechanistic interventions.}
In the content-effects networks, a rank direction was fit at the embedding and after each block using least-squares concept erasure (LEACE)~\citep{BelroseSchneider-JosephRavfogelEtAl23}.
Each item token's coordinate along this rank direction was erased at every layer, in the prompt and throughout the network's reasoning and answer. 
The control manipulation erased a random direction.
Each network's top two context-reading heads, the heads with the most attention from each reasoning step to the context relations containing that step's items, were ablated by replacing their outputs with their mean outputs at the same positions.
The control manipulation ablated two random heads from outside the top four (mean over 20 draws).
To test whether the two strategies compete during learning, the ten pretrained networks of the symbolic distance experiment were finetuned with a learning rate of 0.001 and an uninformative context on 1\% of trials, performing eight reasoning steps before answering.
Their top two context-reading heads were ablated throughout finetuning and testing.
After 1,600 steps, IWL accuracy (best over reasoning budgets) was measured on trained pairs, on non-adjacent pairs containing the first or last item (end item; only AE in the human study, which only had five items), and on non-adjacent pairs among the middle five items (interior; only BD in the human study).
Details are given in Appendix~\ref{app:transformer-methods}.

\paragraph{Data-distribution effects experiments.}
Following~\citet{ChanSantoroLampinenEtAl22} and~\citet{PesnotLerousseauSummerfield26}, we manipulated the diversity of the training data by sampling the $m$th of 100 cue-defined orderings with probability $p(m)\propto m^{-\alpha}$ at 14 values of $\alpha$ from 0 to 10. 
We trained 10 models per condition for 100,000 gradient steps with reasoning chains on 90\% of trials.
We measured IWL by direct answers on trained orderings under an uninformative context and ICL by taking the best accuracy over reasoning-step budgets on a novel ordering shown in context.
We also varied the number of equally frequent cue-defined orderings ($1,2,5,10,100,$ or $1{,}000$), with reasoning chains on every pretraining trial (Fig.~\ref{fig:app-orderings}A).
To distinguish the effects of conflict among different orderings from the raw number of orderings, we finetuned pretrained models for 1,600 gradient steps on either 1 to 10 conflicting orderings of the same items or 1 to 10 orderings of disjoint item sets.
These conditions are compared in Fig.~\ref{fig:app-orderings}B.

\subsection{Human experiments}

\subsubsection{Experiment 1}

\paragraph{Participants.}
We recruited 65 adults through Prolific~\citep{PalanSchitter18}.
Eligibility was restricted to healthy adults aged 18--65 with no history of brain injury.
Participants were paid \$8 per hour plus a performance bonus of up to \$2.67.
Of the 65 participants, 13 did not finish, 11 were removed after exhausting the comprehension-check attempts built into the experiment, and 1 was excluded because accuracy during the second half of training was below 60\%, leaving 40 participants for analysis.
The 40 analyzed participants were aged 21 to 64 years (26 female, 14 male).
All participants provided informed consent under a protocol approved by the Brown University Institutional Review Board.

\paragraph{Procedure.}
Participants completed an online ``alien preference'' task implemented in jsPsych 8.2.3~\citep{deLeeuwGilbertLuchterhandt23}.
Stimuli were 80 unique clip-art images drawn from the PicturePerfectSet~\citep{SaryazdiBannonRodriguesEtAl18} and randomly assigned to experimental roles for each participant.
Participants selected which of two queried objects the aliens preferred by pressing \texttt{F} or \texttt{J}.
Pairwise clues were displayed as a horizontal row: the preferred object appeared above the other object, and an arrow pointed toward it (see Fig.~\ref{fig:sde}A).
Staged instructions and practice covered direct, single-step, and multi-step queries; each comprehension check required 100\% accuracy within two attempts.

Participants then completed 100 training trials on one fixed ordering of seven images.
Every training query compared adjacent items ($d=1$).
Two rank-consistent distractors were inserted into the ordering on each trial, and the displayed window always contained a chain connecting the queried items.
Corrective feedback remained for at least 0.5~s after correct responses and 5~s after errors.
On a random 10\% of training trials, the clue row was masked as in the IWL test, while the query and corrective feedback were unchanged.
Participants next completed three 40-trial test blocks spanning distances $d=1,\ldots,6$.
The ICL test used five disjoint orderings of novel images and showed the full ordering needed to answer each query.
The IWL test queried trained images while replacing every image in the clue display with a featureless gray box, preserving its structure but removing item identity and useful order information.
Following~\citet{PesnotLerousseauSummerfield26}, we also included an arbitrage test querying trained images under a reversed display (see Fig.~\ref{fig:app-human-arbitrage}).
The same approximately distance-balanced set of 40 rank pairs was used in all three blocks within each participant.
ICL and IWL block order was counterbalanced, and arbitrage was always last.
Test trials provided neutral choice confirmation but no accuracy feedback.

\paragraph{Statistical analysis.}
Accuracy was analyzed at the trial level using binomial generalized linear mixed-effects models with a logit link~\citep{BatesMachlerBolkerEtAl15}.
We fit separate ICL and IWL models with linear symbolic distance as a fixed effect and participant-specific random intercepts and distance slopes, and a combined model with fixed effects of distance, test, and their interaction, plus the corresponding participant-specific random effects.
Response-time analyses included correct, non-timeout ICL and IWL trials; response times were natural-log transformed, and trials more than three standard deviations from a participant's mean log response time within a test were removed (4 of 2,478 ICL/IWL trials; 0.16\%).
Log response times were analyzed using linear mixed-effects models~\citep{KuznetsovaBrockhoffChristensen17}.

\subsubsection{Experiment 2}
\label{sec:methods_exp2}

\paragraph{Participants.}
We recruited 132 adults through Prolific~\citep{PalanSchitter18} under the same eligibility criteria and consent protocol as Experiment 1, and assigned each to the congruent or shuffled condition.
After excluding participants who did not finish (14), answered at chance (18), or answered too fast (1), 49 congruent and 50 shuffled participants remained (Appendix~\ref{app:human2-methods}).
The 99 analyzed participants were aged 21 to 64 years (50 female, 48 male, 1 not available).

\paragraph{Procedure.}
Each of 48 questions showed eight statements ordering nine animals by weight, in a scrambled order, and asked which of two animals was heavier according to the statements.
The ends of the ordering were never queried, giving symbolic distances 1--6.
In the shuffled condition, each question used its own reordering of the animals, balanced so that the queried relation agreed with the real-world order on exactly half of the questions at every distance.

\paragraph{Statistical analysis.}
Accuracy and log response times of correct answers were analyzed with mixed-effects models with fixed effects of condition, log symbolic distance, their interaction, and trial number, and participant-specific random intercepts and distance slopes.

\subsection{Large language models}
\label{sec:methods_llms}

\paragraph{Models.} 
We tested 22 open-weight LLMs (1B--70B parameters; Appendix~\ref{app:llm-methods}): 10 reasoning models trained to generate a chain of thought before answering (Qwen3~\citep{YangLiYangEtAl25}, DeepSeek-R1-Distill~\citep{GuoYangZhangEtAl25}, Phi-4-reasoning~\citep{AbdinAgarwalAwadallahEtAl25}, gpt-oss~\citep{OpenAIAgarwalAhmadEtAl25}) and 12 instruction-tuned models that reasoned only when prompted (Qwen2.5~\citep{QwenYangYangEtAl24}, Llama~3~\citep{GrattafioriDubeyJauhriEtAl24}, Gemma~2~\citep{GemmaTeamRivierePathakEtAl24}, GLM-4~\citep{TeamGLMZengXuEtAl24}, Mistral~\citep{JiangSablayrollesMenschEtAl23}, OLMo~2~\citep{TeamOLMoWalshSoldainiEtAl25}).

\paragraph{Task.} 
Each problem was an ordering of 11 items presented only through its 10 adjacent relations, and a query asked whether one item ranked above another.
Analyses used the innermost seven queryable ranks, giving symbolic distances $d = 1, \ldots, 6$.
The IWL test queried familiar items from five real-world dimensions (country population, US state population, city population, animal body mass, and country surface area), with no relations shown, so that only prior knowledge could be used to answer (item selection in Appendix~\ref{app:llm-methods}).
The ICL test queried 11 common names arbitrarily ordered by height, with all 10 relations shown.
Each ICL problem was repeated with 10 random premise orders, and the IWL test was replicated on additional item sets per dimension (Appendix~\ref{app:llm-methods}).

\paragraph{Prompts and scoring.} 
Prompts used each model's chat template, with a short system instruction defining the relation (e.g., \texttt{>} means ``has a larger population''), the relations one per line as \texttt{A > B}, and the query ``Is it true or false that \texttt{A > B}?'' (full example in Appendix~\ref{app:llm-methods}).
Direct answers were read from a single forward pass at the answer position: the total probabilities of \texttt{True}/\texttt{true} were compared with those of \texttt{False}/\texttt{false}, the larger giving the prediction.

\paragraph{Chain of thought.}
The model generated its chain with greedy decoding up to a budget of $B$ new tokens ($B \in \{32, 64, 128, 256, 384, 512, 640, 768, 896, 1{,}024\}$; $B = 0$ is the direct condition). 
After the budget was reached, the cue ``Therefore the answer is:'' was appended and the answer was scored.
Reasoning models used their native thinking block with no additional prompt to reason. 
Instruction-tuned models were explicitly prompted to reason by appending ``Let's think step by step.'' to the query~\citep{KojimaGuReidEtAl22}.
As in the transformer experiments, the reasoning budget required for each distance was defined to be the smallest budget whose accuracy reached 90\% of the maximum over all budgets.
When accuracy was plotted against reasoning budget (Figs.~\ref{fig:sde}P and~\ref{fig:content}N), it was defined to be the best accuracy reached at or below each budget.

\paragraph{Content effects.}
For each dimension, the 10 relations shown were congruent, reversed, or shuffled with respect to the real-world ordering~\citep{LiuPriorBalasubramaniamEtAl25,GeertsLiuChanEtAl25}, and answers were scored against the ordering shown.
A fourth context condition showed 10 true relations among 11 other items of the same dimension that never mentioned the queried pair (uninformative context).
The congruent, reversed, and shuffled conditions were also run with different reasoning budgets.
To vary prior strength, direct answers were also collected on additional item sets per dimension with closer or wider value spacing (Fig.~\ref{fig:content}K; Appendix~\ref{app:llm-methods}).
To resolve ambiguity in the instructions, all context-condition instructions also included ``Base your answer only on the comparisons given --- not on outside knowledge.''

\paragraph{Mechanistic interventions.}
In Qwen3-32B, a rank direction was fit in the same way at every hidden state and erased at every mention of an item.
The 32 of its 4,096 attention heads with the most attention from the composed inferences of its reasoning chains to the context relations mentioning their items were ablated (Appendix~\ref{app:llm-methods}).

%% file: appendix/formalization.tex
\section{Formalization of the geometric and algebraic strategies}
\label{app:formalization}

Here we formalize the geometric and algebraic strategies for transitive inference as two algorithms for answering order queries about a set of premises.
A single ordering is a total order over $n$ items $x_1\prec\cdots\prec x_n$ presented only through adjacent premises, which connect each item to its two immediate neighbors (endpoints have only one).
A query on two items asks their relative order, and the \emph{symbolic distance} between them is the number of premises separating them.

An ordinal transitive-inference \emph{environment} is a family of $M$ such orderings.
For simplicity, we assume that the $M$ orderings are defined over the same $n$ items.
We write $\rankm(x)$ for the position of item $x$ in ordering $m$, so $x\succ y\iff\rankm(x)>\rankm(y)$, and the symbolic distance is $\dist=|\rankm(x)-\rankm(y)|$.
A query is constructed by sampling an ordering $m$ and a pair of interior items $(x,y)$ from that ordering; the task is to return the correct relation $x\succ y$ or $x\prec y$. 
We use $n=9$ (seven queryable items) throughout.

\paragraph{Two algorithms.}

\noindent\emph{Geometric (pre-compile).} 
We can realize the geometric strategy by pre-compiling the rank of every item in an ordering (Alg.~\ref{alg:geo}), enabling inference via constant-time comparisons that are independent of distance.
Ranks can be compiled by walking outward from any starting item.
The ranks are specific to an ordering---different orderings must be compiled separately.

\begin{algorithm}[H]
\caption{Geometric strategy: pre-compile}
\label{alg:geo}
\begin{algorithmic}
\Phase{Pre-compile} (once per ordering $m$): \Comment{$\Theta(n)$}
  \State pick any item $x_0$ of the ordering; $\rankm(x_0)\gets 0$
  \State $x\gets x_0$;\; $r\gets 0$
  \While{$x$ has a greater neighbor $y$}
    \State $r\gets r+1$;\quad $\rankm(y)\gets r$;\quad $x\gets y$ \Comment{walk up to the last item}
  \EndWhile
  \State $x\gets x_0$;\; $r\gets 0$
  \While{$x$ has a lesser neighbor $y$}
    \State $r\gets r-1$;\quad $\rankm(y)\gets r$;\quad $x\gets y$ \Comment{walk down to the first item}
  \EndWhile
\EndPhase
\Phase{Query} $(x,y)$: \Comment{$O(1)$}
  \State \Return $x\succ y$ if $\rankm(x)>\rankm(y)$, else $x\prec y$
\EndPhase
\end{algorithmic}
\end{algorithm}

\noindent\emph{Algebraic (traverse).} We can realize the algebraic strategy by walking from one queried item until reaching either the other queried item or an end of the ordering, at a cost proportional to the number of steps (Alg.~\ref{alg:alg}; Fig.~\ref{fig:sde}E).
Nothing is pre-compiled, so this strategy can answer a query about any ordering.

\begin{algorithm}[!htbp]
\caption{Algebraic strategy: traverse}
\label{alg:alg}
\begin{algorithmic}
\Phase{Pre-compile}:
  \State none
\EndPhase
\Phase{Query} $(x,y)$:
  \If{$x$ and $y$ are adjacent}
    \State \Return the relation given by their premise \Comment{no steps}
  \EndIf
  \State pick a neighbor $c$ of $x$ at random
  \State let $r\in\{\prec,\succ\}$ satisfy $x\mathrel{r}c$, and let $\bar r$ be the opposite relation
  \State $u\gets x$;\; $v\gets c$
  \While{$v\neq y$ and $v$ has a neighbor other than $u$} \Comment{one premise read per step}
    \State $w\gets$ the neighbor of $v$ other than $u$;\quad $u\gets v$;\quad $v\gets w$
  \EndWhile
  \If{$v=y$}
    \State \Return $x\mathrel{r}y$ \Comment{reached the target after $\dist$ steps}
  \Else
    \State \Return $x\mathrel{\bar r}y$ \Comment{reached an end, so $y$ is on the other side of $x$}
  \EndIf
\EndPhase
\end{algorithmic}
\end{algorithm}

\paragraph{Simple noise assumptions reproduce symbolic distance effects.}

Under straightforward assumptions about noise in the inference process, the geometric strategy (Alg.~\ref{alg:geo}) reproduces the symbolic distance effect and the algebraic strategy (Alg.~\ref{alg:alg}) reproduces the reverse symbolic distance effect.
\bigskip

\noindent\emph{Geometric.} A query $(x,y)$ is answered by comparing the two pre-compiled ranks (Alg.~\ref{alg:geo}). 
The decision variable is the rank difference $g=\rankm(x)-\rankm(y)$.
If there is Gaussian noise in the readout, then the response is $x\succ y$ if $g+\eta>0$ and $x\prec y$ otherwise, where $\eta\sim\mathcal{N}(0,\sigma^2)$ ($\sigma=3$ in Figs.~\ref{fig:sde}D and~\ref{fig:content}B,D). In this case, accuracy increases as a function of symbolic distance (Fig.~\ref{fig:sde}D):
\begin{equation*}
P(\text{correct}\mid\dist)=\Phi(\dist/\sigma),
\end{equation*}
where $\Phi$ is the standard normal cumulative distribution function.
Because inference is always a single comparison, inference cost does not grow with distance (Fig.~\ref{fig:sde}E).
\bigskip

\noindent\emph{Algebraic.} A query $(x,y)$ is answered by walking along the ordering (Alg.~\ref{alg:alg}).
The walk ends when it reaches the target, after $\dist$ steps, or an end of the ordering---which comes first depends on the direction chosen.
We assume a limited reasoning budget of $K$ steps, drawn uniformly from $1$ to $K_{\max}$ on each query ($K_{\max}=5$ in Figs.~\ref{fig:sde}D and~\ref{fig:content}B), and that the algorithm falls back to a guess, correct with probability $\tfrac12$, if the walk does not end within the budget.
Writing $L$ for the number of steps the walk takes,
\begin{equation*}
P(\text{correct}\mid\dist)=\tfrac12+\tfrac12\,P(L\le K\mid\dist).
\end{equation*}
Longer walks fail more often, so the algebraic strategy's accuracy \emph{decreases} with distance---the reverse symbolic distance effect (Fig.~\ref{fig:sde}D)---while the expected number of steps grows with distance (Fig.~\ref{fig:sde}E).

\paragraph{Content effects.}

A single system capable of both strategies can combine the advantages of each, using pre-compiled ranks for familiar items and reasoning step by step for unfamiliar items.
When both strategies are available, the information given in context can either agree with prior experience (congruent) or contradict it (incongruent).
When the context is congruent, a system capable of both strategies should answer accurately regardless of the length of the reasoning process. 
However, when the context is incongruent, the system should answer incorrectly if it relies on its pre-compiled ranks, but can overcome this interference by reasoning step by step through the information available in the context. 

Figure~\ref{fig:content}B,C illustrates content effects in a system capable of both strategies.
The system answers from its pre-compiled ranks unless it is given enough time to complete its reasoning process.
With the same reasoning budget $K$ as above, a query is answered correctly if the walk ends within the budget, and otherwise from the pre-compiled ranks, which, as in the geometric strategy, match prior experience with probability $\Phi(\dist/\sigma)$.
Accuracy therefore stays near ceiling when the context is congruent and falls with distance when it is reversed or shuffled (less steeply when shuffled).
When the context is uninformative, accuracy shows the geometric strategy's symbolic distance effect (Fig.~\ref{fig:content}B).
Congruent queries require no reasoning, but incongruent queries do, and the expected number of reasoning steps required grows with distance (Fig.~\ref{fig:content}C).

\paragraph{Learning and inference costs.}

The two strategies trade off learning and inference costs (memory costs also differ, but we do not consider them here).
During learning, the geometric strategy pre-compiles ranks, requiring a pass through all $n$ items, whereas the algebraic strategy does not learn specific orderings at all.
During inference, the geometric strategy's cost is constant in distance, while the algebraic strategy's inference cost grows with distance.
When the algebraic strategy walks toward the target, it takes $\dist$ steps, and when it walks away, it takes steps equal to the distance from the starting item to the end of the ordering in that direction. 
For non-adjacent queries the expected number of steps at distance $\dist$ is $(n-1+\dist)/4$.

The total cost of each strategy therefore depends on the environment.
We assume a total of $Q$ queries, drawn independently from a fixed data distribution.
Orderings $m$ are sampled according to a Zipf distribution with skew $\alpha$: ordering $m \in \{1,\ldots,M\}$ is sampled with probability $p(m)\propto m^{-\alpha}$.
When $\alpha$ is 0, the distribution is uniform over all $M$ orderings.
As $\alpha$ increases, the distribution becomes more skewed, concentrating queries on a few frequent orderings.

The geometric strategy compiles an ordering when it first encounters it, so over $Q$ queries it compiles $D_Q(M,\alpha)=\sum_{m}\big[1-(1-p(m))^{Q}\big]$ distinct orderings in expectation---a count that falls as $\alpha$ rises.
Let $c_I$ be the cost of one elementary inference operation, and let $c_L$ be the cost of learning one ordering.
Averaged over queries, the algebraic walk takes as many steps as the mean distance $\bar\dist$ if adjacent pairs are walked.
The total costs are then:
\begin{equation}
  C_{\text{geo}} = \underbrace{c_L\,D_Q(M,\alpha)}_{\text{learning}}
  + \underbrace{c_I\,Q}_{\text{inference}},
  \qquad
  C_{\text{alg}} = \underbrace{c_I\,Q\,\bar\dist}_{\text{inference}} .
  \label{eq:cost}
\end{equation}
The cost of the geometric strategy is paid at learning time and scales with the number of distinct orderings encountered: when queries are concentrated on a few orderings (large $\alpha$), the learning burden decreases.
The cost of the algebraic strategy is paid at inference time and scales with the mean distance $\bar\dist$: when many queries have large distances, inference becomes more costly.

Figure~\ref{fig:diversity}B shows how these costs affect a learner capable of both strategies. 
We assume that the learner acquires each strategy to the extent that it is not much costlier than the other: strategy $s$ is adopted according to $\ell_s=\sigma\big((m-\log(C_s/C_{s'}))/\tau\big)$, where $\sigma$ is the logistic function, $s'$ is the other strategy, and $m$ and $\tau$ control how much costlier a strategy can be before it is abandoned.
Accuracy on the test that isolates strategy $s$ is given by $\tfrac12+\tfrac12\ell_s$.
As the orderings become more diverse (smaller $\alpha$), accuracy on familiar items (IWL test) falls while accuracy on novel items given in context (ICL test) rises.
Plotted against each other (Fig.~\ref{fig:diversity}D), the two accuracies are both high over an intermediate range of $\alpha$. Figure~\ref{fig:diversity}B,D uses $M=100$, $Q=500$, $c_L=160$, $c_I=1$, $m=0.5$, $\tau=0.15$, and $\alpha$ from 0 to 10.

The geometric strategy can also be applied to a novel ordering given in context, by pre-compiling all ranks from the premises given in context before answering the specific query. 
Because each such ordering is new, this would require reading all $n-1$ premises before making the comparison to answer the query, at a total cost of $c_I\,n$.
This would always be costlier than the algebraic strategy, which has an average cost of $c_I\,\bar\dist$, because $n > \bar\dist$ as no two items are more than $n-1$ premises apart.

%% file: appendix/extended_methods.tex
\section{Extended methods}
\label{app:extended_methods}

Below we describe the methods used for training transformers from scratch, conducting online human experiments, and evaluating large language models (LLMs).

\subsection{Transformers trained from scratch}
\label{app:transformer-methods}

\paragraph{Task and input sequences.}
Transformers learned ordinal relations between items denoted by arbitrary tokens.
Each transitive-inference problem defined a strict total order but exposed that order only through the relations between adjacent items.
A query presented two items and required the model to determine which one had the higher rank.
For example, one possible training sequence for the order \texttt{i0} $>$ \texttt{i1} $>$ \texttt{i2} $>$ \texttt{i3} $>$ \texttt{i4} $>$ \texttt{i5} $>$ \texttt{i6} $>$ \texttt{i7} $>$ \texttt{i8} was:

\begin{quote}
\small\ttfamily
<bos> <cue7> ? i3 i1 ? | i8 i7 -1 | i2 i3 +1 | i1 i0 -1 |\\
i5 i6 +1 | i4 i3 -1 | i7 i6 -1 | i2 i1 -1 | i5 i4 -1 | i3 i1 -1
\end{quote}

Sequences began with \texttt{<bos>} followed by a cue token (e.g., \texttt{<cue7>}) that indexed the particular ordering between items in the sequence.
The query (e.g., \texttt{i3 i1}) appeared both before the context and after it.
Before the context, the query items were presented between two distinct special tokens, both printed \texttt{?}.
The separator \texttt{|} marked the boundary after the query and after each subsequent relation pair.
The eight pairs of items and their relations shown in the context specified the nine-item order.
The positions of these adjacent relations in the context were shuffled.
The two endpoints appeared in the context but were not queried.
Unless indicated otherwise, the symbolic distance (the rank difference between the two query items) was sampled uniformly from 1 to 6.
A valid pair of ranks at that distance was then sampled uniformly, and the order of the two query items was randomized.
Each $+1$ or $-1$ in the context represented the relation ($+1$ encoded $>$ and $-1$ encoded $<$), and scaled a single learned relation embedding.

\paragraph{Chain-of-thought generation.}
During pretraining, sequences sometimes contained demonstrations of chains of thought (CoTs) for the given problem.
For the example problem given above, the complete CoT training sequence would be:

\begin{quote}
\small\ttfamily
<bos> <cue7> ? i3 i1 ? | i8 i7 -1 | i2 i3 +1 | i1 i0 -1 |\\
i5 i6 +1 | i4 i3 -1 | i7 i6 -1 | i2 i1 -1 | i5 i4 -1 |\\
i2 i3 +1 | i2 i1 -1 | i3 i1 -1
\end{quote}

The CoT inserted \texttt{i2 i3 +1} and \texttt{i2 i1 -1} after the in-context relations but before the final query and answer (\texttt{i3 i1 -1}).
Algorithm~\ref{alg:model_cot} gives the CoT-generation procedure used in the main experiments; it implements the algebraic strategy of Alg.~\ref{alg:alg} on the shuffled context, with the first-presented pair playing the role of the random direction.

\begin{algorithm}[!htbp]
\caption{Chain-of-thought generation}
\label{alg:model_cot}
\begin{algorithmic}
\Require shuffled context pairs $C$
\Phase{Query} $(x,z)$:
\State $R\gets[\,]$
\If{a pair $e$ in $C$ contains both $x$ and $z$}
  \State let $\rho\in\{\prec,\succ\}$ satisfy $x\mathrel{\rho}z$, as specified by $e$
\Else
  \State $e\gets$ the earliest pair in $C$ that contains $x$
  \State $c\gets x$
  \While{true}
    \State $v\gets$ the item in $e$ other than $c$
    \State let $r\in\{\prec,\succ\}$ satisfy $c\mathrel{r}v$, as specified by $e$
    \State append $e$ to $R$ in its presented orientation
    \If{$v=z$}
      \State $\rho\gets r$
      \State \textbf{break}
    \ElsIf{no pair in $C$ other than $e$ contains $v$}
      \State $\rho\gets\bar r$ \Comment{$\bar r$ is the relation opposite to $r$}
      \State \textbf{break}
    \EndIf
    \State $e\gets$ the unique pair in $C$ other than $e$ that contains $v$
    \State $c\gets v$
  \EndWhile
\EndIf
\State append the query items $x,z$ to $R$
\State append $\rho$ to $R$
\State \Return $R$
\EndPhase
\end{algorithmic}
\end{algorithm}

The algorithm was designed so that each successive pair could be directly obtained from the current context sequence without consulting the latent global ranks.
For adjacent queries, the answer was read directly from the context pair containing both query items, and no intermediate relation was emitted.
For non-adjacent queries, the pairs emitted at each step copied the context relations in their presented orientations.
If traversal reached the other query item, the answer was the relation emitted at the final step; if instead the traversal reached the opposite endpoint, the answer was the opposite of the one emitted at the final step.
To maintain a fixed sequence length during training, sequences were padded after the first answer by repeating the answer pair, ensuring good performance at reasoning budgets greater than or equal to the number required to produce the answer.

\paragraph{Pretraining distribution.}
Each cue token indexed a fixed ordering over the pool of items.
This was achieved via a seeded pseudorandom score that was fixed for every cue--item combination.
On each pretraining trial, a cue was drawn from the experiment's cue-frequency distribution (see experiment details below), a set of items was sampled without replacement, and those items were placed in the cue-associated order.
After the items were ranked, a query and the complete adjacent-relation context were constructed using the sampling and randomization procedure described above.
During pretraining, a proportion $p(\mathrm{CoT})$ of trials included CoT, specified by the experiment configuration (see details for each experiment below).

\paragraph{Fixed-order learning distribution.}
After pretraining, finetuning repeatedly presented one or more fixed orderings, giving the model the opportunity to memorize them.
Each finetuning trial queried an adjacent pair from one of these orderings and used an answer-only target.
Consistent with Experiment 1 (see details below), distractors were inserted into the sequence so that finetuning retained the same overall format as pretraining.
In order to avoid catastrophic forgetting of the in-context CoT strategy~\citep{McCloskeyCohen89}, 10\% of samples were replaced with pretraining samples.

\paragraph{IWL and ICL tests.}
Held-out cues were never used during training, and in-context learning (ICL) test items were not part of the finetuned ordering or its distractors.
The in-weight learning (IWL) test queried items from a fixed ordering learned during finetuning with an uninformative context composed of unrelated items outside the finetuning main and distractor sets.
The ICL test queried a fresh ordering of novel items under a held-out cue on every trial and provided the complete set of adjacent relations specifying that ordering in the context.
All pairs of queryable items were included in both tests, repeated with 10 random contexts.
Two appendix figures (alternative CoT algorithms and architectures) instead report the ICL test with a cue drawn from the training cues, because the held-out-cue test degrades late in pretraining in some seeds.

\paragraph{Evaluation and scoring.}
At evaluation, the prompt ended after the context, before any CoT or answer emissions.
We used greedy decoding (temperature zero) and varied the number of steps in the reasoning budget $K$ over an experiment-specific range.
For convenience, $K$ was defined as the number of complete pair emissions generated by the model (four tokens each), rather than individual tokens.
At $K=0$, the model predicted the answer relation directly from the prompt.
For $K>0$, the model generated $K$ intermediate pairs, the two query-item tokens were appended, and the model's prediction was scored.
Accuracy was exact agreement between the predicted and correct relation sign.

\paragraph{Architecture and optimization.}
All networks were decoder-only transformers~\citep{VaswaniShazeerParmarEtAl17} with four layers, eight attention heads, a residual width ($d_{\mathrm{model}}$) of 256, and a feedforward width of 1,024 with GELU activations.
Models used rotary position embeddings~\citep{SuAhmedLuEtAl24}, pre-layer normalization, causal scaled dot-product attention, no dropout, and no additive biases in the transformer projections.
Embedding and unembedding matrices were tied~\citep{PressWolf17}.
Relations were represented by a structural \texttt{<rel>} token and a scalar $s_t\in\{-1,+1\}$ at the same sequence position $t$.
At an observed relation position, the representation entering the transformer blocks was $\mathbf{e}_{\langle\mathrm{rel}\rangle}+s_t\mathbf{r}$, where $\mathbf{r}$ denotes the learned relation-direction vector and $\mathbf{e}_{\langle\mathrm{rel}\rangle}$ denotes the embedding of the \texttt{<rel>} token.
The \texttt{<rel>} token indicated the relation position while the sign of $s_t$ distinguished $>$ from $<$.
The same relation-direction vector was used to decode scalar relation outputs:
\begin{equation}
\hat{s}_t=\mathbf{h}_{t-1}^{\mathsf{T}}\mathbf{r}+b,
\end{equation}
where $\mathbf{h}_{t-1}$ is the final normalized hidden state at the sequence position immediately preceding the target relation position $t$.
During CoT, when the model generated a \texttt{<rel>} token, its thresholded scalar output at that position ($+1$ if $\hat{s}_t > 0$, $-1$ if $\hat{s}_t \le 0$) multiplied the relation-direction vector in its input at that token position.
The model was supervised on both the correct token sequence and the correct scalars:
\begin{equation}
\mathcal{L}=-\frac{1}{|\mathcal{T}|}\sum_{t\in\mathcal{T}}\log p_{\theta}\!\left(y_t\mid y_{<t},s_{<t}\right)+\frac{1}{|\mathcal{R}|}\sum_{t\in\mathcal{R}}\left(\hat{s}_t-s_t\right)^2,
\label{eq:model_training_loss}
\end{equation}
where $\mathcal{T}$ is the set of token positions after the context and $\mathcal{R}\subseteq\mathcal{T}$ is the set of \texttt{<rel>} token positions.

Models were optimized with Adam~\citep{KingmaBa15} ($\beta_1=0.9$, $\beta_2=0.95$) using batches of 64 sequences, a constant learning rate of 0.001, zero weight decay, and gradient norms clipped at 1.
Each training run used one NVIDIA RTX 3090, A5000, or A5500 GPU on Brown University's Oscar cluster.

\paragraph{Symbolic distance effects experiments.}
Each problem contained nine ordered items, of which the inner seven were queryable, giving symbolic distances $d=1,\ldots,6$.
Pretraining used 100 uniformly sampled cue-defined orderings over a pool of 300 item tokens and 100,000 gradient steps.
Training queries were sampled uniformly over the six possible distances, and all pretraining trials used CoT targets padded to nine pair emissions.
Each pretrained model was then finetuned for 1,600 steps at a learning rate of 0.0001 on one fixed ordering of seven main items, using adjacent queries and answer-only targets.
Each test condition contained all 42 ordered pairs among the seven queryable ranks, with 10 independently generated contexts per pair, for 420 trials per condition and seed.
The reasoning budget was swept over $K=0,\ldots,8$.
The main-text IWL curve used direct predictions ($K=0$), whereas the ICL curve averaged accuracy at $K=1,\ldots,5$ within seed and distance.
To obtain a measure of the reasoning budget required at test time, the threshold for each seed and distance was defined to be the smallest $K$ whose accuracy was at least 90\% of that seed's maximum accuracy over the full sweep.
Curves show the mean and error bars the standard error across 10 independently initialized models (seeds 0--9).

\paragraph{Content effects experiments.}
For experiments on content effects, one frequent cue, $c^*$, occurred on 20\% of pretraining trials and deterministically selected the same global ordering.
The remaining probability was divided among 99 background cues.
Models received CoT supervision on 90\% of trials and direct answers on 10\%, and 10 models were trained for 100,000 steps with a 300-item pool.
At test, queries under $c^*$ were accompanied by one of four contexts: an \emph{uninformative} context (an IWL test), a context where the ordering was \emph{congruent} with the learned $c^*$ ordering, a context where the ordering was \emph{reversed}, and a context where the ordering was \emph{shuffled}.
In the shuffled condition, both queried items were interior to the ordering shown, and symbolic distance was their distance in that ordering; on exactly half of the trials at every distance (and for every ordered pair of positions), the shown order of the queried pair agreed with the learned ordering, and on the other half it conflicted.
The pair's distance in the learned ordering was drawn uniformly from 1--6 (both ranks interior), independently of the distance shown, and the remaining seven items filled the other positions at random.
In the IWL test condition, responses were scored relative to the learned ordering; in the other three contexts, responses were scored relative to the ordering shown in context.
The main content-effects comparison used $K=0$.
In the congruent, reversed, and shuffled conditions, we also swept $K=0,\ldots,8$ separately by symbolic distance.
The reasoning steps required were computed as in the symbolic distance experiment, except that a seed had no value at a distance when its best accuracy over the sweep was not reliably above chance.
A distance was marked as unreached when this held for more than half of the seeds.
Each condition used all 42 ordered pairs of queryable ranks with 10 context replications per pair and seed.

\paragraph{Probes and interventions.}
We examined the internal representations of the networks from the content-effects experiments, holding $p(\mathrm{CoT})=.9$ during pretraining.
After pretraining, linear probes~\citep{AlainBengio16} were trained via ridge regression to predict each item's rank in the frequent ordering from the model's activity at each layer, and were evaluated on held-out items.
For probes beyond the embedding layer, item activity was recorded under uninformative contexts with the frequent cue.

We then used least-squares concept erasure (LEACE)~\citep{BelroseSchneider-JosephRavfogelEtAl23} to manipulate a rank direction at every layer in networks pretrained with frequent-cue proportion .2.
The direction was fit separately at each layer from the unedited network's states at every item-token occurrence in a set of 500 examples with the frequent cue and an uninformative context, plus one example for each queryable item that the 500 left unqueried.
At a given site, let $\mathbf{h}_n$ denote the state at occurrence $n$, $\boldsymbol{\mu}$ the mean state over all $N$ occurrences, and $z_n=(r_n-\bar r)/\sigma_r$ the standardized rank of its item in the frequent ordering, where $\bar r$ and $\sigma_r$ are the mean and standard deviation of ranks.
We estimated the state covariance $\Sigma$ using Ledoit--Wolf shrinkage~\citep{LedoitWolf04}, with eigenvalues floored at $10^{-8}$ times the largest eigenvalue.
The state--rank covariance $\mathbf{b}$ defined a unit direction $\mathbf{u}$ in whitened space (the \emph{rank direction}):
\[
\mathbf{b}=\frac{1}{N}\sum_{n=1}^{N}(\mathbf{h}_n-\boldsymbol{\mu})z_n,
\qquad
\mathbf{u}=\frac{\Sigma^{-1/2}\mathbf{b}}{\lVert\Sigma^{-1/2}\mathbf{b}\rVert}.
\]
Each item token's coordinate $a$ along this direction could then be changed to $a'$ by editing its state $\mathbf{h}$:
\[
a=\mathbf{u}^{\mathsf T}\Sigma^{-1/2}(\mathbf{h}-\boldsymbol{\mu}),
\qquad
\mathbf{h}'=\mathbf{h}+(a'-a)\Sigma^{1/2}\mathbf{u}.
\]
Setting $a'=g a$ with gains $g=1,0,$ or $-1$ left the rank direction intact, erased it, or inverted it, respectively.
To write a new ordering, we first fit $a\approx\beta_0+\beta_1 z$ over the fit occurrences by least squares and assigned each item a new standardized rank $z^{\mathrm{new}}$ from a random ordering with absolute Spearman rank correlation at most .2 with the frequent ordering.
At write strength $\eta$, we set
\[
a'=(1-\eta)a+\eta\bigl(\beta_0+\beta_1 z^{\mathrm{new}}\bigr),
\qquad 0\leq\eta\leq1.
\]
Matched controls replaced $\mathbf{u}$ at each site with a random unit vector, using the same covariance transformation and, for writing, the same target coordinates.
Edits were applied at every site to every item occurrence in the query, context, generated reasoning chain, and answer positions, in every forward pass, with model weights held fixed. 
Within a forward pass, the sites were edited in order, so that each block read the state edited below it.

To ablate an attention head, we replaced its output with its mean output at the same position, computed over greedy rollouts of 1,024 trials drawn from the pretraining distribution.
Context-reading heads were identified from attention alone, on reversed-context and ICL test trials.
We took the attention from each reasoning step's positions to the relevant relations from the context, and defined a head's score to be the average attention mass on those relevant relations.
The top two scoring heads were defined to be the \emph{context-reading heads}, and were ablated together in each network.
A control manipulation ablated two heads drawn at random from outside the network's top four heads, averaged over 20 draws.
Every head in every network was also ablated on its own in a supplemental experiment (Fig.~\ref{fig:app-representation}D).
Accuracies were pooled over symbolic distances 2--6.

To test whether the two strategies compete during learning, we finetuned the ten pretrained models of the symbolic distance effects experiments again on the same fixed ordering for 1,600 steps, with two changes: the learning rate was 0.001 rather than 0.0001, and an uninformative context replaced the ordering on 1\% rather than 10\% of finetuning trials. 
10\% of trials were still replaced by pretraining trials.
On every finetuning trial whose context showed the ordering, the model reasoned for eight steps.
Loss was only backpropagated from subsequent answers.
Trials with an uninformative context were answered directly, and replay trials were unchanged.
Context-reading heads were scored on the pretrained model as above, from its chains on ICL test trials, and mean outputs were computed from greedy rollouts of 1,024 pretraining trials.
The top two heads were mean-ablated on every finetuning and test trial, but not on replayed pretraining trials.
Every 100 steps, the IWL and ICL tests were run with 50 contexts per ordered pair.
For each model and symbolic distance, accuracy was taken at the best budget over $K=0,\ldots,8$ and then pooled over distances (distances 2--6 for the ICL test).
IWL accuracy was reported separately for trained (adjacent) pairs, for the nine non-adjacent pairs containing the first or last of the seven items (end item), and for the six non-adjacent pairs among the middle five items (interior).
In the human study~\citep{FrankOReillyCurran06}, the corresponding novel pairs were AE and BD of a five-item (A through E) ordering.

\paragraph{Data-distribution effects experiments.}
To vary the effective diversity of pretraining experience while holding the number of available orderings fixed, we sampled the ordering index $m\in\{1,\ldots,100\}$ according to a Zipf distribution~\citep{Zipf49,Piantadosi14} with skew $\alpha$: $p(m)\propto m^{-\alpha}$.
We tested $\alpha\in\{0,0.5,1,1.5,2,2.5,3,3.5,4,5,6,7,8,10\}$, trained 10 models per value for 100,000 gradient steps, and included CoT targets on 90\% of trials.
Sixteen additional cue tokens were held out and never used during training.
ICL was evaluated on a novel ordering under a held-out cue; for every seed and distance, we selected the best accuracy over $K=0,\ldots,8$.
IWL was evaluated at $K=0$ under an uninformative context; both were pooled over distances 2--6.
The cue on each IWL trial was drawn from the training cue distribution.

To separate the effects of including a greater number of cues/orderings from the effects of \emph{conflicting} orderings, we took the 10 models pretrained for the symbolic distance experiment and finetuned them for 1,600 answer-only gradient steps on 1--10 orderings.
In the disjoint condition (``Disjoint orderings'' in Fig.~\ref{fig:app-orderings}B), every ordering used a separate seven-item set under one cue.
In the conflict condition (``Conflicting orderings''), 1--10 cue tokens selected different orderings of the same seven items.
IWL was measured on the first finetuned ordering at $K=0$ and pooled over all query distances; the single ordering is common to the two definitions.

\subsection{Human experiments}
\label{app:human-methods}

\subsubsection{Experiment 1}
\paragraph{Participants.}
We recruited 65 adults through Prolific.
Eligibility was limited to US residents 18--65 years old who were fluent in English, had normal or corrected-to-normal vision, and reported no history of brain injury, dyslexia, color blindness, or cognitive impairment.
Participants were paid \$8 per hour and could earn a performance bonus of up to \$2.67.
All participants gave informed consent under a protocol approved by the Brown University Institutional Review Board.
Of the 65 participants, 13 were excluded due to incomplete data, 11 were excluded due to failed comprehension checks, and 1 was excluded for scoring below 60\% accuracy in the second half of training---leaving 40 participants in the final sample.
Participants in the final sample were aged 21 to 64 years (mean 35.4, SD 9.7); 26 were female and 14 male.

\paragraph{Stimuli and software.}
The task ran in a browser using jsPsych~\citep{deLeeuwGilbertLuchterhandt23} and was deployed with Firebase Hosting.
Visual stimuli were 80 clip-art images selected from the 225-image PicturePerfectSet~\citep{SaryazdiBannonRodriguesEtAl18}.
We retained images with at least 60\% clip-art name agreement, picture--name agreement of at least 4 on a five-point scale, familiarity of at least 2.5 on a five-point scale, labels no longer than 10 characters, and no more than 15 members of any semantic category.
When assigning images to roles for each participant, we kept visually similar images, as measured by the similarity of their CIE-Lab color histograms, out of the same ordering.
Images were cropped to remove whitespace and displayed at equal sizes.
For each participant, the 80 images were pseudorandomly assigned without overlap to seven main items, 16 distractors, 45 novel-ordering items, and 12 practice items.

\paragraph{Display and response format.}
On each trial the two images in the query were displayed side by side, and participants selected an item with the \texttt{F} (left) or \texttt{J} (right) key.
Below the query, ``clues'' were presented as a horizontal row of vertically arranged image pairs (see Fig.~\ref{fig:sde}A).
Within every clue, the higher-ranked image was on top and an arrow pointed toward it.
The positions of clue pairs in the row were shuffled.
A trial allowed up to 60~s for a response and began with a 0.5~s fixation display.

\paragraph{Instructions and practice.}
Participants were instructed that aliens were encountering earthly objects for the first time, and that they would be learning which objects the aliens preferred.
Instruction was staged in three blocks.
The first introduced the two-image query and feedback, followed by four comprehension questions and 10 query-only practice trials.
The second introduced single adjacent-pair clues and the arrow convention, and ended with four comprehension questions and five single-step practice trials.
The third introduced combining multiple clues and included a worked transitive example; it ended with four questions and five multi-step practice trials.
A final four-question check followed the overview of the main experiment.
Each check required 100\% correct and allowed two total attempts.
Failure returned the participant to the beginning of the relevant instruction block; failure on the second attempt ended the experiment.
Each practice block required 70\% accuracy.
After an initial failed attempt, participants could complete up to two five-trial retry blocks; exhausting practice retries did not terminate the study.
Instruction pages had a minimum viewing time of 5~s.

\paragraph{Learning phase.}
Participants completed 100 trials involving a single fixed ordering of seven main images.
Every query compared main items of adjacent rank ($d=1$); positions of the two items within the query were randomized.
Two rank-consistent distractors were inserted into the ordering on each trial. 
Each distractor was assigned to a rank between a pair of adjacent main items or at an end of the ordering.
To discourage familiarity-based heuristics, on each trial one of the non-query main items was replaced by a random distractor (while still ensuring the overall ordering remained consistent).

The number of distractors inserted between the two queried items determined the number of steps an in-context, algebraic strategy would have to perform.
Across learning trials, the number of steps required by such a strategy was 0, 1, or 2 with probabilities .10, .45, and .45, respectively.
Runs of the same chosen number of steps were limited to three trials.

The context displayed five to eight adjacent pairs, and always included the relations required to infer the relation between the queried items.
Consecutive trials were constrained so that the same queried pair did not appear more than twice in a row.
During the learning phase, participants received feedback on every trial: after an error, the incorrectly selected image was highlighted in red and participants had to wait for at least 5~s before they could continue; after a correct response, the selected image was highlighted in green for at least 0.5~s.
A break was offered after every 20 learning trials.
Exactly 10 of the 100 learning trials were randomly selected for an IWL-style display.
On those trials the query and feedback were unchanged, but the items in context were covered by gray boxes.
These trials were meant to familiarize participants with the masked test display.

\paragraph{Test phase.}
The test phase consisted of three 40-trial blocks: IWL, ICL, and arbitrage.
The order of the ICL and IWL test blocks was counterbalanced; the arbitrage block always occurred last.
Each block contained the same set of 40 query pairs: all 21 possible pairs, plus 19 repeated pairs chosen to approximately balance symbolic distances 1--6.
No feedback was given during testing.
Instead, the selected image was highlighted in gray and participants had to wait for 1~s before they could continue.

The ICL block used five mutually disjoint novel seven-item orderings.
Each ordering was augmented by two distractors that were assigned to the endpoints; the complete nine-item adjacent chain was displayed on every trial.
The same ordering was prevented from appearing on consecutive trials.
The IWL block queried the seven trained images, but the items in context were covered by gray boxes.
The arbitrage block also queried trained images, but the relations in context were reversed, introducing conflict between the context and prior knowledge, and between the IWL and ICL strategies.
A 10-item post-task questionnaire asked about strategy, difficulty, reliance on clues versus memory, and display comprehension.

\paragraph{Statistical analysis.}
All statistical analyses were implemented in R.
Random effects were chosen from a fixed sequence in order of most complex to simplest.
We report the first model that was fit without warnings, errors, or singular fits.

\emph{Accuracy.}
Accuracy was analyzed with binomial generalized linear mixed-effects models using a logit link.
All 1,600 ICL and 1,600 IWL trials were included, with timeouts contributing incorrect responses.
Separate models for ICL and IWL included symbolic distance as a fixed effect and participant-specific random intercepts and distance slopes: {\small\texttt{is\_correct $\sim$ symbolic\_distance + (1 + symbolic\_distance $\mid$ participant)}}.
A combined model tested the difference between the two distance effects using fixed effects of symbolic distance, test (\texttt{phase}), and their interaction, together with the corresponding participant-specific random effects: {\small\texttt{is\_correct $\sim$ symbolic\_distance * phase + (1 + symbolic\_distance * phase $\mid$ participant)}}.
The logistic models were fit by maximum likelihood with \texttt{lme4::glmer}~\citep{BatesMachlerBolkerEtAl15} using the \texttt{bobyqa} optimizer and a maximum of 200,000 function evaluations.

\emph{Response times.}
Response-time analyses were restricted to correct, non-timeout ICL and IWL trials, yielding 1,252 ICL trials and 1,226 IWL trials before outlier removal.
Response times in milliseconds were transformed with the natural logarithm.
Within each participant and test, trials with an absolute $z$ score greater than 3 in log response time were removed; the grouping means and sample standard deviations were computed separately for every participant--test combination.
This procedure removed one ICL and three IWL trials, leaving 1,251 ICL and 1,223 IWL trials (2,474 total) for analysis.
Separate linear mixed-effects models for ICL and IWL included symbolic distance as a fixed effect and participant-specific random intercepts and distance slopes: {\small\texttt{log\_rt $\sim$ symbolic\_distance + (1 + symbolic\_distance $\mid$ participant)}}.
A combined model tested the distance-by-test interaction with the corresponding participant-specific random effects: {\small\texttt{log\_rt $\sim$ symbolic\_distance * phase + (1 + symbolic\_distance * phase $\mid\mid$ participant)}}.
The response-time models were fit by restricted maximum likelihood with \texttt{lmerTest::lmer}~\citep{KuznetsovaBrockhoffChristensen17}, again using \texttt{bobyqa} with a maximum of 200,000 function evaluations.

\subsubsection{Experiment 2}
\label{app:human2-methods}
\paragraph{Participants.}
We recruited 132 adults through Prolific.
Eligibility was limited to US residents 18--65 years old who were fluent in English, had normal or corrected-to-normal vision, and reported no history of brain injury, dyslexia, color blindness, or cognitive impairment; anyone who had taken part in Experiment 1 or in pilot versions of this experiment was excluded.
Participants were paid \$8 per hour and could earn a performance bonus of up to \$3.56.
All participants gave informed consent under a protocol approved by the Brown University Institutional Review Board.
Participants were assigned to the congruent or shuffled condition from a pre-generated list of randomly ordered pairs of the two conditions.
Data were collected in two waves.
In the first wave (50 participant slots), each participant took the next entry of the list.
Because more shuffled than congruent participants quit or answered at chance in the first wave, the second wave recruited more to the shuffled condition, to reach about 50 retained participants per condition.
Of the 132 participants, 14 did not finish (2 congruent, 12 shuffled).
Of those who finished, we excluded participants whose answers beat chance (one-sided binomial test, $p < .05$) neither with respect to the in-context statements order nor with respect to real-world order (1 congruent, 17 shuffled).
This rule removes random responders but keeps participants who answered from real-world knowledge alone, who score 50\% against the statements in the shuffled condition.
We also excluded participants who pressed the same key on at least 80\% of questions (2, both also at chance) or answered at least 25\% of questions in under 0.5~s (1 shuffled).
The final sample comprised 49 congruent and 50 shuffled participants.
Participants in the final sample were aged 21 to 64 years (mean 40.9, SD 10.6); 50 were female and 48 male, and sex was not available for 1.
One participant in the shuffled condition reloaded the page mid-experiment; only their second, complete session was analyzed.
To ensure that our results were not due to a bias in participant exclusions, we reran the main analyses without the exclusion for answering at chance, and the condition effects remained significant (Appendix~\ref{app:extended-results-human2}). 

\paragraph{Stimuli and software.}
The task ran in a browser in full-screen mode on a desktop or laptop computer, using jsPsych 8.2.3~\citep{deLeeuwGilbertLuchterhandt23}, and was deployed with Firebase Hosting.
Questions always asked which of two animals weighed more, according to the statements.
Thirty-three animals were grouped into nine weight classes of three to five animals, by median body mass across three databases (typical adult mass for the six insects), so that every animal in one class was clearly lighter than every animal in the next.
Each question used one animal from each class, so the nine animals' order by weight was unambiguous.
Within a participant, no pair of animals was queried twice.

\paragraph{Display and response format.}
Each question showed, from top to bottom, the question ``According to the statements, which is \textbf{heavier}?'', the two answer choices, and eight statements, one per line, of the form ``The horse is heavier than the dog.''
The statements specified an ordering of all nine animals but were listed in a scrambled order.
Participants pressed \texttt{F} for the left answer or \texttt{J} for the right, with up to 50~s per question and no visible countdown.
After an answer, the chosen answer was outlined and ``You chose $\langle$answer$\rangle$.'' appeared, without feedback about whether the chosen answer was correct. 
After at least 0.8~s, participants pressed the space bar to continue, and a 0.5~s blank screen preceded the next question.

\paragraph{Instructions.}
The same minimal instructions were used in both conditions.
Participants had to spend a minimum of 4~s per instruction screen before they could move on.
Instructions stated that ``Your goal is to answer according to the given statements,'' described the keys and the 50~s limit, asked participants to answer as quickly and accurately as possible, explained that bonuses would depend on both speed and accuracy, and stated that feedback would show the chosen answer but not whether it was correct.
There were no practice problems or comprehension checks.

\paragraph{Design.}
Each participant answered 48 questions in four blocks of 12, with a break after each of the first three blocks.
The two animals at the ends of each ordering were never queried, giving symbolic distances $d=1,\ldots,6$, with two questions per distance in each block.
Questions were ordered so that no more than four consecutive questions had the correct answer on the same side and no more than two shared a distance.
In the shuffled condition, each question used its own reordering of the nine animals.
Real-world distances were paired with shown distances so that the two distances were nearly uncorrelated.
At every shown distance and every real-world distance the queried relation agreed with the real-world order on exactly half of the questions.
Each reordering was drawn uniformly from those that placed the required animals at the queried positions.
On each question, three to five of the eight statements agreed with real-world order.

\paragraph{Questionnaire.}
After the questions, participants answered an optional questionnaire: how they had found their answers (free text); how many of the statements had been true of the real world; how often they had answered from real-world knowledge instead of the statements; whether they had written anything down; the difficulty of the questions; and any technical or other problems.
A debrief explained that some statements may have been false on purpose.

\paragraph{Statistical analysis.}
All statistical analyses were implemented in R.
Every model had fixed effects of condition, log symbolic distance ($\ln d$, centered), their interaction, and trial number (per 10 questions, centered).
Random effects were chosen from a fixed sequence of models from most complex to simplest.
For each analysis, we report the first model that converged without warnings, errors, or singular fits.
Models were fit with \texttt{lme4} using the \texttt{bobyqa} optimizer with up to 200,000 function evaluations, and distance slopes within each condition were estimated from the fitted models with \texttt{emmeans}.

\emph{Accuracy.}
Accuracy (agreement with the statements' ordering) was analyzed with a binomial generalized linear mixed-effects model with a logit link, fit by maximum likelihood, on all 4,752 questions of the retained participants, with timeouts (1 congruent, 19 shuffled) counted as incorrect: {\small\texttt{is\_correct $\sim$ condition * log\_distance + trial + (1 + log\_distance $\mid$ participant)}}.

\emph{Response times.}
Response-time analyses were restricted to correct, non-timeout answers (4,010).
Response times were natural-log transformed, and answers more than three standard deviations from the participant's mean log response time were removed (22; 0.55\%), leaving 3,988.
The linear mixed-effects model {\small\texttt{log\_rt $\sim$ condition * log\_distance + trial + (1 + log\_distance $\mid$ participant)}} was fit by restricted maximum likelihood with \texttt{lmerTest}~\citep{KuznetsovaBrockhoffChristensen17}.

\subsection{Large language models}
\label{app:llm-methods}

\paragraph{Models.}
We evaluated 22 open-weight LLMs ranging from 1B to 70B parameters.
Ten were reasoning models, trained to generate a chain of thought before answering: Qwen3 (1.7B, 4B, 8B, 14B, and 32B; \citealp{YangLiYangEtAl25}), DeepSeek-R1-Distill (Llama-8B, Qwen-14B, and Qwen-32B; \citealp{GuoYangZhangEtAl25}), Phi-4-reasoning~\citep{AbdinAgarwalAwadallahEtAl25}, and gpt-oss-20b~\citep{OpenAIAgarwalAhmadEtAl25}.
The other 12 were instruction-tuned models that reasoned only when prompted to: Qwen2.5-Instruct (7B and 32B; \citealp{QwenYangYangEtAl24}), Llama-3.1-Instruct (8B and 70B) and Llama-3.2-Instruct (1B and 3B; \citealp{GrattafioriDubeyJauhriEtAl24}), Gemma-2-9B-it~\citep{GemmaTeamRivierePathakEtAl24}, GLM-4-0414 (9B and 32B; \citealp{TeamGLMZengXuEtAl24}), Mistral-7B-Instruct-v0.3~\citep{JiangSablayrollesMenschEtAl23}, and OLMo-2-Instruct (7B and 32B; \citealp{TeamOLMoWalshSoldainiEtAl25}).
Qwen3-32B, shown as an example in the main text, can switch its thinking on or off through its chat template, so its answers with and without a chain of thought came from the same weights and prompt.
All models were run in bfloat16 (gpt-oss-20b dequantized from its released MXFP4 weights) using the \texttt{transformers} library~\citep{WolfDebutSanhEtAl20} on NVIDIA RTX PRO 6000 Blackwell Max-Q GPUs on Brown University's Oscar cluster.

\paragraph{Task and prompts.}
As in the transformers trained from scratch and in the humans, each problem used a single ordering and premises with adjacent pairs of items.
Orderings included 11 items total, but only the inner seven items were queried, giving symbolic distances $d=1,\ldots,6$.
Orderings of 11 rather than 9 were used because we found that some LLMs exhibited strong edge effects, performing better in context on pairs that had at least one item close to the edge of the ordering.
Prompts were formatted according to each model's specific template.
A short system instruction described the meaning of the relations (e.g., ``is taller''), and the 10 adjacent relations were listed in the user turn in scrambled order, followed by the query.
An example Qwen3-32B ICL test prompt is shown below:

\begin{verbatim}
<|im_start|>system
In the following question, > means "is taller" and < means "is shorter".
Answer whether the statement is true or false.<|im_end|>
<|im_start|>user
Bob > Mike
Sue > Joe
Tom > Dan
Amy > Sue
Joe > Mary
Anna > John
Dan > Lisa
Mike > Tom
Lisa > Anna
Mary > Bob

Is it true or false that Bob > Mary?<|im_end|>
<|im_start|>assistant
<think>
</think>
\end{verbatim}

The query always used \texttt{>}, so whether the queried statement was true depended on the order of the two items.
Every pair was asked once as a true and once as a false statement, so a model that always gave the same answer scored exactly .5.
The instruction specified the relation of each real-world dimension: ``has a larger population'' for country, US state, and city population; ``is heavier'' for animal body mass; ``has a larger area'' for country surface area; and ``is taller'' for the given names, with the opposite relation for \texttt{<}.
For Gemma-2 and Phi-4-reasoning, the instruction was placed in the user turn, and the date that Llama-3 chat templates insert was fixed to the template default.
The prompt format was chosen to maximize overall direct ICL test accuracy in a preliminary sweep over 54 prompts, which crossed six relation templates (from \texttt{A > B} to natural-language phrasings) with different item sets, instruction preambles, separators, and ways of writing the relations.

\paragraph{Familiar and novel items.}
The IWL test and the content-effects experiments used familiar items from five real-world dimensions: country population, US state population, city population, animal body mass, and country surface area (Table~\ref{tab:llm-items}).
Values were taken from one source per dimension: the Wikipedia list of countries by population, the US Census Bureau's 2024 population estimates, Demographia's World Urban Areas (2025), the PanTHERIA database of adult body mass, and the World Bank's surface-area indicator.
Items were spaced to place queries at $d=1$ near chance and at $d=6$ near ceiling, based on a pilot in which Qwen3-32B was queried without context or reasoning on items spanning a wide range of values.
We describe the spacing of an item set by the ratio between the values of items six ranks apart: this ratio was 17.8 for country population, animal body mass, and country surface area and 5.6 for US state population, and city populations were evenly spaced.
Countries and cities with few English Wikipedia page views were excluded, as were ambiguous names.
Additional item sets with the same spacing but disjoint items were used to replicate the IWL test (Table~\ref{tab:llm-items}), and one further disjoint set per dimension supplied the uninformative context.
To vary prior strength (Fig.~\ref{fig:content}K), we also built 19 item sets with other spacings, in which items six ranks apart differed in value by factors from 2 to 56.2.
The ICL test used novel items with no known ordering: 11 common English given names (John, Mary, Tom, Sue, Bob, Anna, Mike, Lisa, Joe, Amy, and Dan) assigned to heights at random in each problem and, in a second framing, 11 invented city names ordered by population (Ostrelia, Kavendale, Morvath, Tessalin, Brunmark, Halvora, Quillsby, Zendria, Pelmont, Corvane, and Ilbrook), which used the same instruction as country population.

\begin{table}[!htbp]
\centering
\small
\caption{\textbf{Main LLM item sets.}
Items are listed from smallest (rank 1) to largest (rank 11); the two endpoints were never queried.
Spacing gives the target ratio between adjacent items; replicate sets are the additional disjoint item sets, with the same spacing, used to replicate the IWL test.}
\label{tab:llm-items}
\vspace{4pt}
\begin{tabular}{llllll}
\toprule
Rank & Country & US state & City & Animal & Country \\
 & population & population & population & body mass & surface area \\
\midrule
1 & Jamaica & New Mexico & Vienna & wolf & Tajikistan \\
2 & Panama & Kansas & Milan & cheetah & Romania \\
3 & Singapore & Oklahoma & Chicago & jaguar & Japan \\
4 & Belgium & Alabama & Chennai & gorilla & Kenya \\
5 & Netherlands & Indiana & Ho Chi Minh City & elk & Tanzania \\
6 & Ukraine & New Jersey & Mexico City & zebra & Mongolia \\
7 & Spain & Illinois & Kolkata & buffalo & Algeria \\
8 & Germany & New York & Beijing & giraffe & India \\
9 & Mexico & Florida & Mumbai & hippopotamus & Australia \\
10 & Brazil & Texas & Delhi & rhinoceros & United States \\
11 & United States & California & Jakarta & elephant & Russia \\
\midrule
Spacing & $\times 1.62$ & $\times 1.33$ & even & $\times 1.62$ & $\times 1.62$ \\
Replicate sets & 2 & 2 & 0 & 1 & 2 \\
\bottomrule
\end{tabular}
\end{table}

\paragraph{IWL and ICL tests.}
The IWL test queried familiar items with no relations shown, so that only knowledge stored in the weights could be used to answer.
A query without context has only one possible prompt, so the IWL test was repeated on each of a dimension's item sets (42 trials per set at $d=1,\ldots,6$).
The single-model panels and Fig.~\ref{fig:sde}O pooled all item sets of a dimension, and Fig.~\ref{fig:sde}M and Figs.~\ref{fig:app-llm-sde-qwen3}--\ref{fig:app-llm-sde-llama-olmo} used the first two sets of each dimension.
The ICL test showed all 10 adjacent relations among novel items.
Because LLMs are sensitive to the order in which premises are presented~\citep{ChenChiWangEtAl24}, each ICL distance was run with 10 random premise orders, each with its own random ordering, for 420 trials at $d=1,\ldots,6$.
The single-model panels show the given names, and the cross-model panels average the two novel framings.

\paragraph{Content effects experiments.}
For each dimension, the main item set was queried with all 10 adjacent relations shown, either \emph{congruent} with the real-world ordering, \emph{reversed} relative to it, or \emph{shuffled}~\citep{LiuPriorBalasubramaniamEtAl25,GeertsLiuChanEtAl25}, and answers were scored against the ordering shown.
Each shuffled problem assigned the 11 items to ranks at random (one assignment per distance and premise order, shared by all models), and symbolic distance was measured in the ordering shown.
In the \emph{uninformative} condition, the context showed 10 true relations among 11 other items of the same dimension (written with \texttt{>} or \texttt{<} at random).
In all four context conditions, the instruction was extended by the sentence ``Base your answer only on the comparisons given --- not on outside knowledge.''
As in the ICL test, each problem was repeated with 10 random premise orders.

\paragraph{Reasoning and scoring.}
Without reasoning, the answer was read from a single forward pass.
For reasoning models, the thinking block was closed before the answer: Qwen3's chat template left it empty, an empty block was inserted for the DeepSeek-R1-Distill models and Phi-4-reasoning, and gpt-oss-20b was moved directly to its final-answer channel.
With reasoning, reasoning models generated their own chain of thought, and instruction-tuned models were prompted by appending ``Let's think step by step.'' to the query~\citep{KojimaGuReidEtAl22}.
We used greedy decoding (temperature zero) and varied the reasoning budget $B$, the maximum number of generated tokens, over $B\in\{0,32,64,128,256,384,512,640,768,896,1{,}024\}$, where $B=0$ is a direct answer.
For each trial, one chain was generated up to 1,024 tokens, and each smaller budget used its first $B$ tokens, which is the chain that greedy decoding would produce with that budget.
The chain was followed by the cue ``Therefore the answer is:'', and the answer was scored as without reasoning; this resembles budget forcing~\citep{MuennighoffYangShiEtAl25}, except that the thinking block was not closed.
The prediction was whichever of \texttt{True} and \texttt{False} had the higher total probability at the answer position, summing over capitalized and lowercase forms with and without a leading space.
Accuracy was agreement between the prediction and the queried statement's truth under the reference ordering: the real-world ordering in the no-context and uninformative conditions, and the ordering shown in context otherwise.
Every trial yielded an answer, and no trials were excluded.

\paragraph{Analysis.}
Accuracy at each symbolic distance was the proportion of correct trials, shown with 68\% Wilson intervals.
As in the transformer experiments, the reasoning budget required at each distance was the smallest budget whose accuracy reached 90\% of the maximum over budgets.
A distance whose maximum accuracy was not reliably above chance had no required budget, and uncertainty in the required budget was the standard deviation over 1,000 bootstrap resamples of trials.

The cross-model panels summarized each model with a few numbers.
In Fig.~\ref{fig:sde}M, the distance effect was the least-squares slope of accuracy on distance, averaged over the five dimensions for the IWL test and over the two novel framings for the ICL test.
In Fig.~\ref{fig:sde}O, each model was placed by the difference between its mean required budgets on the ICL and IWL tests and by the slope of its required ICL budget on distance; ICL values were averaged over the two framings and IWL values over the dimensions with a required budget at three or more distances, and models without such a budget on either ICL framing were omitted.
In Fig.~\ref{fig:content}K, each point was one model and item set, and the relation between no-context and reversed-context accuracy was summarized by Spearman's rank correlation.
In Fig.~\ref{fig:content}M, each point was one model and dimension; the content effect was congruent minus reversed accuracy averaged over distances and budgets, and recovery was the least-squares slope of reversed-context accuracy on budget (per 1,000 tokens).
Figures~\ref{fig:sde}M and~\ref{fig:content}K excluded the four models whose no-context accuracy, pooled over the first two item sets of every dimension (378 trials), was not above chance (one-sided binomial test, $p<.05$): Llama-3.2-1B, Llama-3.2-3B, Phi-4-reasoning, and Qwen3-1.7B.

\paragraph{Interventions.}
The interventions in Figs.~\ref{fig:mechanisms}C and~\ref{fig:app-llm-representation} were run on Qwen3-32B for three dimensions: country, city, and US state population.
They were scored on the IWL test and reversed-context trials with direct answers, and on reversed-context trials and the ICL test (given names) after a reasoning budget of 1,024 tokens.
Accuracies were pooled over symbolic distances 2--6 (90 trials per bar: three of the 10 premise orders with context, or three item sets on the IWL test, two for cities), and error bars show the binomial standard error over trials.
A rank direction was fit at every hidden state (the token embeddings and the output of each of the 64 blocks) with the same closed form as for the transformers, from the unedited model's representations at the last subtoken of every mention of the dimension's familiar items (123 countries and 48 cities with at least 1.2 million annual English Wikipedia page views, and all 50 US states).
Edits were applied at the last subtoken of every item mention, at every hidden state, in the prompt, the reasoning chain, and the answer pass.
Writes toward another ordering followed the same procedure as for the transformers; answers after reasoning were collected only at write strengths 0 and 1.
Attention heads were scored on the 252 reasoning trials of the reversed context and the ICL test, by the attention mass from the tokens of every composed inference in the chain (a relation between items at least two ranks apart) onto the context relations mentioning either item.
The 32 highest-scoring of the model's 4,096 heads were ablated together by replacing each head's output with its mean output over all positions, and the control ablated 32 heads drawn at random from outside the top 64.
These procedures differed from the transformers' in four respects.
More heads were ablated (32 of 4,096 rather than 2 of 32), because ablating 16 heads impaired reasoning only for US states.
Each head had one mean output rather than one per position, because positions do not align across trials, whereas the transformers were sensitive to position because every sequence length was always exactly the same for them.
A single random set of heads was drawn, because each draw requires regenerating every reasoning chain.
Only composed inferences were scored, whereas every reasoning step of the transformers restates one relation given in context and was scored.
The city population items in these runs were spaced by a constant ratio rather than evenly, unlike the cities in the other LLM experiments.

%% file: appendix/extended_results.tex
\renewcommand{\floatpagefraction}{0.8}
\section{Extended results}
\label{app:extended-results}

Below we report additional results from transformers trained from scratch, human experiments, and LLMs.

\subsection{Transformers trained from scratch}
\label{app:extended-results-transformers}

\begin{figure}[!htbp]
  \centering
  \includegraphics[width=0.95\textwidth]{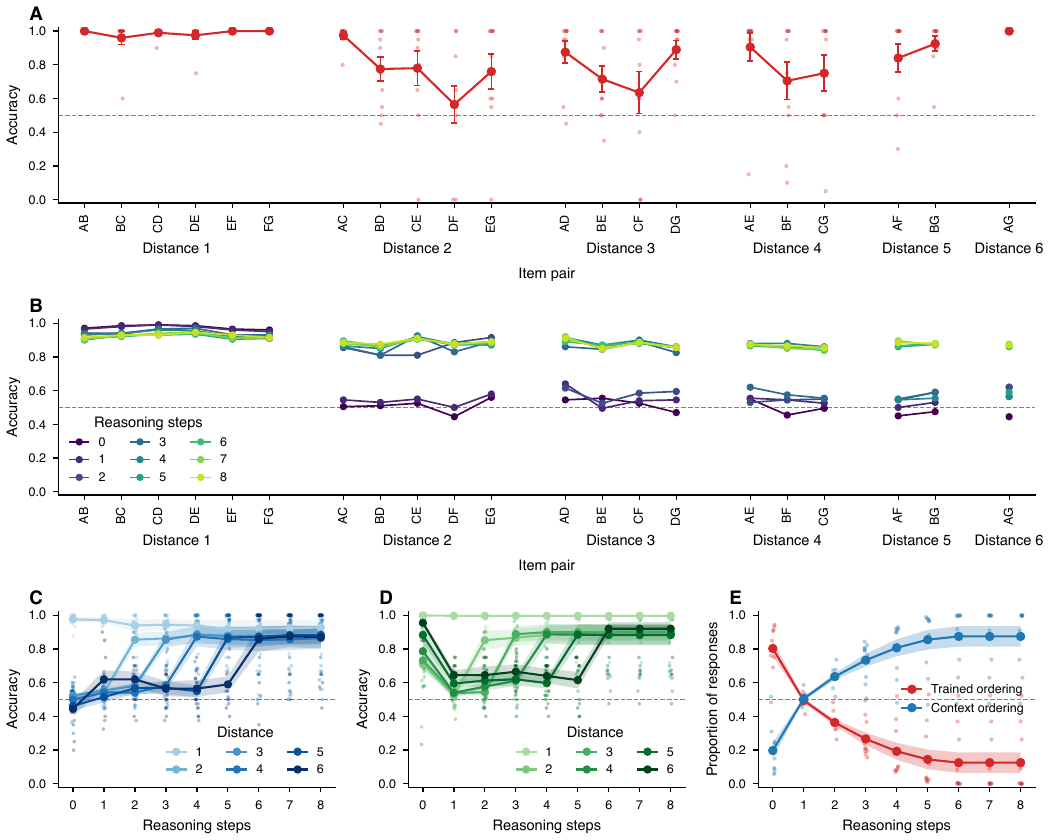}
  \caption{\textbf{Additional symbolic-distance results.}
  (\textbf{A})~IWL test accuracy for each pair of finetuned items with no reasoning steps, grouped by symbolic distance; A--G denote ranks in the learned ordering.
  (\textbf{B})~ICL test accuracy for each pair of ranks in a novel ordering shown in context, with one line per reasoning-step budget.
  (\textbf{C})~The same ICL test as a function of reasoning steps, with one line per symbolic distance.
  (\textbf{D})~Baseline test accuracy for finetuned items accompanied by a context that was congruent with the learned ordering.
  (\textbf{E})~Proportions of arbitrage responses consistent with the finetuned ordering or the reversed ordering shown in context, pooled over distances 2--6.
  Lines and bars show means across 10 seeds; bands and error bars show $\pm1$~SE; faint dots show individual seeds where plotted; dashed lines indicate chance.}
  \label{fig:app-basic-tests}
\end{figure}

\paragraph{Additional symbolic-distance results.}
To provide a fuller view of the symbolic-distance results in Fig.~\ref{fig:sde}, we examined the same networks' performance on individual item pairs, and with different numbers of reasoning steps (Fig.~\ref{fig:app-basic-tests}).
On the IWL test (Fig.~\ref{fig:app-basic-tests}A), the directly trained adjacent pairs showed a strong memorization effect and some terminal-item effects~\citep{LipplKayJensenEtAl24,JensenMunozAlkanEtAl15}.
On the ICL test (Fig.~\ref{fig:app-basic-tests}B), adjacent ICL queries were also answered highly accurately, and did not require intermediate reasoning.
For non-adjacent ICL queries, the full sweep over reasoning steps showed that increasingly distant pairs required more reasoning steps (Fig.~\ref{fig:app-basic-tests}C).

With one or more reasoning steps, the same pattern appeared in a baseline test, which queried the same finetuned items as the IWL test but with an informative context (Fig.~\ref{fig:app-basic-tests}D).
We also used an \emph{arbitrage} test~\citep{PesnotLerousseauSummerfield26,GeertsLiuChanEtAl25} in which the same finetuned items appeared with relations given in context that were reversed with respect to what was learned during finetuning, placing the IWL and ICL-CoT strategies in conflict (similar to the reversed condition in the content-effects experiments).
As expected, without reasoning, responses primarily followed the finetuned (IWL) ordering, whereas additional reasoning shifted responses toward the (ICL) ordering shown in context, though not completely (Fig.~\ref{fig:app-basic-tests}E).

\paragraph{Alternative chain-of-thought algorithms.}
In the main experiments, models' CoTs were supervised on one particular reasoning algorithm (Alg.~\ref{alg:model_cot}). 
Here, we asked whether the reverse SDE depended on the choice of algorithm for generating CoTs.
Using the same pretraining task with a 300-item pool and no finetuning, we trained 40 models with four different algorithms: the \emph{main} algorithm (Alg.~\ref{alg:model_cot}) emitted pairs given in context, starting from the first query item and following the relations; \emph{inferred relations} instead emitted the cumulative inferred relation between that starting query item and the item reached so far; \emph{descend first} began the walk toward lower-ranked items rather than along the first context pair containing the starting query item; and \emph{backtrack at dead ends} restarted from the queried item in the opposite direction after reaching an endpoint.

\begin{figure}[!htbp]
  \centering
  \includegraphics[width=\textwidth]{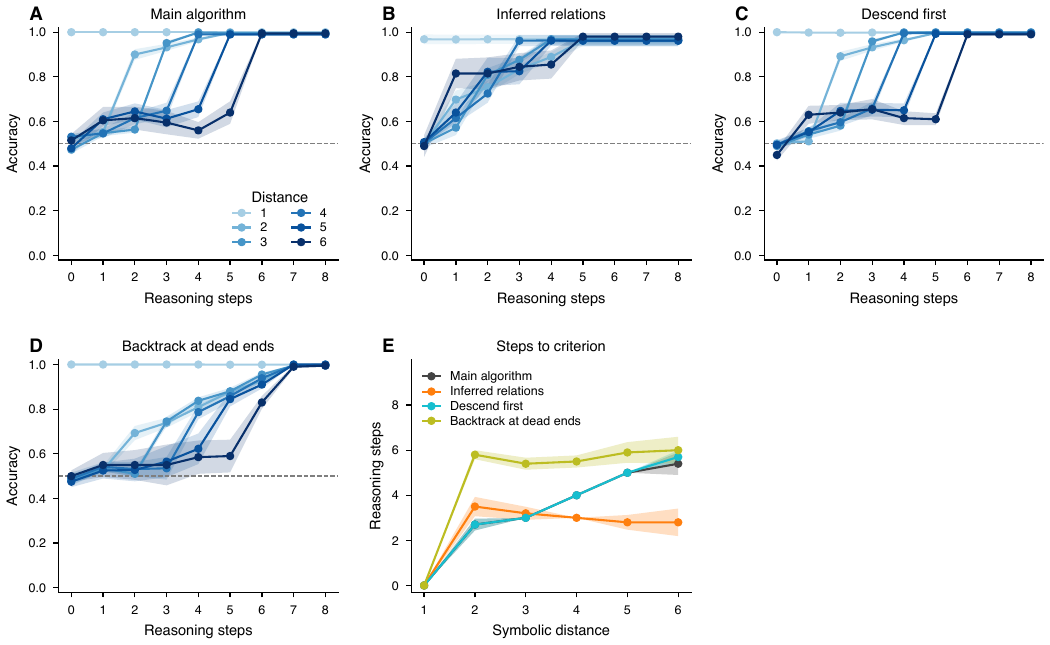}
  \caption{\textbf{Alternative chain-of-thought algorithms.}
  (\textbf{A--D})~ICL test accuracy as a function of reasoning steps after 100,000 pretraining steps for the main, inferred-relations, descend-first, and backtrack-at-dead-ends algorithms, respectively.
  (\textbf{E})~The smallest number of reasoning steps required to reach 90\% of each seed's maximum accuracy, by distance and algorithm.
  Lines show means across 10 seeds and bands show $\pm1$~SE; dashed lines indicate chance.}
  \label{fig:app-cot-algorithms}
\end{figure}

Three of the four algorithms reproduced the main result: more distant comparisons required more reasoning steps (Fig.~\ref{fig:app-cot-algorithms}).
The inferred-relations algorithm instead reversed this pattern, with larger-distance problems reaching criterion in slightly fewer steps.
This suggests that models trained to produce inferred relations may learn a shortcut that allows them to infer larger-distance relations before shorter-distance ones.

\begin{figure}[p]
  \centering
  \setlength{\abovecaptionskip}{3pt}
  \includegraphics[width=\textwidth]{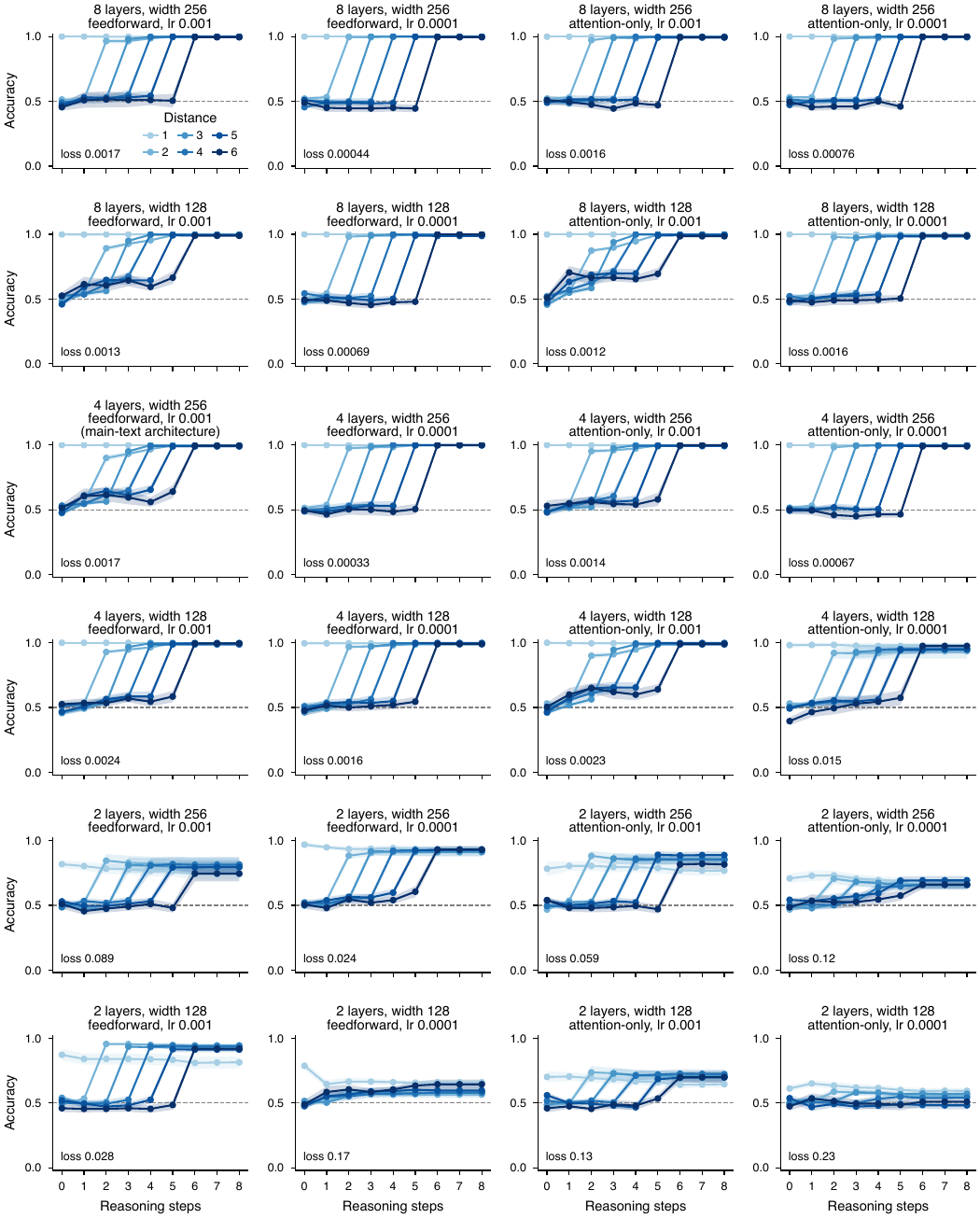}
  \caption{\textbf{Architecture and learning-rate sweep.}
  Each panel shows ICL test accuracy against reasoning steps for one of the 24 architectures after 100,000 pretraining steps; the printed loss is the mean training loss over the last 5,000 steps.
  Lines and bands show mean $\pm1$~SE across 10 seeds; dashed lines indicate chance.}
  \label{fig:app-architectures}
\end{figure}

\paragraph{Architecture robustness.}
To test whether the main CoT results depended on architecture or learning rate, we pretrained 10 initializations for each of 24 combinations of number of layers, residual width ($d_{model}$; feedforward width $4d_{model}$), inclusion of feedforward blocks, and learning rate.
Every architecture that reached a low training loss showed the same pattern, where more reasoning steps were required for larger distances (Fig.~\ref{fig:app-architectures}).
The only exceptions were two-layer networks that did not learn the task as well.

\begin{figure}[!htbp]
  \centering
  \includegraphics[width=\textwidth]{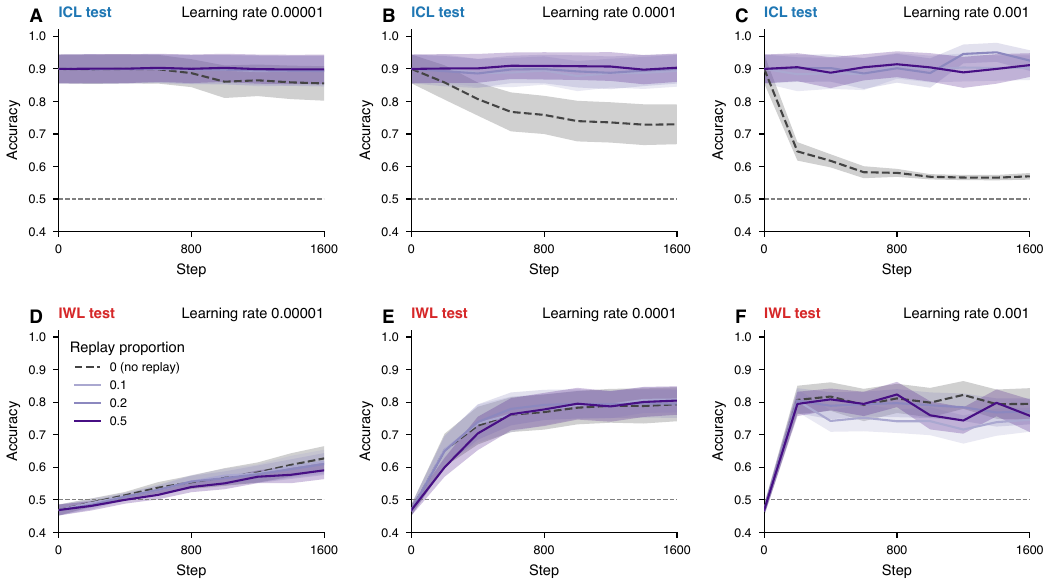}
  \caption{\textbf{Replay during finetuning.}
  Columns show different finetuning learning rates and colors show different levels of replay.
  The top row reports ICL test accuracy (best over reasoning steps); the bottom row reports direct IWL test accuracy with no reasoning steps.
  Accuracy is averaged over symbolic distances 2--6.
  Lines and bands show mean $\pm1$~SE across 10 checkpoints; dashed horizontal lines indicate chance.}
  \label{fig:app-finetune-replay}
\end{figure}

\paragraph{Replay prevents catastrophic forgetting.}
In the main symbolic-distance experiment, fresh pretraining trials replaced 10\% of finetuning examples to prevent the well-known problem of catastrophic forgetting~\citep{McCloskeyCohen89,KirkpatrickPascanuRabinowitzEtAl17} of the ICL-CoT strategy acquired during pretraining.
Here we demonstrate that this replay mechanism prevented catastrophic forgetting (Fig.~\ref{fig:app-finetune-replay}).
Starting from the same 10 pretrained checkpoints, we crossed replay proportions $q\in\{0,.1,.2,.5\}$ with finetuning learning rates $10^{-5}$, $10^{-4}$, and $10^{-3}$ and evaluated performance every 200 gradient steps.

Without replay, the larger learning rates rapidly established good IWL performance, but degraded ICL performance via catastrophic forgetting.
Even modest replay preserved the ICL-CoT strategy while allowing substantial IWL.
These results are consistent with prior work on replay~\citep{Robins95} and complementary learning systems~\citep{McClellandMcNaughtonOReilly95,KumaranHassabisMcClelland16,RussinZolfagharParkEtAl21} in neural networks and in humans. 

\begin{figure}[p]
  \centering
  \includegraphics[width=\textwidth,height=0.83\textheight,keepaspectratio]{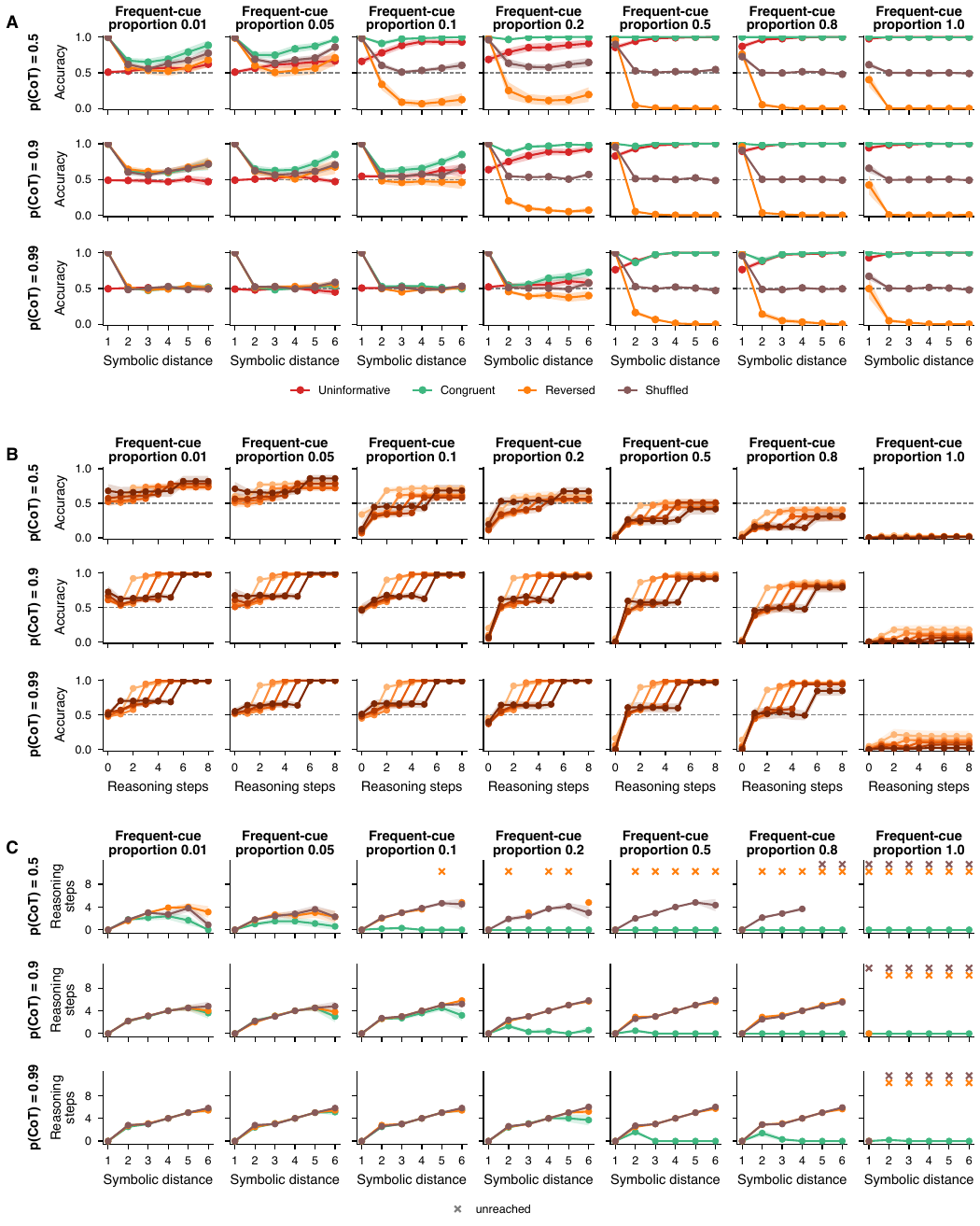}
  \caption{\textbf{Content effects across pretraining distributions.}
  Columns vary the proportion of pretraining trials using the frequent cue $c^*$, and rows vary the probability of CoT supervision, $p(\mathrm{CoT})$.
  (\textbf{A})~Direct accuracy by symbolic distance in the IWL test and under congruent, reversed, or shuffled contexts.
  (\textbf{B})~Reversed accuracy as a function of reasoning steps, with one line per non-adjacent symbolic distance.
  (\textbf{C})~Number of reasoning steps required to reach 90\% of maximum accuracy, by symbolic distance, under congruent, reversed, and shuffled contexts.
  Crosses above the axis mark distances at which most seeds never rose above chance.
  Lines and bands show mean $\pm1$~SE across 10 seeds; dashed lines indicate chance.}
  \label{fig:app-content-grid}
\end{figure}

\paragraph{Content effects across pretraining distributions.}
The content-effects results reported in the main text were obtained using a frequent-cue proportion of .2 and a CoT proportion of $p(\mathrm{CoT})=.9$.
To gain a more comprehensive understanding of the conditions under which these effects occur, we repeated that design across frequent-cue proportions $.01,.05,.1,.2,.5,.8,$ and 1 and $p(\mathrm{CoT})$ values $.5,.9,$ and $.99$ (Fig.~\ref{fig:app-content-grid}).
Results showed that as the frequent ordering became more prevalent, the effect of prior experience strengthened, driving behavior in incongruent conditions without reasoning.
Greater CoT supervision during training generally improved the ability of the model to recover good performance on incongruent trials with reasoning, but simultaneously reduced the model's performance when the context was uninformative at lower frequent-cue proportions (Fig.~\ref{fig:app-content-grid}A, red lines).
This suggests a tradeoff where flexible reasoning can \emph{prevent} learning information that would otherwise be easy to memorize.
However, when the $c^*$ ordering was used on every pretraining trial (Fig.~\ref{fig:app-content-grid}B, rightmost column), the model did not develop a flexible reasoning strategy that could overcome the prior when the context was incongruent, even when CoT supervision was high.
Thus, sufficient CoT supervision allowed the model to overcome interference on incongruent trials, but sufficient diversity in the training distribution was required for reasoning to remain flexible.

\paragraph{Additional mechanism results.}
To supplement the results shown in Fig.~\ref{fig:mechanisms}B, we performed additional experiments investigating the mechanisms underlying each strategy in the transformers trained from scratch (Fig.~\ref{fig:app-representation}; procedures in Appendix~\ref{app:transformer-methods}).
Figure~\ref{fig:app-representation}A shows accuracy of linear decoders (probe accuracy) trained on representations from each layer, including the models' embeddings, in transformers trained with different frequent-cue proportions (i.e., the proportion of trials on which the most frequent cue and corresponding ordering were used). 
As expected, as the frequent cue became more prevalent, the learned ranks became more decodable from the model's item embeddings and activations in all layers.

In the main text (Fig.~\ref{fig:mechanisms}), we report results for erasing the rank direction, pooled over distances 2--6. 
Figure~\ref{fig:app-representation}B shows results across symbolic distances, for erasing but also inverting the rank direction.
As expected, erasing the rank direction reduced IWL performance at every distance, and removed the corresponding SDE (Fig.~\ref{fig:app-representation}B, green).
Inverting the rank direction drove responses below chance and reversed the distance effect (Fig.~\ref{fig:app-representation}B, pink).
We also experimented with writing a random unrelated ordering along the rank direction (Fig.~\ref{fig:app-representation}C).
Doing so increasingly shifted direct answers toward that specific unrelated ordering, whereas agreement after eight reasoning steps remained near chance.
These results are consistent with the hypothesis that the model employs a geometric strategy that uses learned ranks encoded in representational geometry when possible, but is also capable of using an algebraic strategy that is relatively independent of the rank direction.

\begin{figure}[!htbp]
  \centering
  \includegraphics[width=\textwidth]{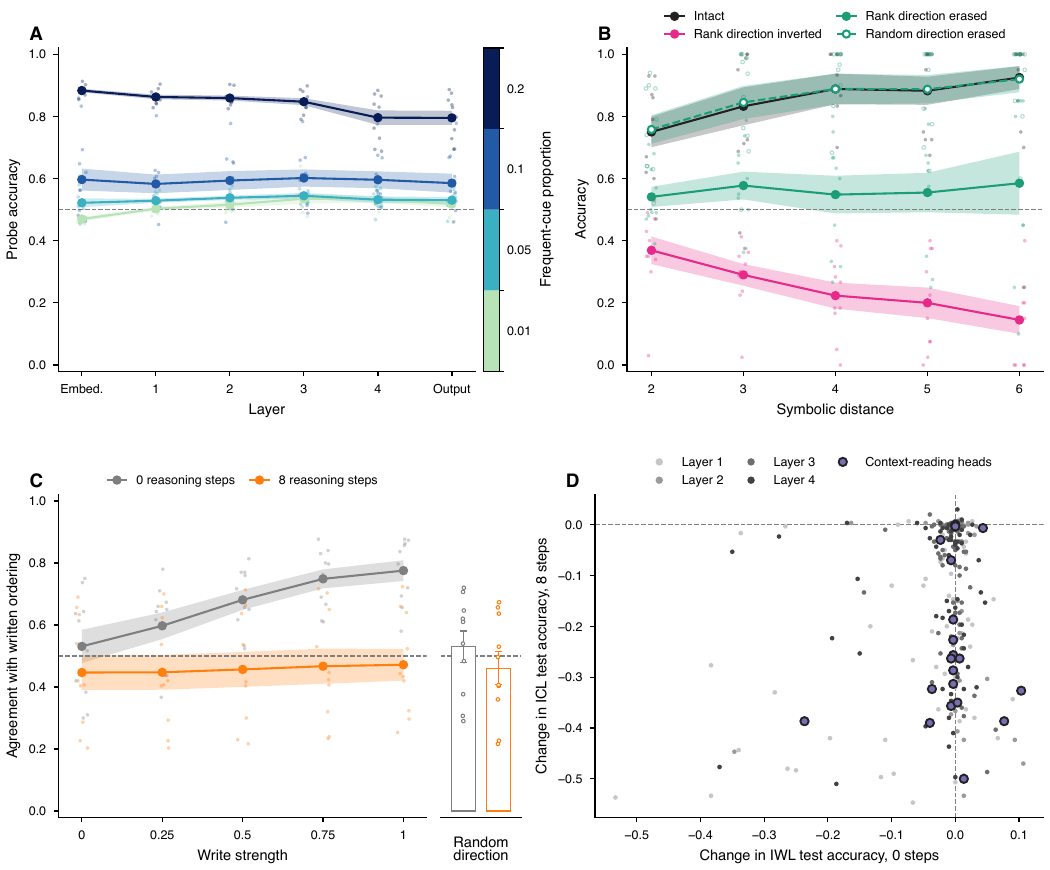}
  \caption{\textbf{Additional mechanism results.}
  (\textbf{A})~Pairwise ordering accuracy of linear rank probes evaluated on held-out items, by layer and the frequent-cue proportion used during training ($p(\mathrm{CoT})$ fixed at .9).
  (\textbf{B})~Direct IWL accuracy at distances 2--6 when the rank direction is intact, erased, or inverted, or a matched random direction is erased.
  (\textbf{C})~Agreement with a random unrelated ordering written along the rank direction on reversed-context trials, with zero or eight reasoning steps.
  Write strength ranges from untouched (0) to full replacement (1); the control bars show full-strength writes along a random direction.
  (\textbf{D})~Change in IWL-test accuracy (zero reasoning steps) and ICL-test accuracy (eight reasoning steps) when each attention head is ablated alone, for every head of every network, shaded by layer; each network's two context-reading heads are outlined.
  Lines and bars show means across 10 seeds; bands and error bars show $\pm1$~SE; dots show individual seeds; dashed lines indicate chance (in D, no change).}
  \label{fig:app-representation}
\end{figure}

The main text reports the result of ablating the top two context-reading heads (Fig.~\ref{fig:mechanisms}B). 
Figure~\ref{fig:app-representation}D shows the change in IWL (zero reasoning steps), and ICL (eight reasoning steps) test performance after ablating every head in the network individually. 
Many attention heads reduced ICL test accuracy with little effect on IWL accuracy, including many of the context-reading heads (shown in purple), as shown by the density of points clustered around the vertical 0 line (no change in IWL accuracy). 
Many heads did not affect either, and only some reduced IWL accuracy without ICL accuracy, or reduced both. 
These results are consistent with the idea that the geometric strategy operates mostly via encodings of rank in representational geometry (and therefore does not depend as much on actions of specific attention heads), while the algebraic strategy is largely independent of representational geometry and is instead supported by specific attention heads that control the model's dynamic reasoning process. 

\begin{figure}[!htbp]
  \centering
  \includegraphics[width=\textwidth]{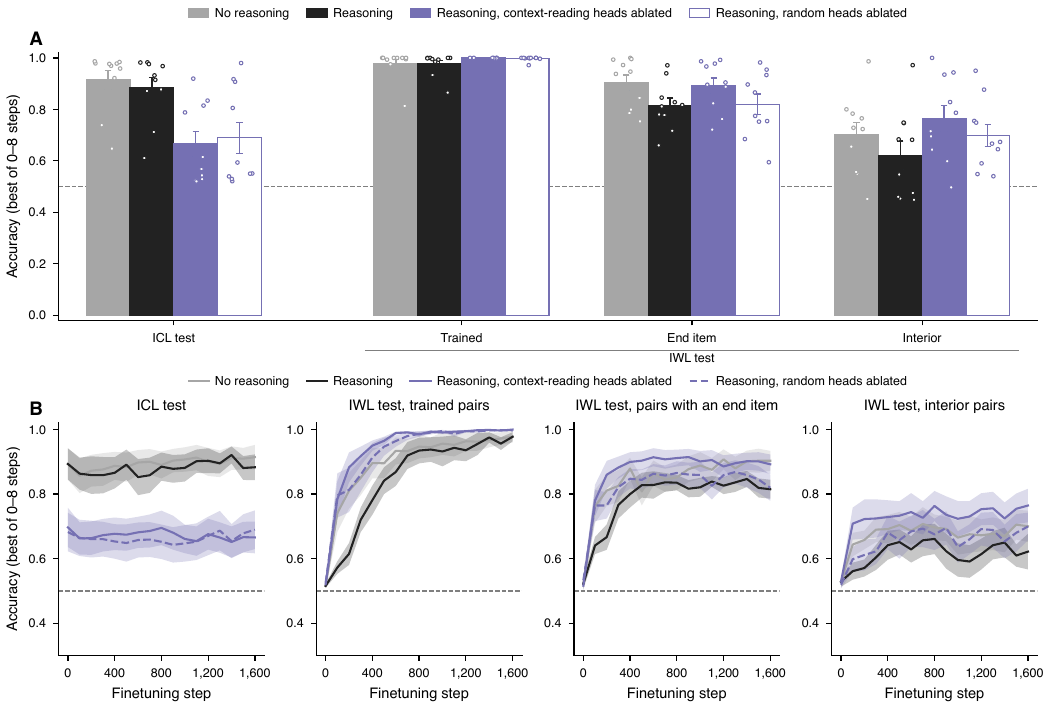}
  \caption{\textbf{Ablating context-reading heads during finetuning.}
  (\textbf{A})~Accuracy on the ICL test and on the IWL test for networks finetuned with or without reasoning, and with reasoning but ablating two context-reading heads or two random heads. IWL accuracy is split by trained pairs, non-adjacent pairs with an end item, and interior non-adjacent pairs. Ablated networks were tested with the same heads ablated.
  (\textbf{B})~The same measures over finetuning.
  Accuracy is the best of 0--8 reasoning steps at test; the ICL test pools distances 2--6.
  Bars and lines show means across 10 seeds; bands and error bars show $\pm1$~SE; dots show individual seeds; dashed lines indicate chance.}
  \label{fig:app-finetuning-reasoning}
\end{figure}

In the main text (Fig.~\ref{fig:mechanisms}E), we report the effect of ablating the context-reading heads while networks reasoned during finetuning.
Figure~\ref{fig:app-finetuning-reasoning} shows results for networks finetuned without reasoning and for networks that reasoned while two random heads were ablated, drawn once per network from outside its top four heads.
Reasoning during finetuning reduced IWL test accuracy on non-adjacent pairs relative to finetuning without reasoning (lower in 9 of 10 networks on both end-item and interior pairs; Fig.~\ref{fig:app-finetuning-reasoning}A).
Surprisingly, ablating random heads in the models from the SDE experiments impaired performance on the ICL test about as much as ablating the context-reading heads, suggesting that the mechanism underlying reasoning in these models was more distributed across heads than in the content-effects models (Fig.~\ref{fig:app-representation}D).
Ablating these heads also increased accuracy on interior pairs, though less than context-reading heads: ablating the context-reading heads gave higher accuracy than ablating random heads on end-item pairs in 8 of 10 networks and on interior pairs in 7 of 10.
The advantage of ablating the context-reading heads appeared within the first 100 steps and persisted throughout finetuning (Fig.~\ref{fig:app-finetuning-reasoning}B).

\begin{figure}[!htbp]
  \centering
  \includegraphics[width=\textwidth]{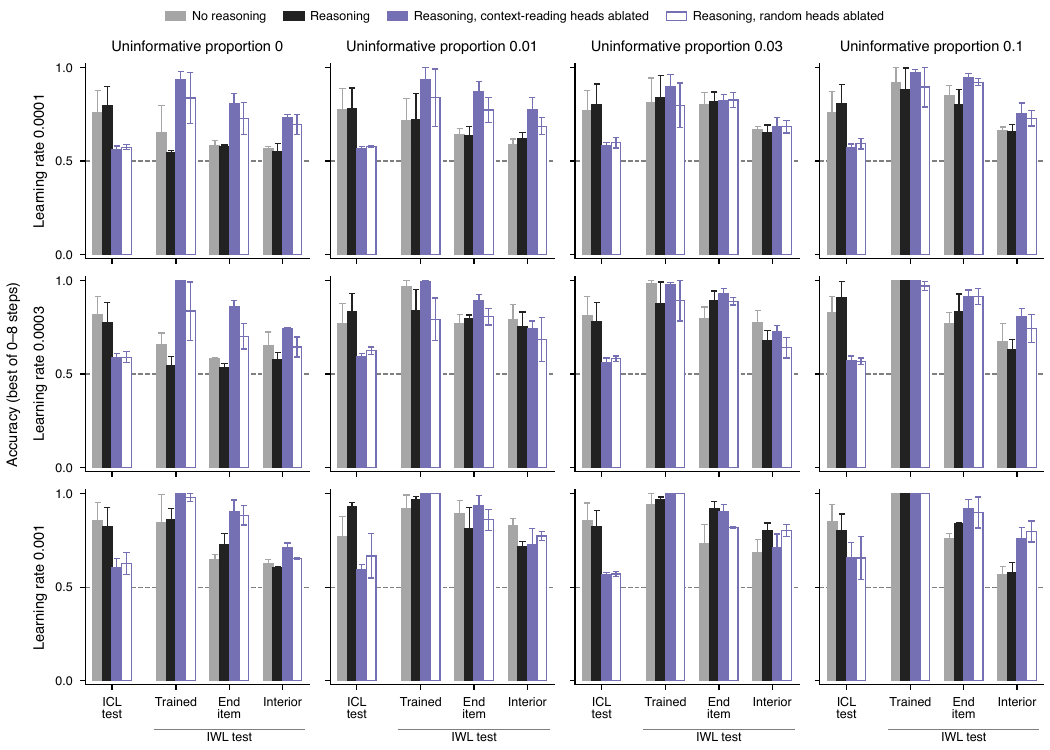}
  \caption{\textbf{Ablation during finetuning across learning rates and proportions of uninformative context trials.}
  As in Fig.~\ref{fig:app-finetuning-reasoning}A, accuracy for each learning rate (rows) and proportion of finetuning trials with an uninformative context (columns).
  Bars show means across 3 seeds; error bars show $\pm1$~SE; dashed lines indicate chance.}
  \label{fig:app-finetuning-reasoning-hyperparameters}
\end{figure}

The learning rate (0.001) and proportion of uninformative trials (1\%) in this experiment differ from those used elsewhere in the paper; they were chosen after inspecting three networks (seeds 0--2) in an exploratory version of the grid in Fig.~\ref{fig:app-finetuning-reasoning-hyperparameters}.
In the seven networks not used for this choice (seeds 3--9), ablating the context-reading heads lowered ICL test accuracy in all seven and raised accuracy on end-item and interior pairs in six of seven each.
Across the grid (three seeds per setting), ablating the context-reading heads lowered ICL test accuracy at all 12 settings and raised IWL test accuracy at most of them (trained pairs 11, end-item pairs 11, interior pairs 10), including the setting used elsewhere in the paper (learning rate 0.0001, 10\% uninformative trials).
In contrast, reasoning without ablation did not consistently lower IWL test accuracy relative to finetuning without reasoning across settings.

\FloatBarrier

\begin{figure}[!htbp]
  \centering
  \includegraphics[width=\textwidth]{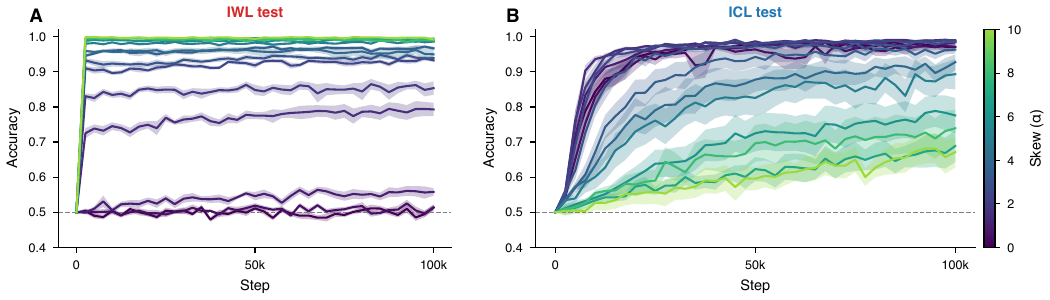}
  \caption{\textbf{Learning over pretraining by skew.}
  (\textbf{A},~\textbf{B})~Test performance varied with data diversity. 
  IWL generalization performance emerges rapidly with high skew (high $\alpha$), while ICL-CoT generalization emerges when orderings are more diverse (low $\alpha$).
  Lines and bands show mean $\pm1$~SE across 10 seeds; dashed lines indicate chance.}
  \label{fig:app-skew-learning}
\end{figure}

\paragraph{Additional data-distribution manipulations.}
In the main skew experiment, IWL performance emerged within the first evaluations when a few orderings dominated pretraining, whereas ICL performance emerged later, and more slowly the more skewed the distribution (Fig.~\ref{fig:app-skew-learning}).

\begin{figure}[!htbp]
  \centering
  \includegraphics[width=\textwidth]{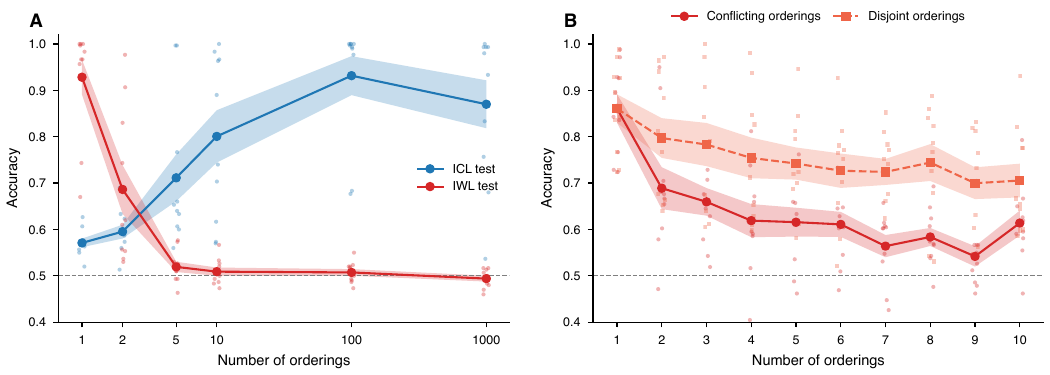}
  \caption{\textbf{Additional data-distribution manipulations.}
  (\textbf{A})~Final ICL and IWL test accuracy after pretraining on different numbers of equally frequent orderings; ICL accuracy is the best over different numbers of reasoning steps, and IWL uses no reasoning steps. Both are averaged over distances 2--6.
  (\textbf{B})~Direct IWL test accuracy after finetuning on different numbers of conflicting orderings over shared items vs orderings over disjoint sets of items. The point at one ordering is shared by both lines.
  Lines and bands show mean $\pm1$~SE across 10 seeds; dashed lines indicate chance.}
  \label{fig:app-orderings}
\end{figure}

We tested the main data-distribution effect in two additional ways (Fig.~\ref{fig:app-orderings}).
First, rather than changing the skew over a fixed set of orderings, we pretrained 10 models for 100,000 gradient steps on 1, 2, 5, 10, 100, or 1,000 equally frequent orderings over a 300-item pool.
ICL was measured using a held-out cue and by taking the maximum over reasoning steps.
As expected, increasing the number of orderings shifted performance from IWL to ICL.

Second, we finetuned 10 pretrained models for 1,600 gradient steps on 1--10 orderings over either the same items (conflict) or disjoint sets of items (no conflict).
Conflicting orderings degraded IWL performance faster than equally numerous disjoint orderings, suggesting that incompatible orderings among the same set of items affect IWL more than the raw number of orderings alone.

\FloatBarrier

\paragraph{Chain-of-thought improves sample efficiency.}
The main results show how a step-by-step algebraic strategy can decompose unfamiliar multi-step problems into familiar single-step operations~\citep{PrystawskiLiGoodman23,WiesLevineShashua23a,LiLiuZhouEtAl24,MerrillSabharwal24,RussinKimPavlickEtAl25}.
Here, we ask whether this CoT strategy results in greater overall sample efficiency than memorizing all orderings independently (IWL) or acquiring a generalizable direct ICL strategy that can be accomplished in a single forward pass~\citep{GargTsiprasLiangEtAl22,BrinkmannSheshadriLevosoEtAl24,GeertsLiuChanEtAl25}.
Using the pretraining setup described in Appendix~\ref{app:extended_methods}, we varied $p(\mathrm{CoT})$ from 0 to 1 in increments of .1, evaluated ICL at initialization and every 2,500 gradient steps for 100,000 total gradient steps, and averaged over 10 seeds.
Because the IWL strategy and direct ICL strategy do not require intermediate reasoning steps, we took the maximum accuracy over all reasoning steps for each model and each seed.
 
\begin{figure}[!htbp]
  \centering
  \includegraphics[width=\textwidth]{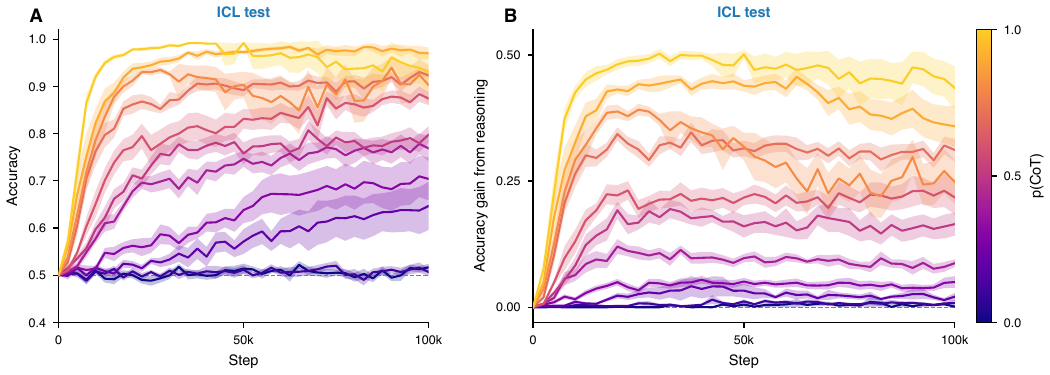}
  \caption{\textbf{Chain-of-thought supervision improves sample efficiency.}
  (\textbf{A})~ICL test learning curves for values $p(\mathrm{CoT})=0,.1,\ldots,1$, scored at the best reasoning budget within seed and distance.
  (\textbf{B})~The corresponding reasoning gap, computed within seed and training step as best-over-budget accuracy minus accuracy with no reasoning steps.
  Both panels average distances 2--6.
  Lines and bands show mean $\pm1$~SE across 10 seeds; dashed lines indicate chance in A and zero gain in B.}
  \label{fig:app-cot-proportion}
\end{figure}

At higher proportions of CoT trials during training, the model generalized to the ICL test in fewer training steps (Fig.~\ref{fig:app-cot-proportion}), consistent with the hypothesis that reasoning during training can improve sample efficiency~\citep{WiesLevineShashua23a,LeeSreenivasanLeeEtAl24,PrystawskiLiGoodman23}.
Unsurprisingly, these models also showed a larger ``reasoning gap''~\citep{PrystawskiLiGoodman23}, as measured by the difference between performance with and without additional reasoning steps at test time.
Previous work has shown that it is possible for transformers to learn to implement an ICL strategy in a single forward pass~\citep{GeertsLiuChanEtAl25}, but in our setup, models trained at a very low proportion of CoT trials did not train as well.

\FloatBarrier

\begin{figure}[!htbp]
  \centering
  \includegraphics[width=\textwidth]{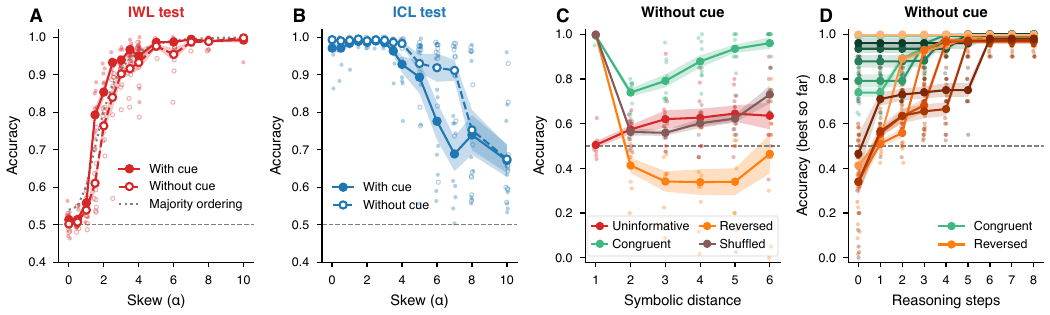}
  \caption{\textbf{Experiments without an explicit cue.}
  (\textbf{A},~\textbf{B})~IWL- and ICL-test accuracy against ordering-frequency skew, with and without the cue, at $p(\mathrm{CoT})=.9$.
The dotted line in A is the IWL test accuracy of answering each pair by the frequency-weighted majority of the training orderings.
  (\textbf{C})~Direct accuracy by symbolic distance under four contexts, without the cue, at frequent-ordering proportion .2 and $p(\mathrm{CoT})=.9$.
  (\textbf{D})~Congruent- and reversed-context accuracy at the same settings, taking the best accuracy up to each number of reasoning steps within seed and distance; darker lines indicate larger distances.
  Lines and bands show mean $\pm1$~SE across 10 seeds; faint dots show individual seeds; dashed horizontal lines indicate chance.}
  \label{fig:app-no-cue}
\end{figure}

\paragraph{Learning without an explicit cue.}
To bring our setup closer to prior studies~\citep{ChanSantoroLampinenEtAl22,PesnotLerousseauSummerfield26}, we repeated the ordering-frequency and content-effects experiments with the same token in the cue position on every training and test trial.
All other procedures were unchanged.
The data-distribution effects were similar, although without the cue, IWL performance was lower at intermediate levels of skew and ICL performance remained high over a broader range (Fig.~\ref{fig:app-no-cue}A,B).
Content effects were also broadly similar, with weaker direct IWL performance and less interference on reversed-context trials (Fig.~\ref{fig:app-no-cue}C,D).

\FloatBarrier

\subsection{Human experiments}
\subsubsection{Experiment 1}
\label{app:extended-results-human}

These analyses extend the main results from Experiment 1 (Fig.~\ref{fig:sde}J,K).

\paragraph{Statistical tables.}
Tables~\ref{tab:exp1-iwl-accuracy}--\ref{tab:exp1-rt-by-test} show the statistical tests reported in the main text.

\begin{table}[!htbp]
\centering
\small
\caption{\textbf{Experiment 1: IWL test accuracy (log-odds).}}
\label{tab:exp1-iwl-accuracy}
\vspace{4pt}
\input{tables/experiment1_iwl_accuracy}
\end{table}

\begin{table}[!htbp]
\centering
\small
\caption{\textbf{Experiment 1: ICL test accuracy (log-odds).}}
\label{tab:exp1-icl-accuracy}
\vspace{4pt}
\input{tables/experiment1_icl_accuracy}
\end{table}

\begin{table}[!htbp]
\centering
\small
\caption{\textbf{Experiment 1: accuracy on both tests (log-odds).} Test coded ICL = 0, IWL = 1; distance not centered.}
\label{tab:exp1-accuracy-by-test}
\vspace{4pt}
\input{tables/experiment1_accuracy_by_test}
\end{table}

\begin{table}[!htbp]
\centering
\small
\caption{\textbf{Experiment 1: IWL test response times (log ms).}}
\label{tab:exp1-iwl-rt}
\vspace{4pt}
\input{tables/experiment1_iwl_rt}
\end{table}

\begin{table}[!htbp]
\centering
\small
\caption{\textbf{Experiment 1: ICL test response times (log ms).}}
\label{tab:exp1-icl-rt}
\vspace{4pt}
\input{tables/experiment1_icl_rt}

\vspace{2.5em}
\caption{\textbf{Experiment 1: response times on both tests (log ms).} Test coded ICL = 0, IWL = 1; distance not centered.}
\label{tab:exp1-rt-by-test}
\vspace{4pt}
\input{tables/experiment1_rt_by_test}
\end{table}

\begin{figure}[!htbp]
  \centering
  \includegraphics[width=\textwidth]{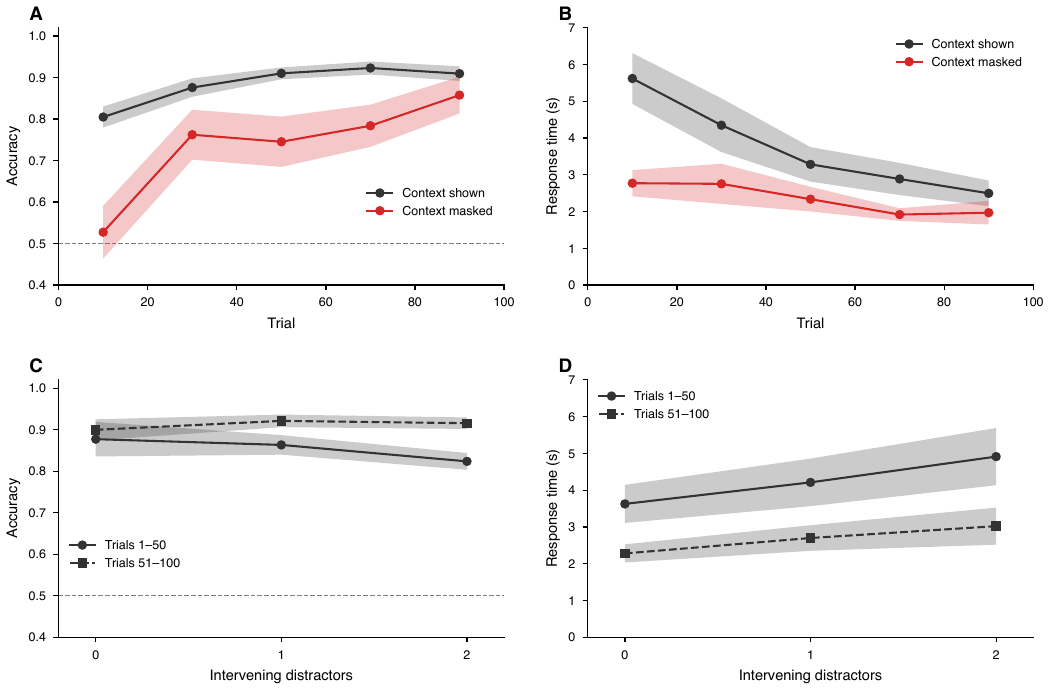}
  \caption{\textbf{Learning phase.}
  (\textbf{A},~\textbf{B})~Accuracy and response time in 20-trial bins with the context shown or masked.
  (\textbf{C},~\textbf{D})~Accuracy and response time by intervening distractors on trials where context was shown, comparing early and late trials.
  Lines and bands show mean $\pm1$~SE across participants; response-time points average participant medians.
  Dashed horizontal lines indicate chance.}
  \label{fig:app-human-learning}
\end{figure}

\paragraph{Learning phase.}
To examine how learning shaped participants' use of the geometric and algebraic strategies, we compared trials with the context shown (90\% of trials) or masked (10\% of trials; Fig.~\ref{fig:app-human-learning}A,B).
When the context was shown, accuracy was high from the beginning and increased over the course of learning, while response times fell.
With the context masked, accuracy rose from near chance (53\%) to 86\%, and responses were relatively fast throughout.

We also compared the effect of the number of intervening distractors (i.e., the number of reasoning steps required by an in-context strategy) across the two halves of the learning phase on trials where the context was shown (Fig.~\ref{fig:app-human-learning}C,D).
Early in learning (trials 1--50), accuracy decreased with more intervening distractors, suggesting reliance on the algebraic strategy.
In the second half of learning (trials 51--100), accuracy was high and relatively insensitive to the number of intervening distractors, suggesting participants were relying more on the geometric strategy.
Response times in the second half of the learning phase still showed some dependence on the number of distractors, but were faster than in the first half.

\begin{figure}[!htbp]
  \centering
  \includegraphics[width=\textwidth]{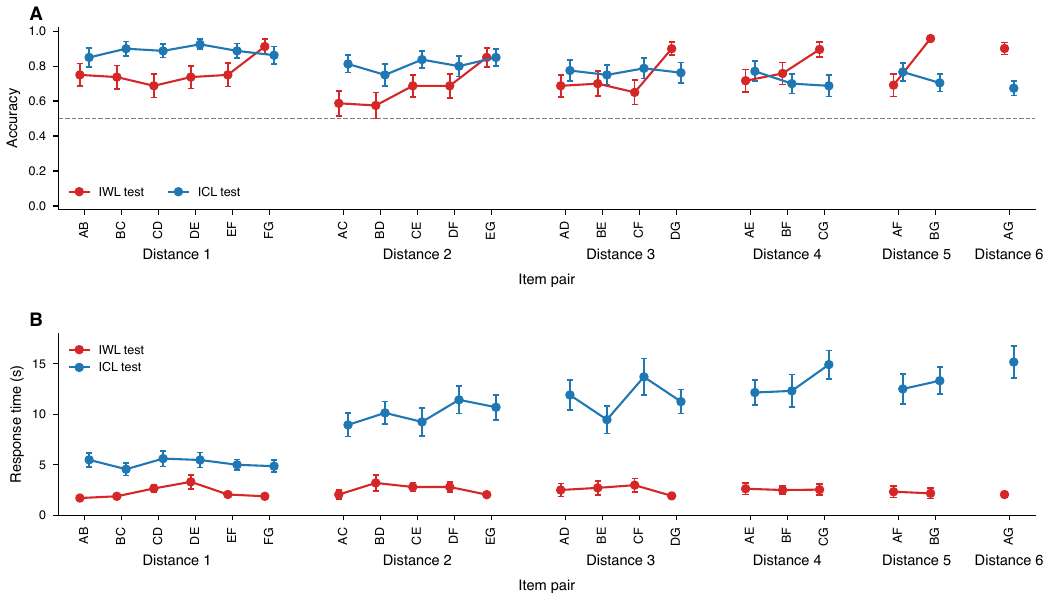}
  \caption{\textbf{Additional human symbolic-distance results.}
  (\textbf{A})~Accuracy and (\textbf{B})~response time for each item pair on the IWL and ICL tests, grouped by symbolic distance.
  A--G denote ranks from highest to lowest.
  Points and error bars show mean $\pm1$~SE across participants; response-time points average participant medians.
  The dashed line marks chance.}
  \label{fig:app-human-pairs}
\end{figure}

\begin{figure}[!htbp]
  \centering
  \includegraphics[width=\textwidth]{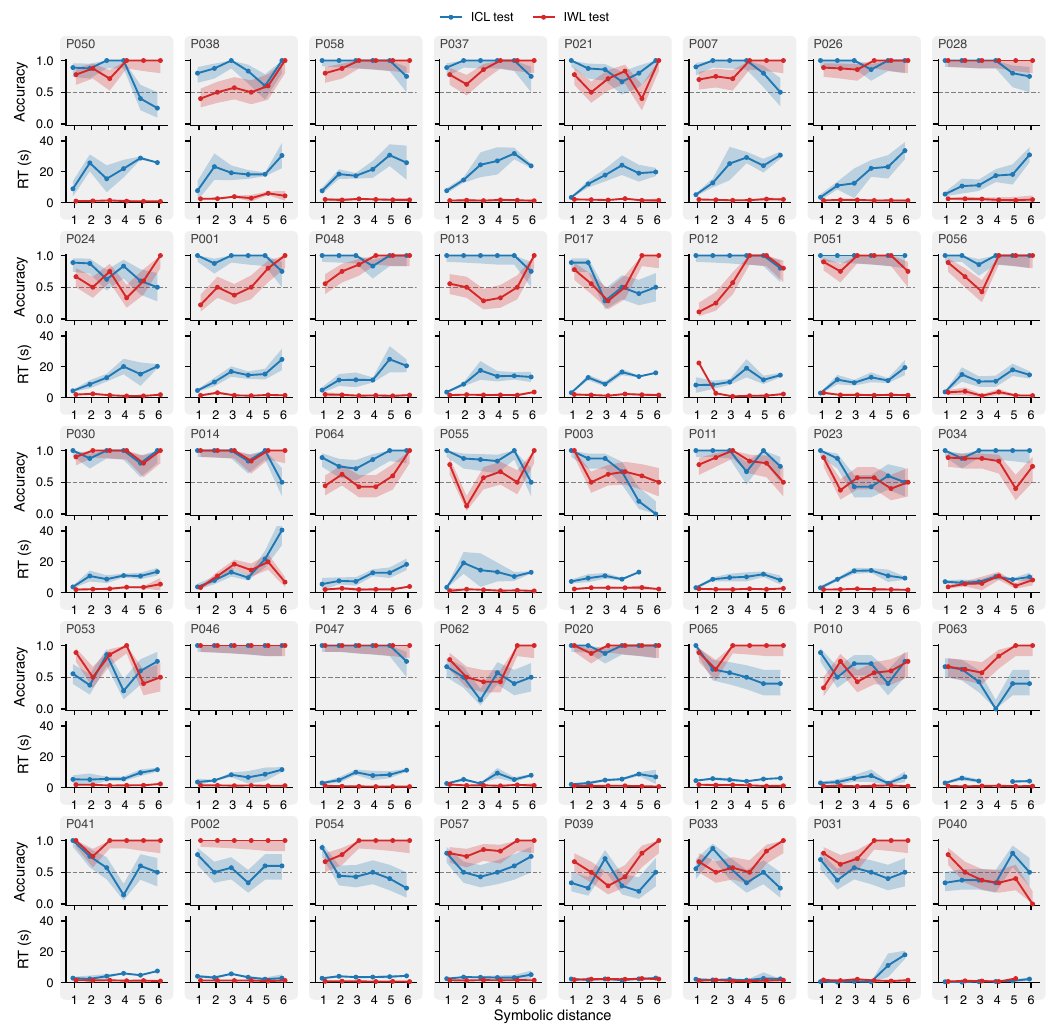}
  \caption{\textbf{Individual participants in Experiment 1.}
  Accuracy (top) and response time (bottom) by symbolic distance on the IWL and ICL tests for each participant, ordered from slowest to fastest median ICL test response time.
  Response-time points are medians over correct trials; bands show 68\% confidence intervals (accuracy) and $\pm1$ bootstrap SD (response time) over each participant's trials.
  The dashed line marks chance.}
  \label{fig:app-human-participants}
\end{figure}

\begin{figure}[!htbp]
  \centering
  \includegraphics[width=\textwidth]{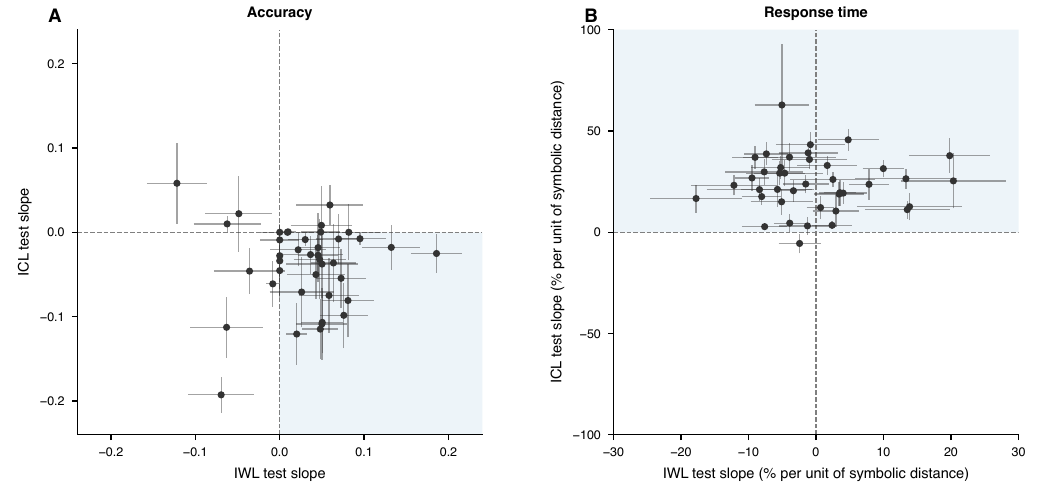}
  \caption{\textbf{Individual symbolic-distance slopes in Experiment 1.}
  Each participant's slope of (\textbf{A})~accuracy and (\textbf{B})~log response time (shown as percent change) against symbolic distance on the IWL and ICL tests.
  Shaded regions mark an SDE on the IWL test with a reverse SDE on the ICL test in \textbf{A} and response times increasing with distance on the ICL test in \textbf{B}.
  Error bars show $\pm1$ bootstrap SD.}
  \label{fig:app-human-slopes}
\end{figure}

\paragraph{Additional symbolic-distance results.}
To provide a fuller view of the distance effects in Fig.~\ref{fig:sde}J,K, we examined accuracy and response times for all 21 item pairs, including the adjacent pairs that were directly trained (Fig.~\ref{fig:app-human-pairs}).
On the IWL test, in addition to the symbolic distance effect (SDE), we also observed a memorization effect and a partial terminal-item effect.
The memorization effect can be seen in that the directly trained adjacent pairs had generally higher accuracy than the distance-two pairs. 
On pairs including terminal items, accuracy was high for every pair containing the lowest rank item G, but not for pairs containing the highest rank item A.
On the ICL test, participants showed a strong reverse symbolic distance effect in both accuracy and response time: accuracy generally declined with symbolic distance, and response times increased from approximately 5~s for adjacent pairs to 15~s for the longest-distance pair (AG), consistent with an algebraic strategy.
IWL response times remained between 1.7 and 3.3~s across all pairs and did not depend much on distance.
Figure~\ref{fig:app-human-participants} shows accuracy and response time by symbolic distance for each participant, and Fig.~\ref{fig:app-human-slopes} shows each participant's slope of accuracy and log response time against symbolic distance on each test.
The accuracy slope was positive on the IWL test and negative on the ICL test for 22 of the 40 participants, and response times increased with distance on the ICL test for 39 of the 40 participants.

\begin{figure}[!htbp]
  \centering
  \includegraphics[width=\textwidth]{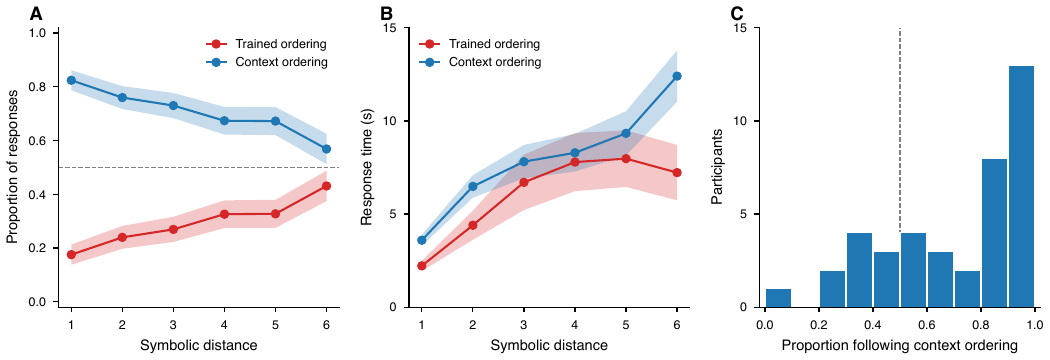}
  \caption{\textbf{Human arbitrage test.}
  (\textbf{A})~Proportions of responses following the trained ordering or the reversed ordering shown in context, by symbolic distance.
  (\textbf{B})~Response times split by response type.
  Lines and bands show mean $\pm1$~SE across participants; response-time points average participant medians.
  (\textbf{C})~Individual differences in context-following proportions.}
  \label{fig:app-human-arbitrage}
\end{figure}

\begin{samepage}
\paragraph{Arbitrage test.}
To examine competition between the geometric and algebraic strategies in humans, we analyzed results from the arbitrage test~\citep{PesnotLerousseauSummerfield26}, where the relations shown in context reversed the ordering seen during the learning phase, placing prior knowledge and contextual information in conflict.
Because there were only two responses available on each trial (the training-consistent and context-consistent responses), we could not determine from response data alone whether a given choice reflected success of one of the strategies or failure of the other.
Overall, responses were strongly biased toward the context ordering (72.5\%), but this bias decreased with distance (Fig.~\ref{fig:app-human-arbitrage}A).
Responses consistent with the trained ordering were faster overall, but both response types slowed with symbolic distance (Fig.~\ref{fig:app-human-arbitrage}B).
Of the 40 participants, 15 chose the context-consistent response on at least 90\% of trials (Fig.~\ref{fig:app-human-arbitrage}C).
This pattern of results suggests that participants were biased toward the algebraic strategy on the arbitrage test, and produced more errors (i.e., responses consistent with the trained rather than the context ordering) on more distant queries.
This bias may have resulted from the emphasis on using information in context in the instructions.
\par
\end{samepage}

\FloatBarrier

\subsubsection{Experiment 2}
\label{app:extended-results-human2}

These analyses extend the main results from Experiment 2 (Fig.~\ref{fig:content}H,I).

\paragraph{Statistical tables.}
Tables~\ref{tab:exp2-accuracy} and~\ref{tab:exp2-rt} show the statistical tests reported in the main text.

\begin{table}[!htbp]
\centering
\small
\caption{\textbf{Experiment 2: accuracy (log-odds).} Condition coded congruent $-0.5$, shuffled $+0.5$.}
\label{tab:exp2-accuracy}
\vspace{4pt}
\input{tables/experiment2_accuracy}
\end{table}

\begin{table}[!htbp]
\centering
\small
\caption{\textbf{Experiment 2: response times (log ms).} Condition coded congruent $-0.5$, shuffled $+0.5$.}
\label{tab:exp2-rt}
\vspace{4pt}
\input{tables/experiment2_rt}
\end{table}

\begin{figure}[!htbp]
  \centering
  \includegraphics[width=\textwidth]{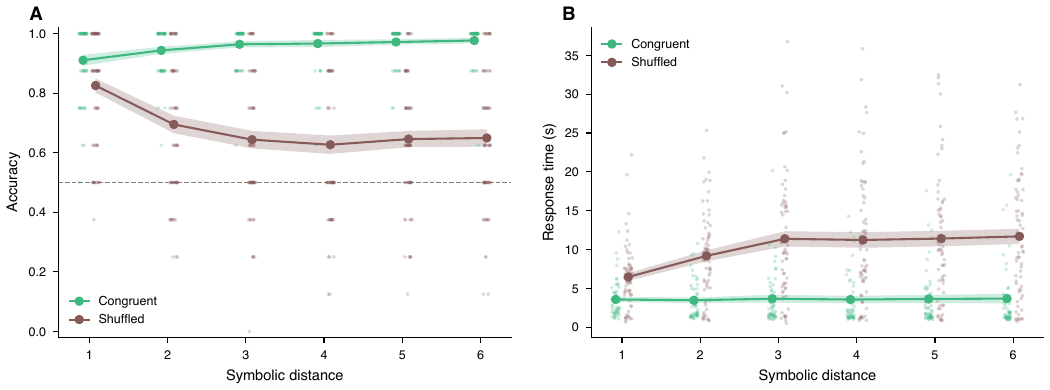}
  \caption{\textbf{Experiment 2 results for participants who finished the task.}
  (\textbf{A})~Accuracy and (\textbf{B})~response time by symbolic distance for participants who finished (49 congruent, 66 shuffled), including shuffled participants excluded from the main analysis for answering at chance.
  Lines and bands show mean $\pm1$~SE across participants; dots show individual participants; response-time points average participant medians.
  The dashed line marks chance.}
  \label{fig:app-human2-robustness}
\end{figure}

\paragraph{Robustness.}
Because more shuffled than congruent participants quit or answered at chance (Appendix~\ref{app:human2-methods}), we tested whether the condition effects in Fig.~\ref{fig:content}H,I depended on which participants were included in the analysis.
Figure~\ref{fig:app-human2-robustness} shows the same results for every participant who finished, applying only the exclusions for pressing the same key on at least 80\% of questions (two participants) and for answering under 0.5~s on at least 25\% of questions (one more participant), including shuffled participants who were excluded from the main analysis for answering at chance. 
This left 49 participants in the congruent condition and 66 in the shuffled condition.
When statistical models were refit on this sample, the main effects of condition and the condition-by-distance interactions on accuracy and response time all remained significant ($p < .001$).

Because we assigned more participants to the shuffled condition after the first wave to offset its higher rates of quitting and answering at chance, the conditions were unevenly split across the two waves in the main analysis (first wave: 30 congruent, 13 shuffled; second wave: 19 congruent, 37 shuffled).
We therefore refit the models with wave and its interactions with condition and distance, which estimates each of the four effects within each wave and averages the two estimates with equal weight.
All four remained significant ($p < .001$), were significant within each wave separately ($p < .001$), and did not differ reliably between the waves (all $p \geq .327$).
They also remained significant ($p < .001$) without the seven congruent participants recruited at the end of the second wave, when only the congruent condition was being assigned.

Leaving out one participant at a time never changed the sign or significance of any effect (99 refits per model), Lee bounds for the greater dropout in the shuffled condition (on participant means and slopes; 2,000 bootstrap samples) excluded zero for every effect, and both interactions remained significant with symbolic distance coded as logarithmic, linear, two-piece (a step at distance two, then linear), or categorical.
The significant distance slopes within each condition also held in the refits on all finishers, with wave, and without the seven late-recruited congruent participants; the congruent response-time slope, which was not significant in the main analysis ($p = .050$), ranged from $p = .047$ to $.069$ across these refits.

\FloatBarrier

\paragraph{Agree and conflict questions.}
In the shuffled condition, the queried relation agreed with the real-world ordering on half of the questions at each distance (agree questions) and conflicted with it on the other half (conflict questions).
To examine how prior knowledge contributed to the lower accuracy and reverse SDE in the shuffled condition (Fig.~\ref{fig:content}H,I), we compared the two types of questions (Fig.~\ref{fig:app-human2-agree-conflict}), fitting the models described in Appendix~\ref{app:human2-methods} to the shuffled condition with question type in place of condition.
Models that included participant-specific effects of question type had singular fits, so the tests of question type may be anti-conservative; we therefore also report participant-level tests (sign-flip tests of each participant's least-squares coefficients).
Accuracy was lower on conflict than on agree questions at every distance (.76 vs .95 at distance one; .56--.61 vs .83--.87 at distances 2--6; $b = -1.749$, $\mathrm{SE} = 0.117$, $z = -14.94$, $p < .001$), and this difference held on all finishers, with wave, and in participant-level tests.
Accuracy declined with distance on both types of questions, and the decline did not differ reliably between them (question type $\times$ distance: $p = .783$), although participant-level tests detected a slightly larger decline on conflict questions ($p = .018$).
Response times increased with distance on both types of questions.
Correct answers to conflict questions were slower than correct answers to agree questions ($b = 0.084$, $\mathrm{SE} = 0.029$, $t(1668.44) = 2.93$, $p = .003$), but this difference did not hold in participant-level tests ($p = .535$) or when all answered questions were analyzed ($p = .772$), and it differed between the two waves ($p = .002$).
Overall, these results are consistent with the hypothesis that prior knowledge interfered with performance in the shuffled condition.

\begin{figure}[!htbp]
  \centering
  \includegraphics[width=\textwidth]{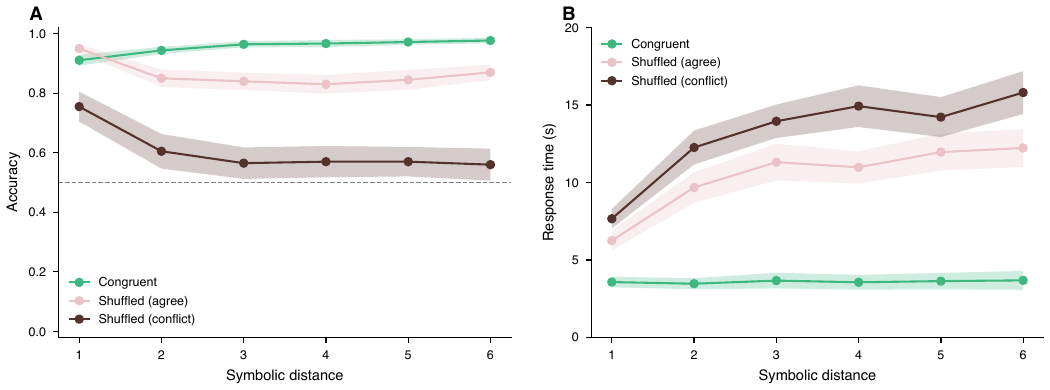}
  \caption{\textbf{Agree and conflict questions.}
  (\textbf{A})~Accuracy and (\textbf{B})~response time by symbolic distance in the congruent condition and on shuffled-condition questions whose queried relation agreed or conflicted with the real-world ordering.
  Lines and bands show mean $\pm1$~SE across participants; response-time points average participant medians over correct answers.
  The dashed line marks chance.}
  \label{fig:app-human2-agree-conflict}
\end{figure}
\FloatBarrier

\begin{figure}[!htbp]
  \centering
  \includegraphics[width=\textwidth]{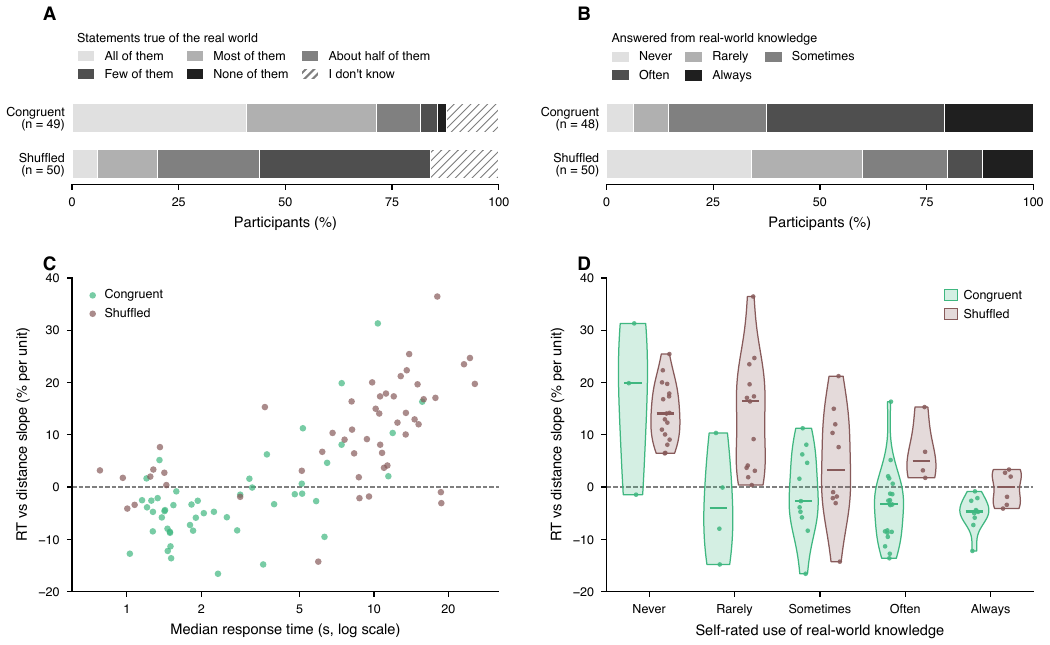}
  \caption{\textbf{Strategies and self-reports.}
  (\textbf{A})~Judged proportion of true statements and (\textbf{B})~self-rated use of real-world knowledge, by condition.
  (\textbf{C})~Each participant's median response time against their RT vs distance slope: the percent change in response time per unit of symbolic distance, $100(e^{b}-1)$, where $b$ is the distance coefficient of a least-squares regression of log response time on symbolic distance and question number over the participant's correct answers.
  (\textbf{D})~The same slope by answer in B; violins show the distribution, points show individual participants, and lines mark medians.}
  \label{fig:app-human2-strategies}
\end{figure}

\begin{figure}[!htbp]
  \centering
  \includegraphics[width=\textwidth]{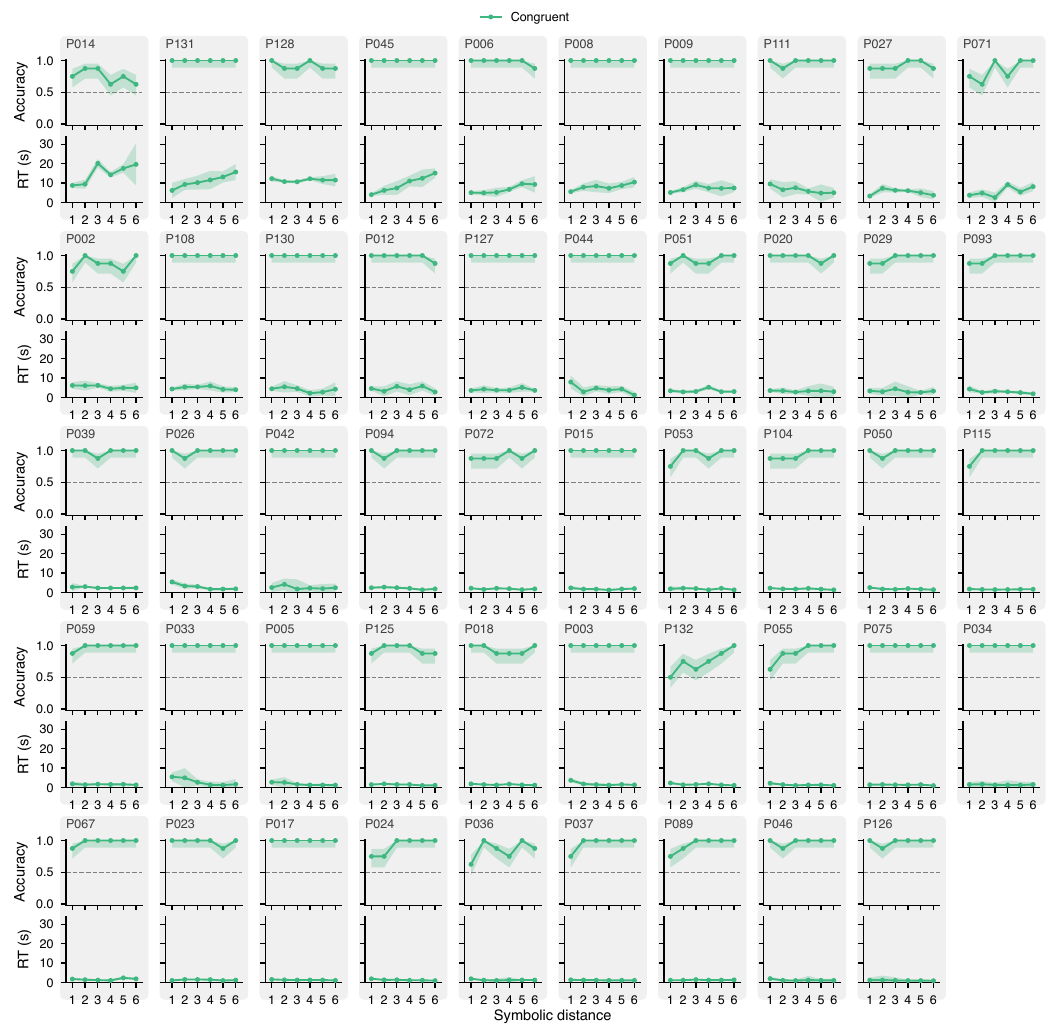}
  \caption{\textbf{Individual participants in the congruent condition of Experiment 2.}
  Accuracy (top) and response time (bottom) by symbolic distance for each participant, ordered from slowest to fastest median response time.
  Response-time points are medians over correct answers; bands show 68\% confidence intervals (accuracy) and $\pm1$ bootstrap SD (response time) over each participant's questions.
  The dashed line marks chance.}
  \label{fig:app-human2-participants-congruent}
\end{figure}

\begin{figure}[!htbp]
  \centering
  \includegraphics[width=\textwidth]{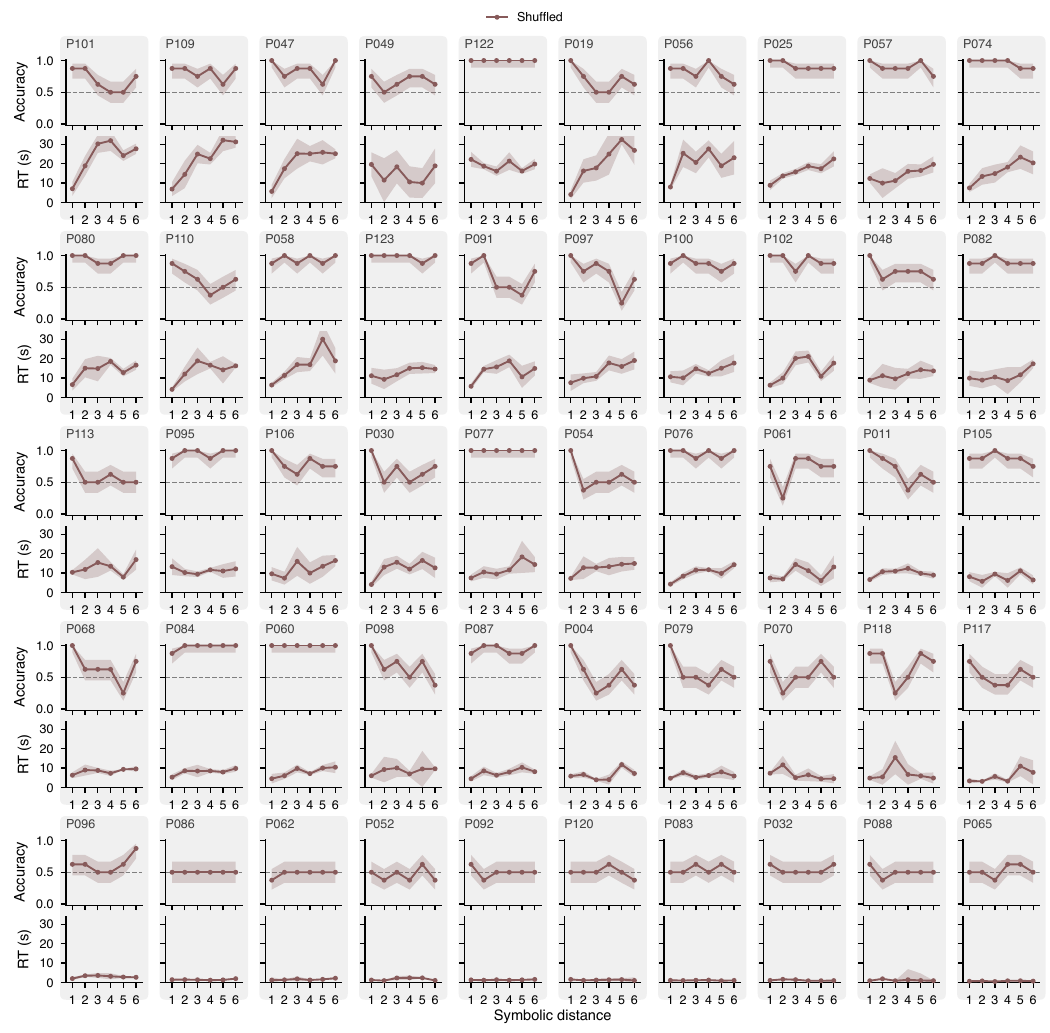}
  \caption{\textbf{Individual participants in the shuffled condition of Experiment 2.}
  Accuracy (top) and response time (bottom) by symbolic distance for each participant, ordered from slowest to fastest median response time.
  Response-time points are medians over correct answers; bands show 68\% confidence intervals (accuracy) and $\pm1$ bootstrap SD (response time) over each participant's questions.
  The dashed line marks chance.}
  \label{fig:app-human2-participants-shuffled}
\end{figure}

\paragraph{Strategies and self-reports.}
The results in Fig.~\ref{fig:content}H,I suggest that participants answered from prior knowledge (the geometric strategy) when the statements were congruent with it, and chained through the statements (the algebraic strategy) when they were not.
To examine strategy use in individual participants, we compared questionnaire answers (Appendix~\ref{app:human2-methods}) with each participant's response times (Fig.~\ref{fig:app-human2-strategies}).
In the questionnaire, a much higher proportion of participants in the congruent condition answered \emph{All of them} or \emph{Most of them} when asked how many of the statements were true (Fig.~\ref{fig:app-human2-strategies}A), and a much higher proportion of participants in the congruent condition reported answering from real-world knowledge \emph{Often} or \emph{Always} (Fig.~\ref{fig:app-human2-strategies}B).
To measure how much each participant's response times depended on distance, we regressed log response time on symbolic distance and question number over the participant's correct answers (outliers removed as in Appendix~\ref{app:human2-methods}), and expressed the distance coefficient $b$ as the percent change in response time per unit of symbolic distance, $100(e^{b}-1)$ (the RT vs distance slope).
Across congruent participants, the median response time was 1.9~s and the median slope was $-3.3\%$; across shuffled participants, they were 10.3~s and $+9.6\%$ (Fig.~\ref{fig:app-human2-strategies}C).
Within each condition, slower participants had more positive slopes (Spearman's $\rho = .48$ congruent, $.61$ shuffled), and both conditions included fast participants with flat or negative slopes and slow participants with positive slopes.
Figures~\ref{fig:app-human2-participants-congruent} and~\ref{fig:app-human2-participants-shuffled} show accuracy and response time by symbolic distance for each participant in the two conditions, ordered from slowest to fastest median response time.
Slopes also tracked self-reports (Fig.~\ref{fig:app-human2-strategies}D): shuffled participants who reported \emph{Never} or \emph{Rarely} answering from real-world knowledge had the most positive slopes (medians 14\% and 16\%, compared with 0--5\% for the other answers), and among congruent participants, median slopes were negative within every answer choice except \emph{Never}.
These results suggest that both strategies were used in both conditions, but that the condition shifted the balance between them, with many more participants in the congruent condition relying on real-world knowledge.

\FloatBarrier

\clearpage
\subsection{Large language models}
\label{app:extended-results-llm}

\paragraph{Additional symbolic-distance results.}
To provide a fuller view of the LLM results in Fig.~\ref{fig:sde}L,M, here we show direct accuracy for all 22 models and five dimensions (Figs.~\ref{fig:app-llm-sde-qwen3}--\ref{fig:app-llm-sde-llama-olmo}; procedures in Appendix~\ref{app:llm-methods}).
On the IWL test, accuracy was higher at distance six than at distance one in 82 of the 110 model--dimension combinations, consistent with a symbolic distance effect.
On the ICL test, accuracy was lower at distance six than at distance one in 15 of the 22 models, consistent with a reverse symbolic distance effect.

\begin{figure}[H]
  \centering
  \includegraphics{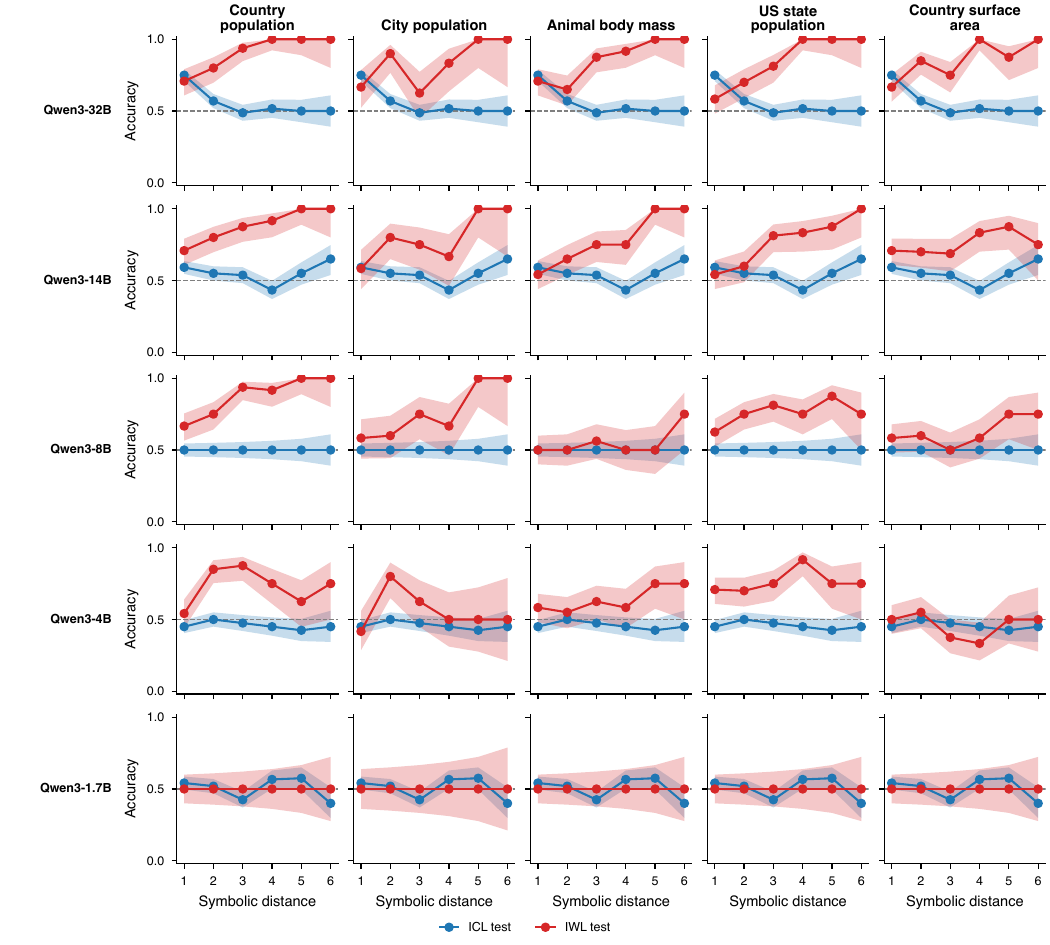}
  \caption{\textbf{LLM symbolic-distance results: Qwen3 models.}
  Direct accuracy on the IWL and ICL tests, with one panel per model (row) and dimension (column).
  All plots use the innermost seven ranks of 11-item orderings.
  Bands show 68\% confidence intervals over trials; dashed lines indicate chance.}
  \label{fig:app-llm-sde-qwen3}
\end{figure}

\begin{figure}[p]
  \centering
  \includegraphics{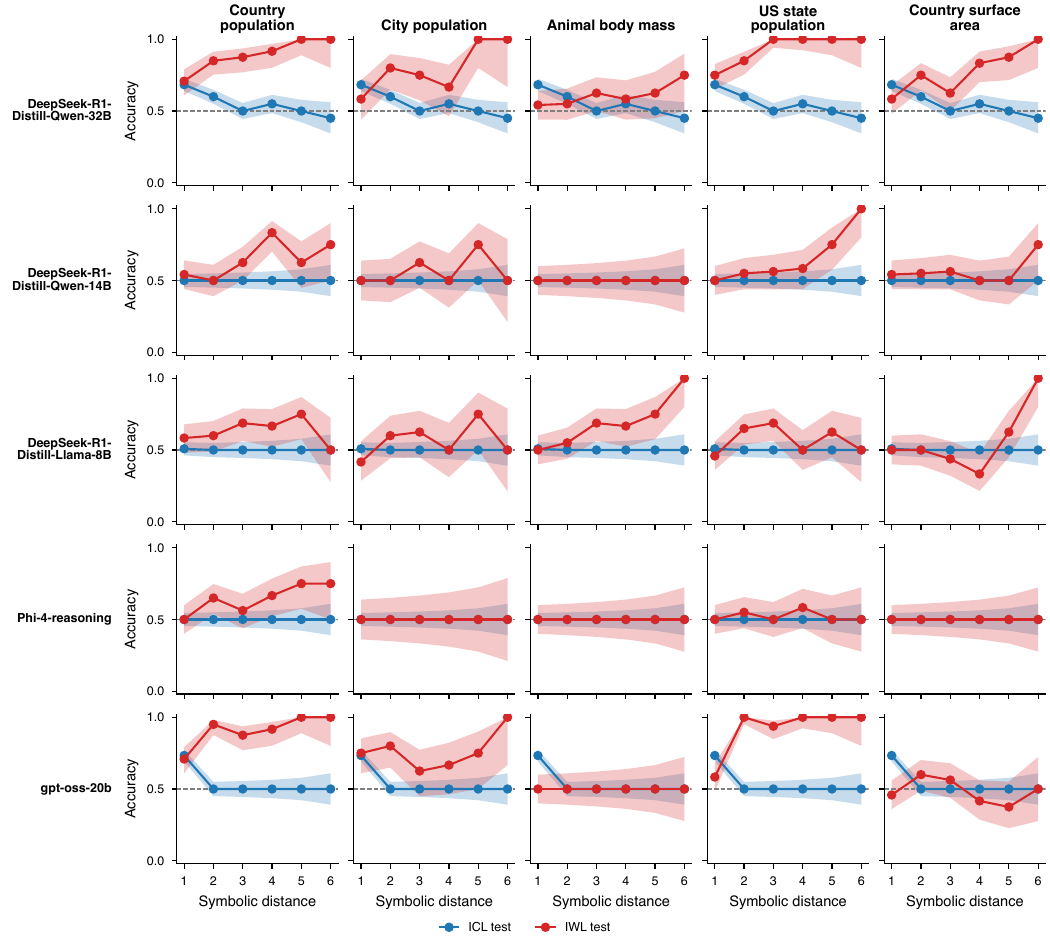}
  \caption{\textbf{LLM symbolic-distance results: DeepSeek, Phi, and gpt-oss models.}
  Direct IWL and ICL test accuracy, formatted as in Fig.~\ref{fig:app-llm-sde-qwen3}.
  The ICL measurement repeats across columns.
  Bands show 68\% confidence intervals over trials; dashed lines indicate chance.}
  \label{fig:app-llm-sde-deepseek-phi-gptoss}
\end{figure}

\begin{figure}[p]
  \centering
  \includegraphics{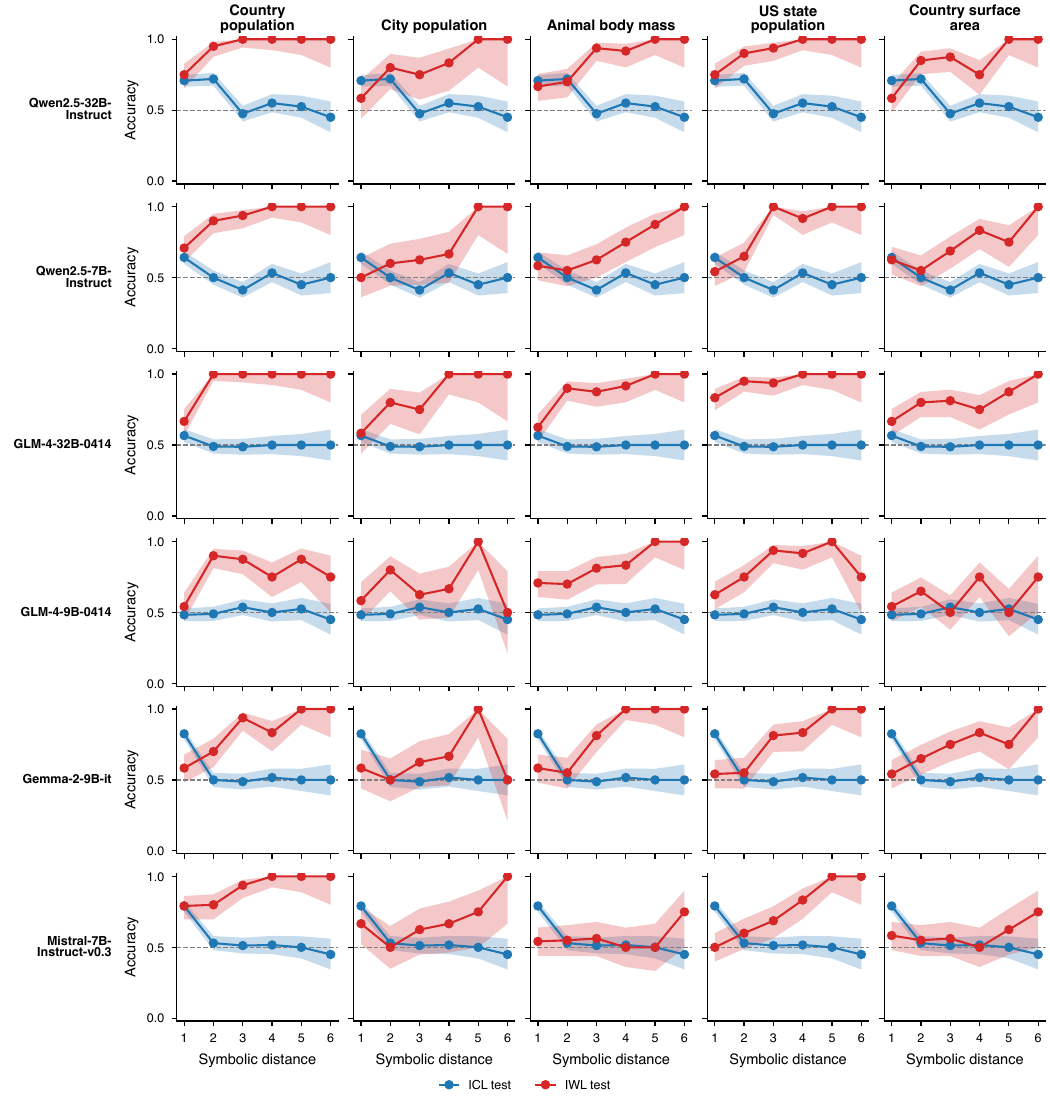}
  \caption{\textbf{LLM symbolic-distance results: Qwen2.5, GLM, Gemma, and Mistral models.}
  Direct IWL and ICL test accuracy, formatted as in Fig.~\ref{fig:app-llm-sde-qwen3}.
  The ICL measurement repeats across columns.
  Bands show 68\% confidence intervals over trials; dashed lines indicate chance.}
  \label{fig:app-llm-sde-qwen25-glm-gemma-mistral}
\end{figure}

\begin{figure}[p]
  \centering
  \includegraphics{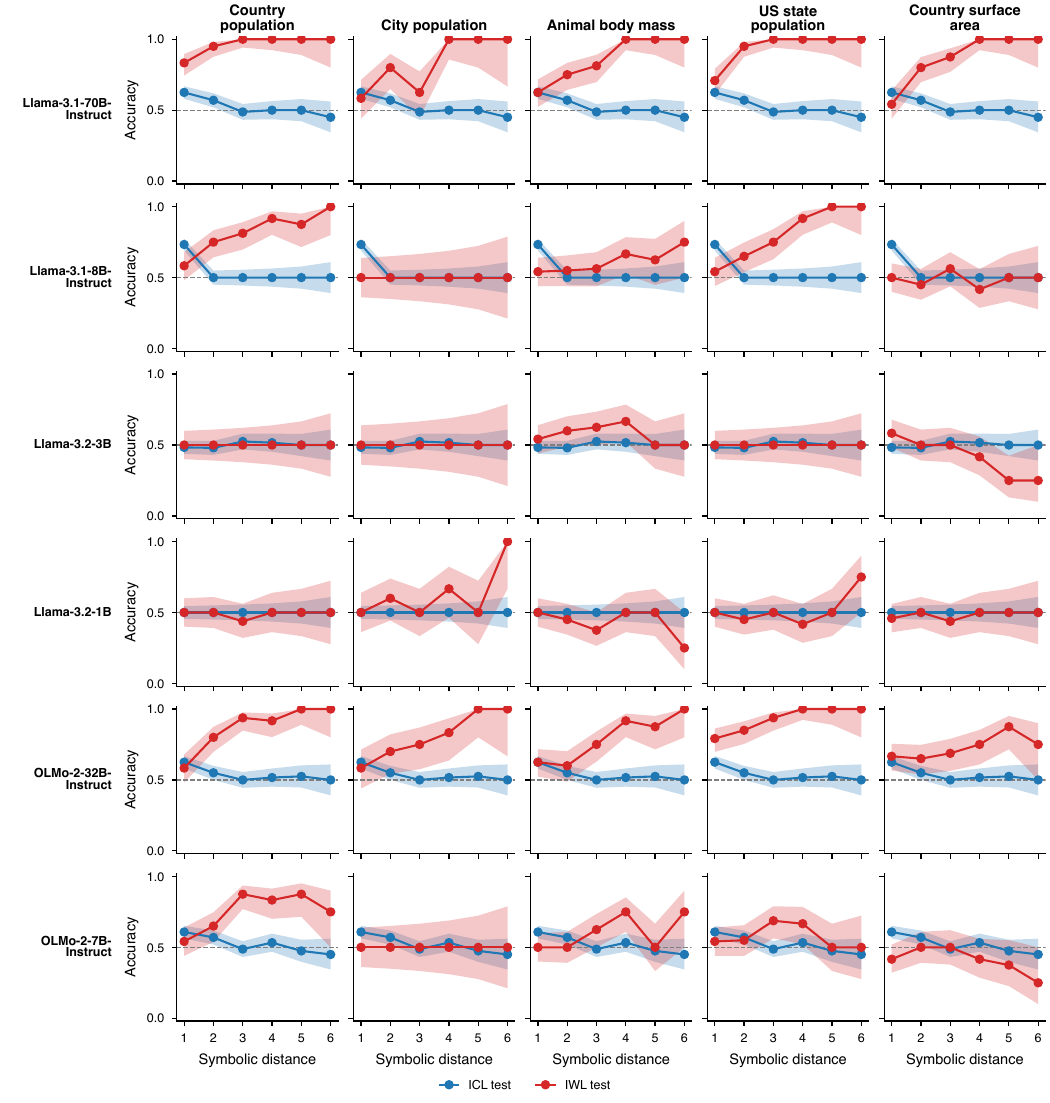}
  \caption{\textbf{LLM symbolic-distance results: Llama and OLMo models.}
  Direct IWL and ICL test accuracy, formatted as in Fig.~\ref{fig:app-llm-sde-qwen3}.
  The ICL measurement repeats across columns.
  Bands show 68\% confidence intervals over trials; dashed lines indicate chance.}
  \label{fig:app-llm-sde-llama-olmo}
\end{figure}
\FloatBarrier

\begin{figure}[H]
  \centering
  \includegraphics{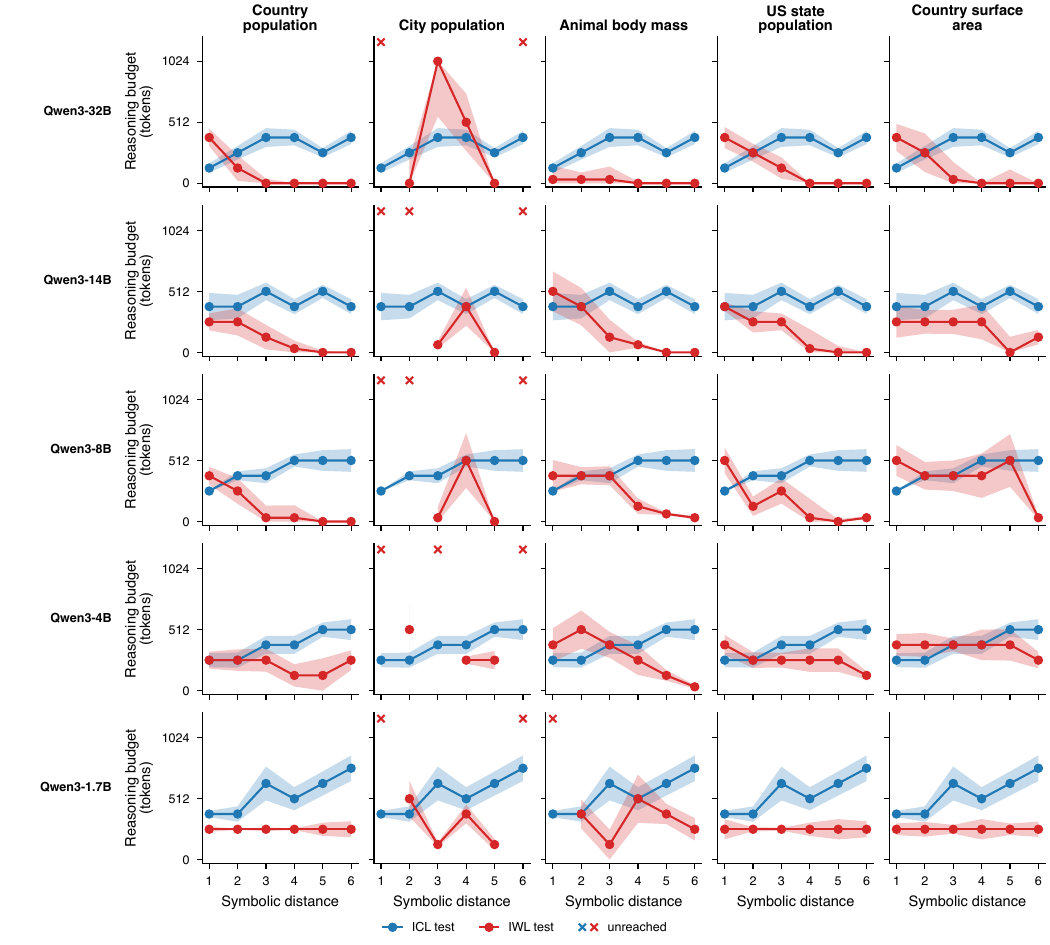}
  \caption{\textbf{LLM reasoning requirements: Qwen3 models.}
  Smallest reasoning budget reaching 90\% of each series' maximum accuracy, by distance; zero denotes a direct answer.
  IWL estimates pool all available item sets per dimension; the ICL measurement repeats across columns.
  Crosses mark distances where accuracy was not reliably above chance at any tested budget (the 95\% Wilson lower bound on maximum accuracy did not exceed .5), so no budget is reported.
  City population has only two IWL trials at distance six, so this criterion cannot be met even at perfect accuracy.
  Bands show $\pm1$ bootstrap SE over trials, bounded by zero and 1,024 tokens.}
  \label{fig:app-llm-budget-qwen3}
\end{figure}

\begin{figure}[p]
  \centering
  \includegraphics{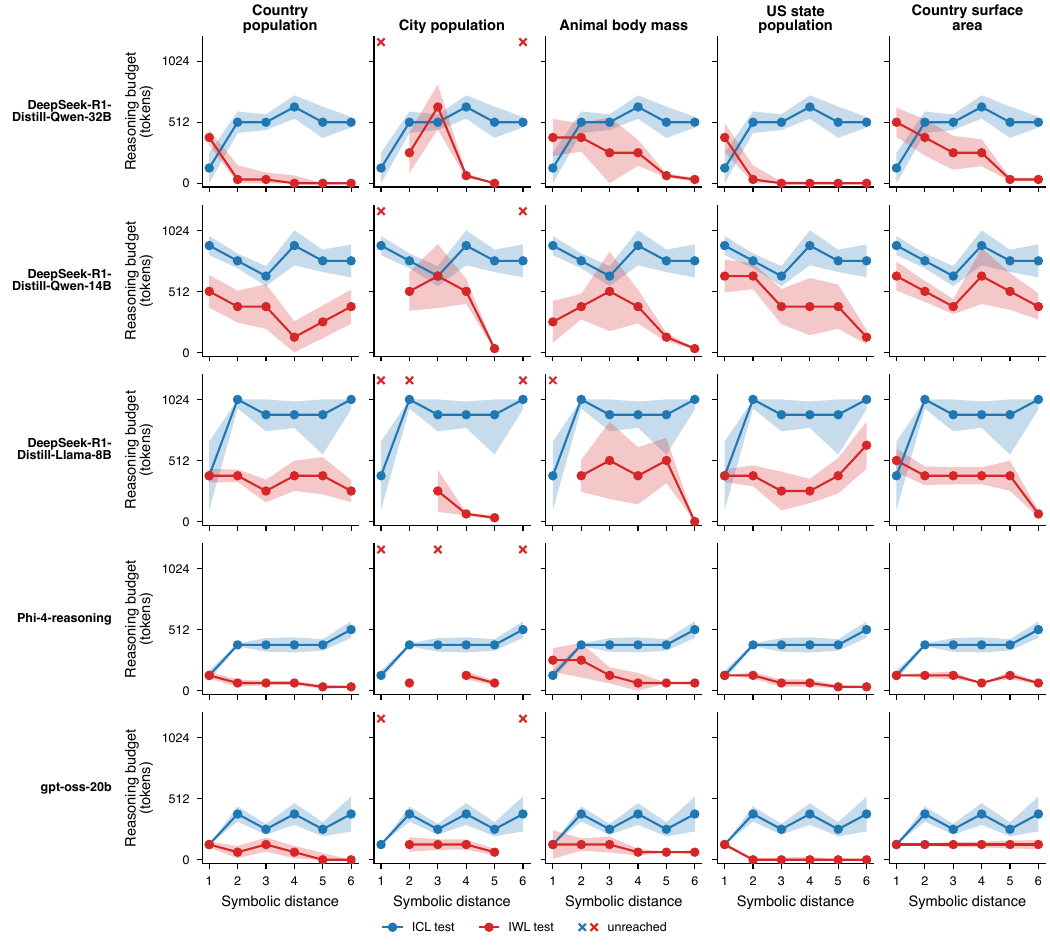}
  \caption{\textbf{LLM reasoning requirements: DeepSeek, Phi, and gpt-oss models.}
  Budget estimates are defined as in Fig.~\ref{fig:app-llm-budget-qwen3}; the ICL measurement repeats across columns.
  Bands show $\pm1$ bootstrap SE over trials.
  Crosses mark distances where accuracy was not reliably above chance at any tested budget, so no budget could be estimated.}
  \label{fig:app-llm-budget-deepseek-phi-gptoss}
\end{figure}

\begin{figure}[p]
  \centering
  \includegraphics{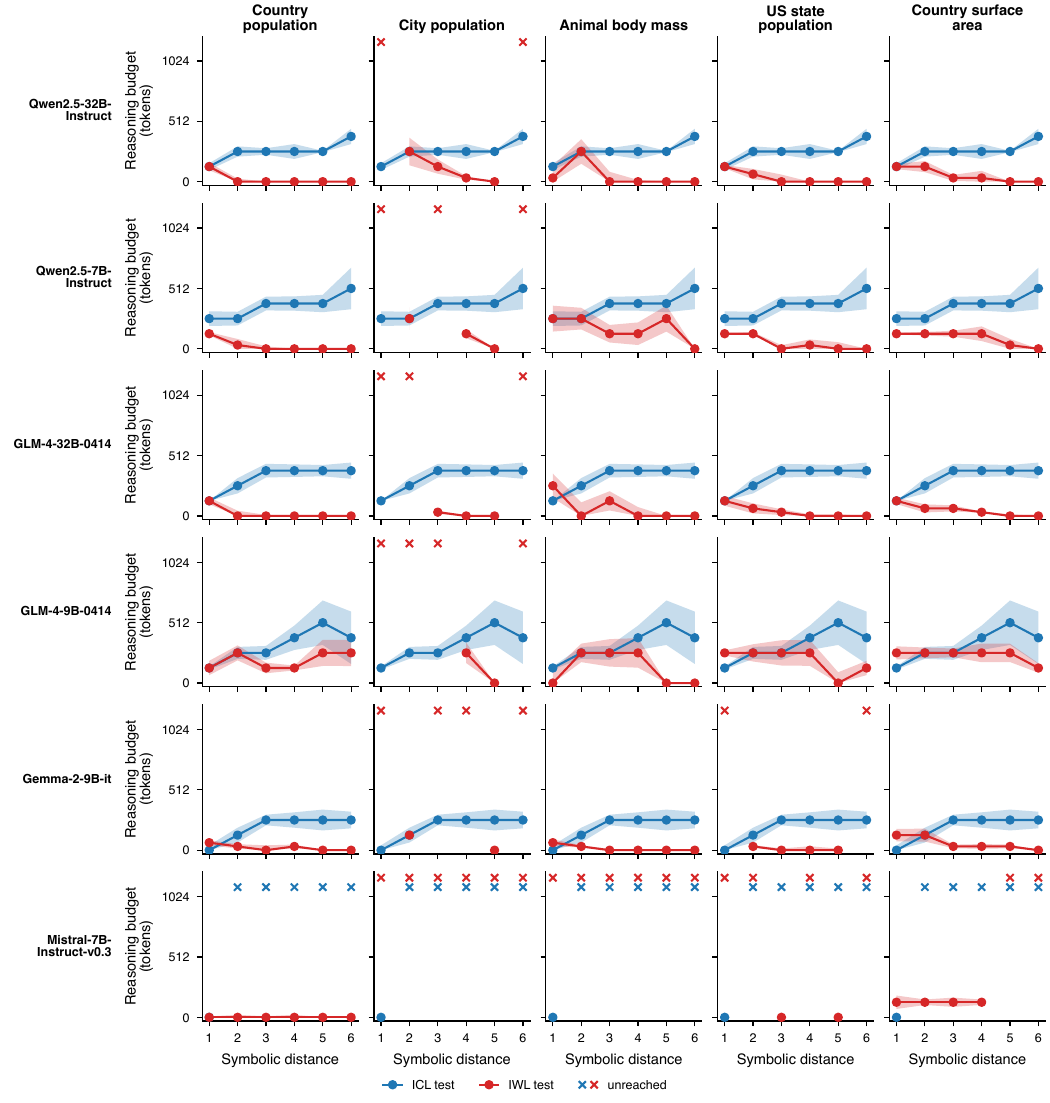}
  \caption{\textbf{LLM reasoning requirements: Qwen2.5, GLM, Gemma, and Mistral models.}
  Budget estimates are defined as in Fig.~\ref{fig:app-llm-budget-qwen3}; the ICL measurement repeats across columns.
  Bands show $\pm1$ bootstrap SE over trials.
  Crosses mark distances where accuracy was not reliably above chance at any tested budget, so no budget could be estimated.}
  \label{fig:app-llm-budget-qwen25-glm-gemma-mistral}
\end{figure}

\begin{figure}[p]
  \centering
  \includegraphics{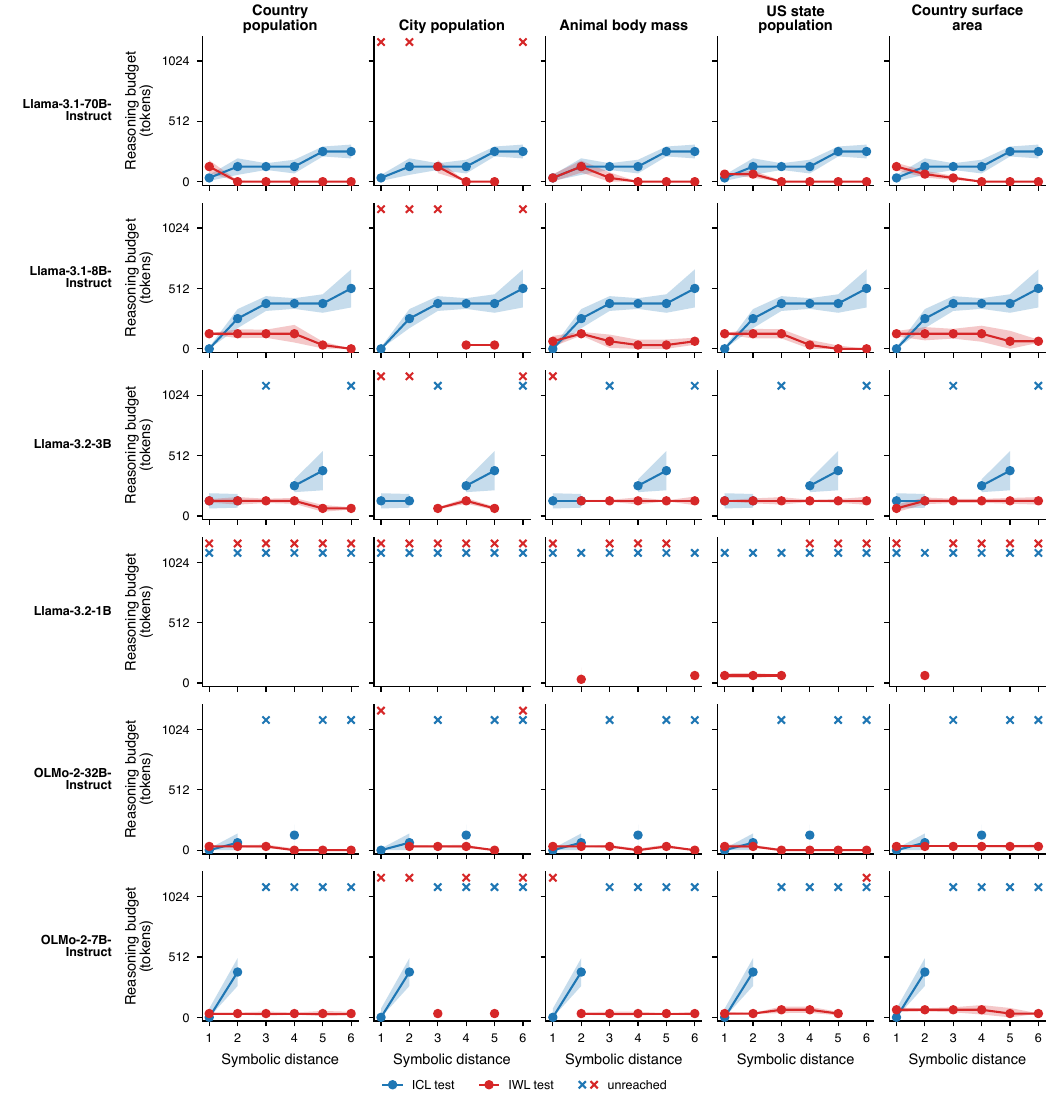}
  \caption{\textbf{LLM reasoning requirements: Llama and OLMo models.}
  Budget estimates are defined as in Fig.~\ref{fig:app-llm-budget-qwen3}; the ICL measurement repeats across columns.
  Bands show $\pm1$ bootstrap SE over trials.
  Crosses mark distances where accuracy was not reliably above chance at any tested budget, so no budget could be estimated.}
  \label{fig:app-llm-budget-llama-olmo}
\end{figure}

\begin{figure}[p]
  \centering
  \includegraphics{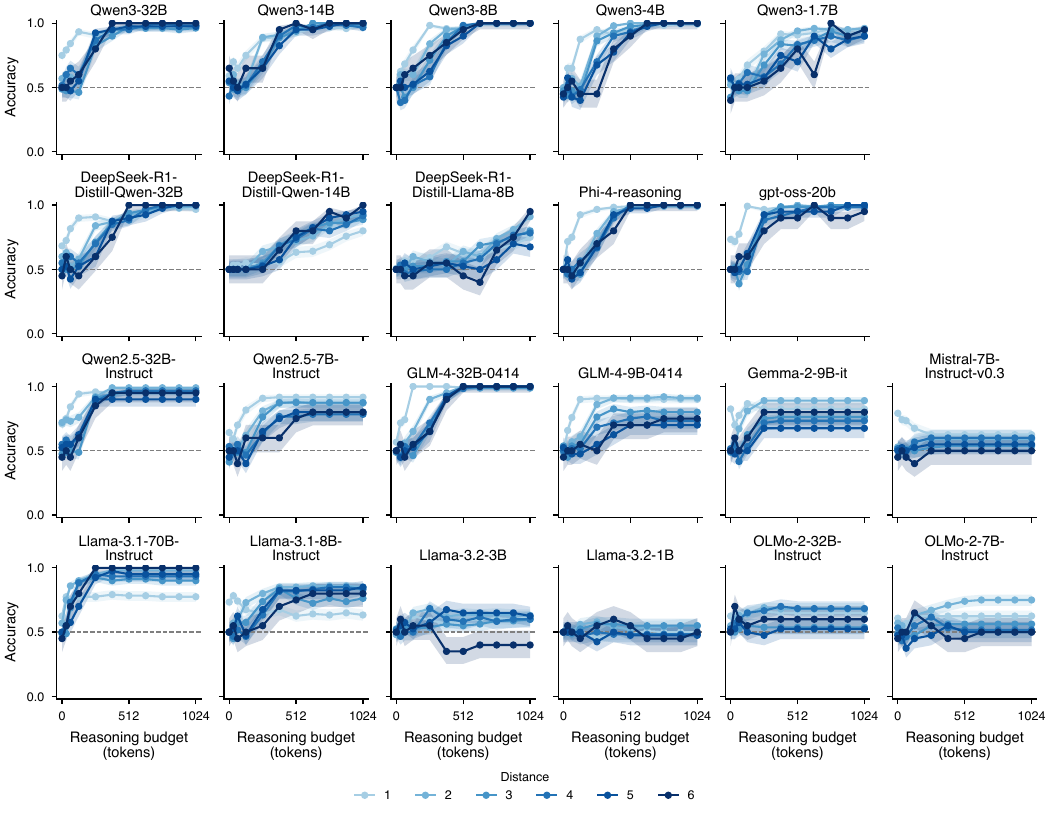}
  \caption{\textbf{LLM ICL accuracy by reasoning budget.}
  ICL accuracy by reasoning budget, with one panel per model. Models trained to reason are in the top two rows and instruction-tuned models prompted to reason are in the bottom two.
  Darker lines indicate larger symbolic distances (1--6).
  Bands show 68\% confidence intervals over trials; dashed lines indicate chance.}
  \label{fig:app-llm-icl-budget}
\end{figure}

\paragraph{Reasoning requirements across models and dimensions.}
We also examined how the reasoning budget required to reach 90\% of maximum accuracy varied with distance on the IWL and ICL tests (Figs.~\ref{fig:app-llm-budget-qwen3}--\ref{fig:app-llm-budget-llama-olmo}).
Crosses mark cases where the model did not perform reliably above chance at any tested budget, so no reasoning budget could be estimated.
Among models with an estimate at both distances, the ICL budget was larger at distance six than at distance one in 15 of 17 models.
The IWL budget instead decreased in 61 of 75 model--dimension combinations with estimates at both distances.
Figure~\ref{fig:app-llm-icl-budget} shows ICL accuracy at each reasoning budget for all 22 models.

\FloatBarrier

\begin{figure}[H]
  \centering
  \includegraphics{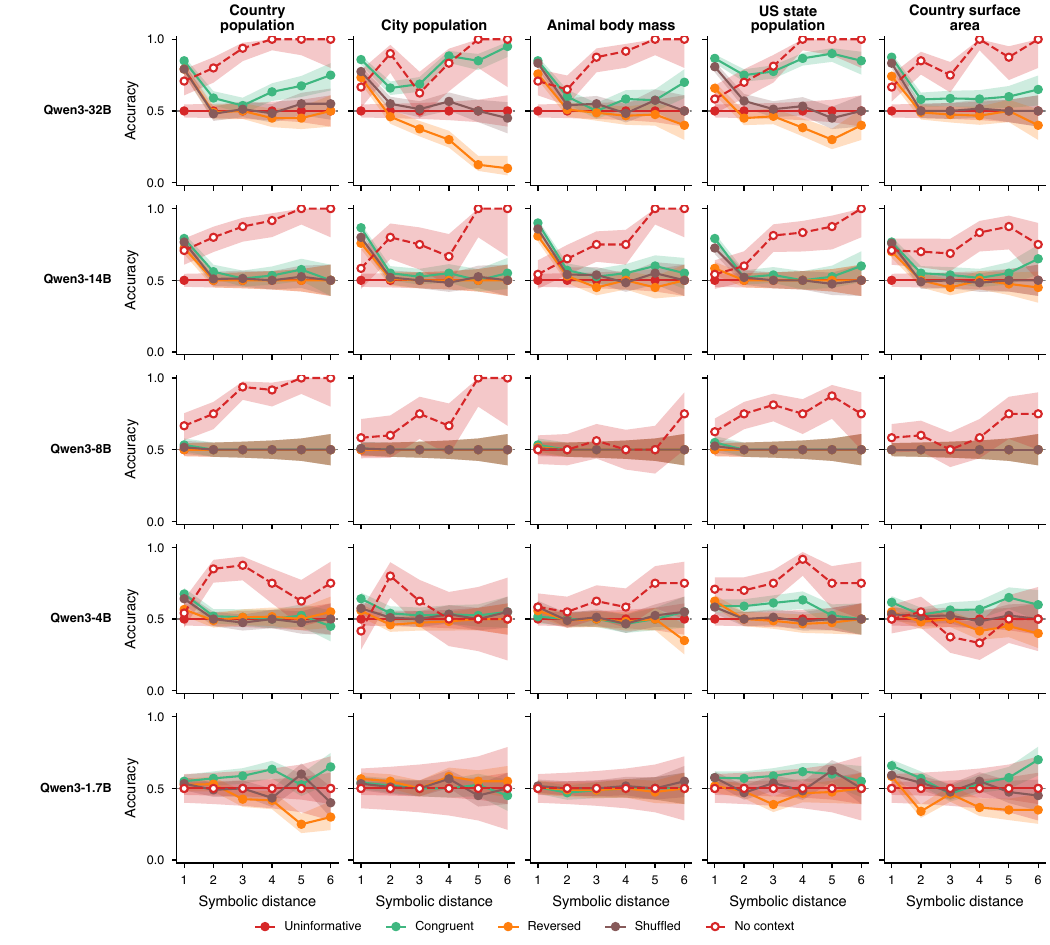}
  \caption{\textbf{LLM content effects: Qwen3 models.}
  Direct accuracy by distance under the five context conditions.
  All four conditions with context use the instruction to rely only on the comparisons given; the no-context control uses the original instruction.
  Congruent, reversed, and shuffled trials are scored against the context ordering; the other conditions are scored against the real-world ordering.
  Bands show 68\% confidence intervals over trials; dashed horizontal lines indicate chance.}
  \label{fig:app-llm-content-qwen3}
\end{figure}

\begin{figure}[p]
  \centering
  \includegraphics{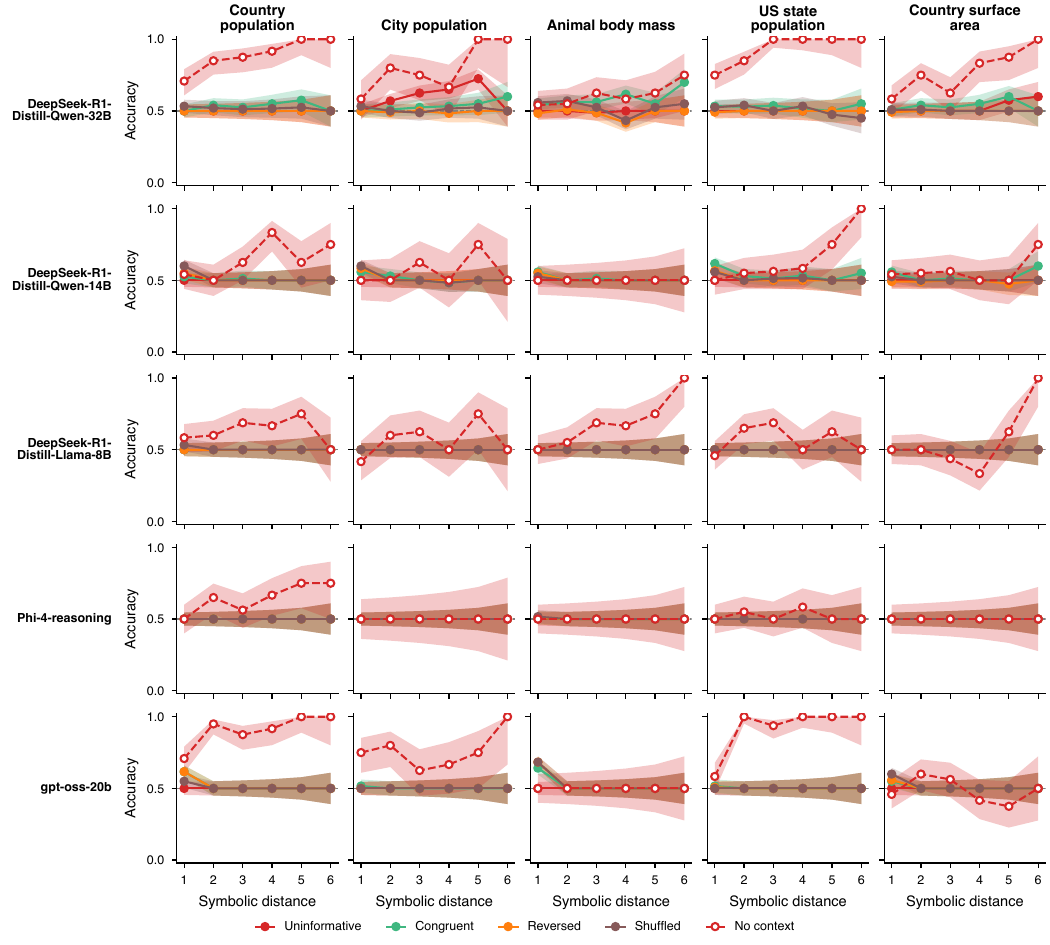}
  \caption{\textbf{LLM content effects: DeepSeek, Phi, and gpt-oss models.}
  Direct accuracy under the five context conditions, with the instruction and scoring described in Fig.~\ref{fig:app-llm-content-qwen3}.
  Bands show 68\% confidence intervals over trials; dashed horizontal lines indicate chance.}
  \label{fig:app-llm-content-deepseek-phi-gptoss}
\end{figure}

\begin{figure}[p]
  \centering
  \includegraphics{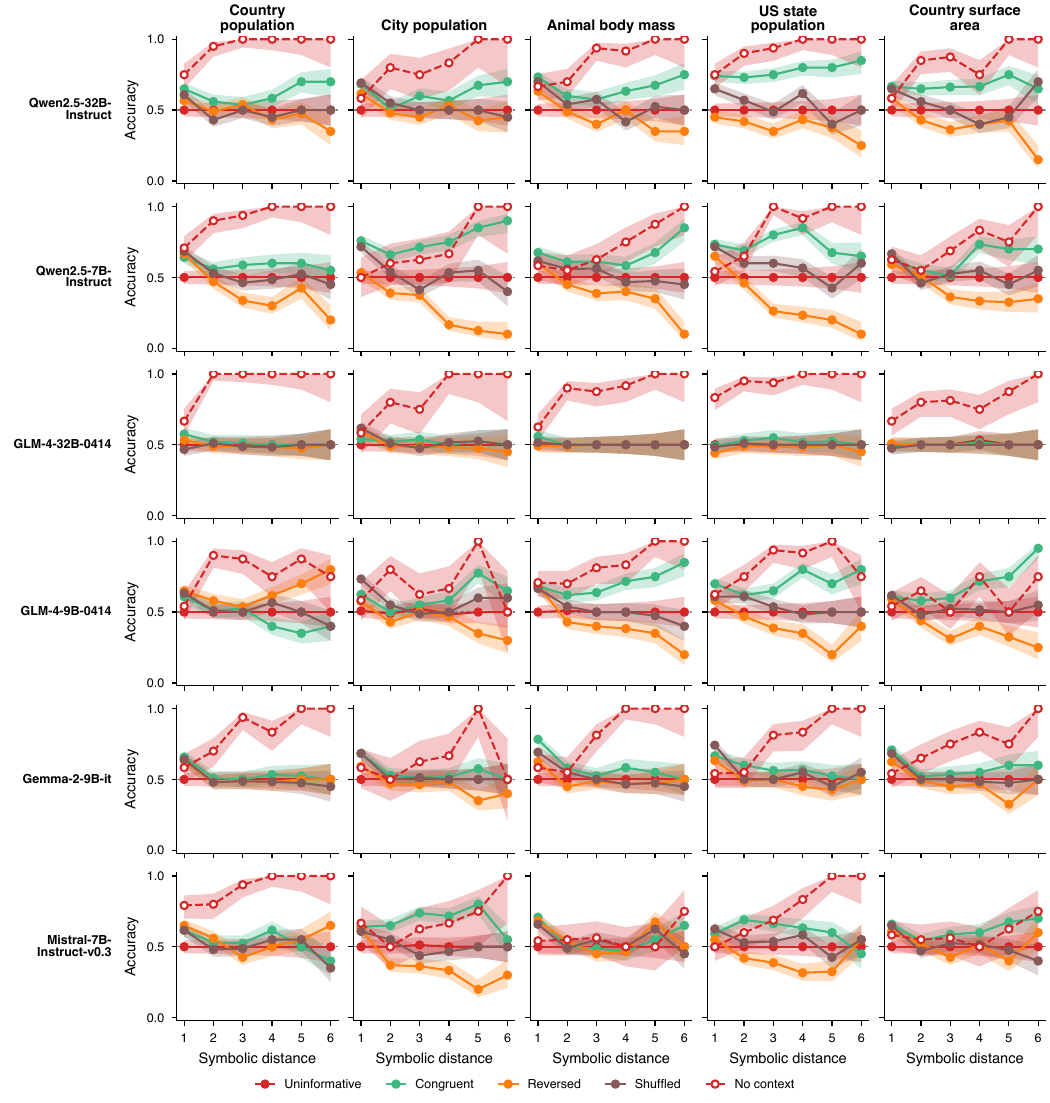}
  \caption{\textbf{LLM content effects: Qwen2.5, GLM, Gemma, and Mistral models.}
  Direct accuracy under the five context conditions, with the instruction and scoring described in Fig.~\ref{fig:app-llm-content-qwen3}.
  Bands show 68\% confidence intervals over trials; dashed horizontal lines indicate chance.}
  \label{fig:app-llm-content-qwen25-glm-gemma-mistral}
\end{figure}

\begin{figure}[p]
  \centering
  \includegraphics{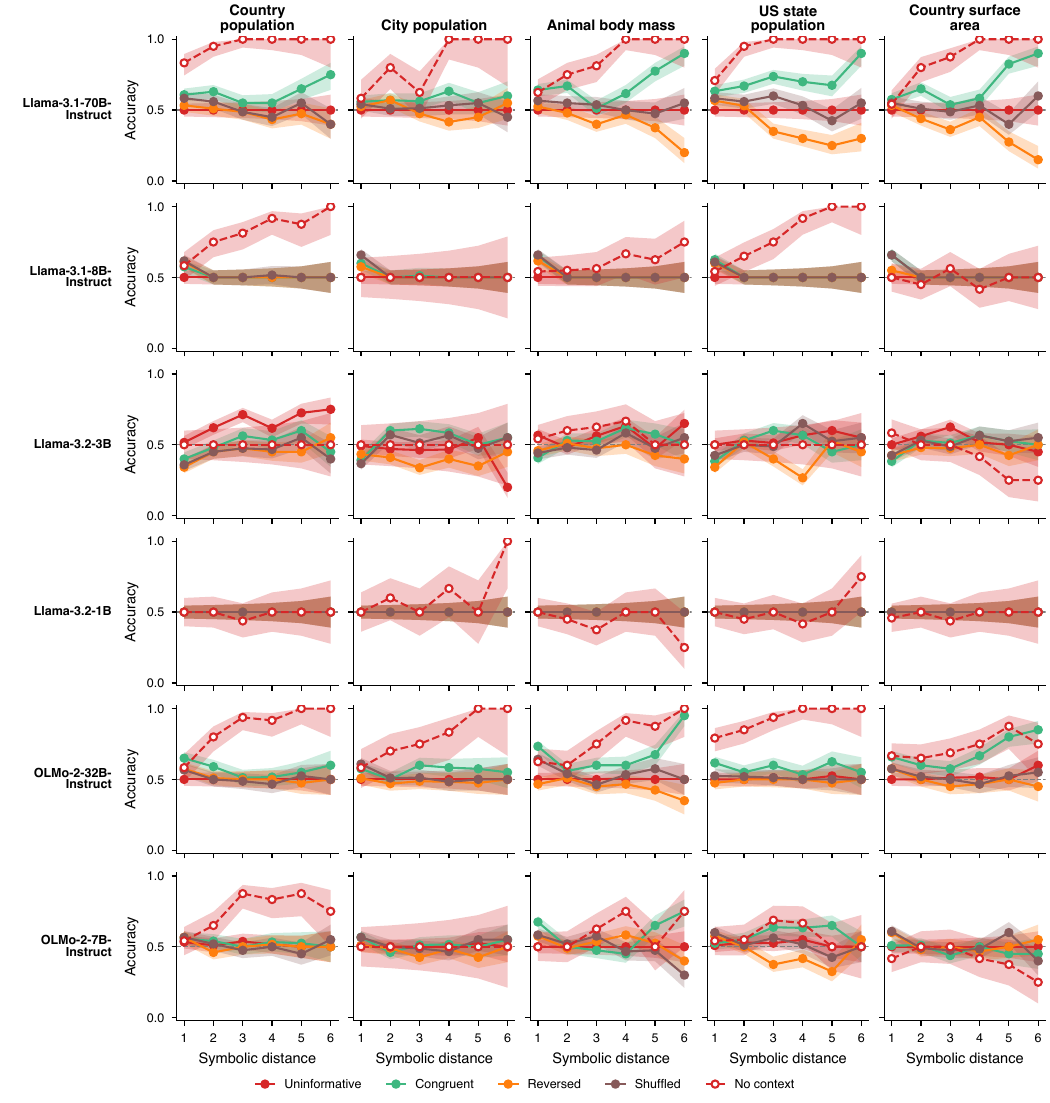}
  \caption{\textbf{LLM content effects: Llama and OLMo models.}
  Direct accuracy under the five context conditions, with the instruction and scoring described in Fig.~\ref{fig:app-llm-content-qwen3}.
  Bands show 68\% confidence intervals over trials; dashed horizontal lines indicate chance.}
  \label{fig:app-llm-content-llama-olmo}
\end{figure}

\paragraph{Content effects across models and dimensions.}
Figures~\ref{fig:app-llm-content-qwen3}--\ref{fig:app-llm-content-llama-olmo} extend the content-effects results in Fig.~\ref{fig:content}J to all models and dimensions.
Models were explicitly instructed to answer from the comparisons given in context.
Nevertheless, direct accuracy was higher with congruent than reversed contexts in 81 of 110 model--dimension combinations, consistent with interference from prior knowledge when it conflicted with the context.

\FloatBarrier

\begin{figure}[H]
  \centering
  \includegraphics{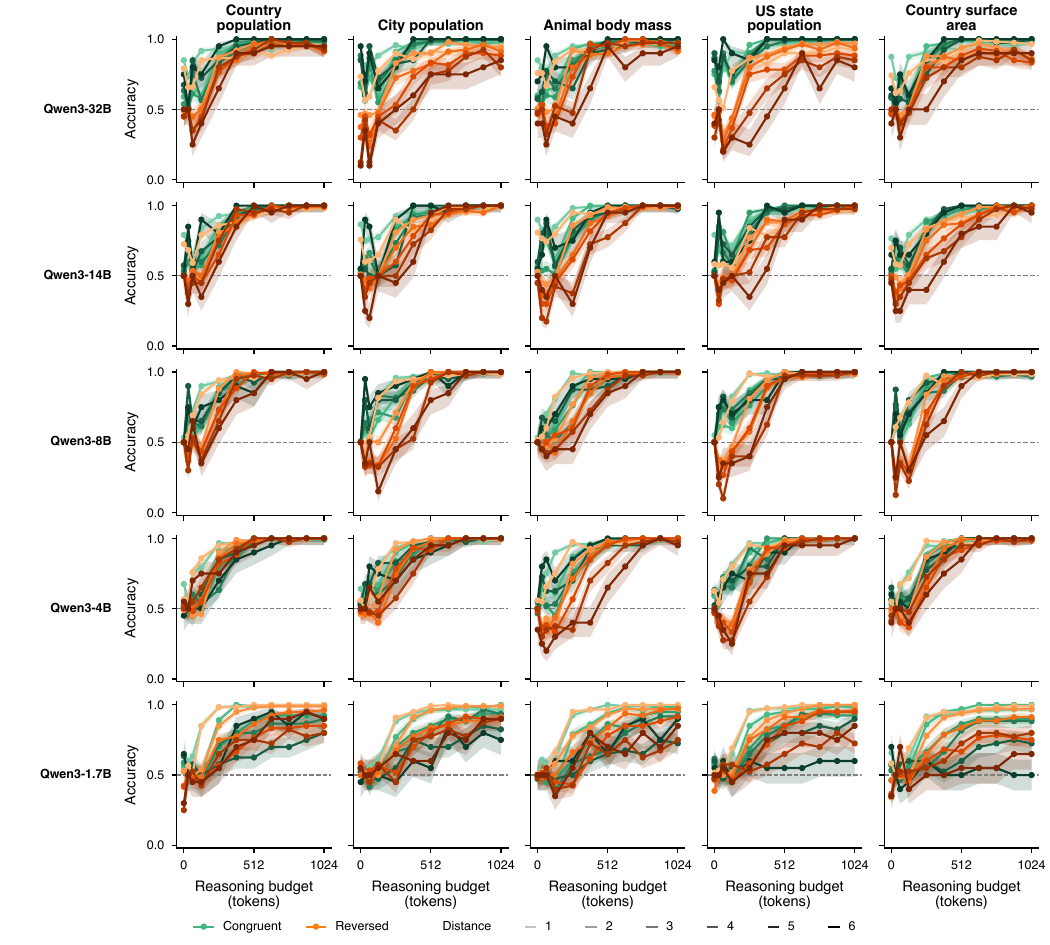}
  \caption{\textbf{Recovery with reasoning: Qwen3 models.}
  Context-consistent accuracy on congruent and reversed trials by reasoning budget.
  Darker lines indicate larger symbolic distances (1--6).
  Zero denotes direct answers; positive budgets score prefixes of the same generated trace, up to 1,024 tokens.
  Lines show accuracy at each budget, without taking a running maximum.
  Each series contains 120 trials at distance one, decreasing to 20 at distance six.
  Bands show 68\% confidence intervals over trials; dashed lines indicate chance.}
  \label{fig:app-llm-recovery-qwen3}
\end{figure}

\begin{figure}[p]
  \centering
  \includegraphics{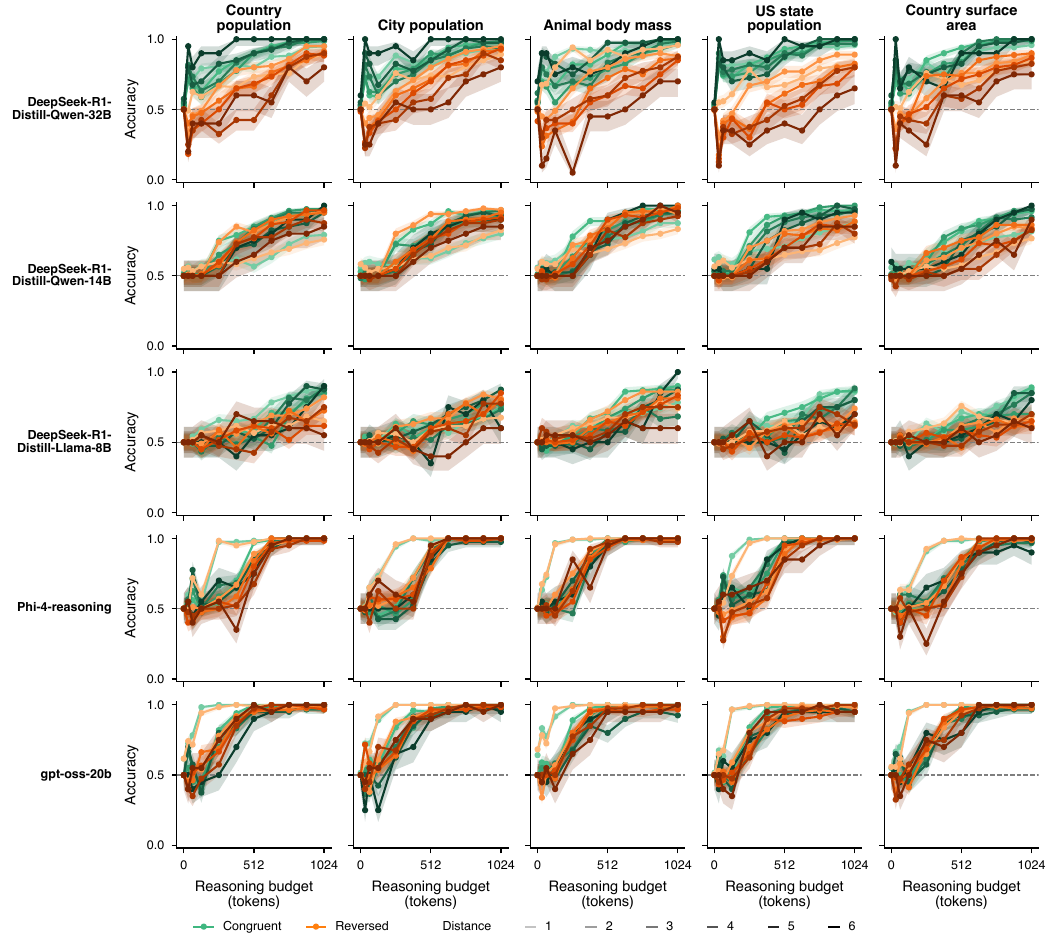}
  \caption{\textbf{Recovery with reasoning: DeepSeek, Phi, and gpt-oss models.}
  Congruent- and reversed-context accuracy by reasoning budget, formatted as in Fig.~\ref{fig:app-llm-recovery-qwen3}.
  Darker lines indicate larger distances.
  Bands show 68\% confidence intervals over trials; dashed lines indicate chance.}
  \label{fig:app-llm-recovery-deepseek-phi-gptoss}
\end{figure}

\begin{figure}[p]
  \centering
  \includegraphics{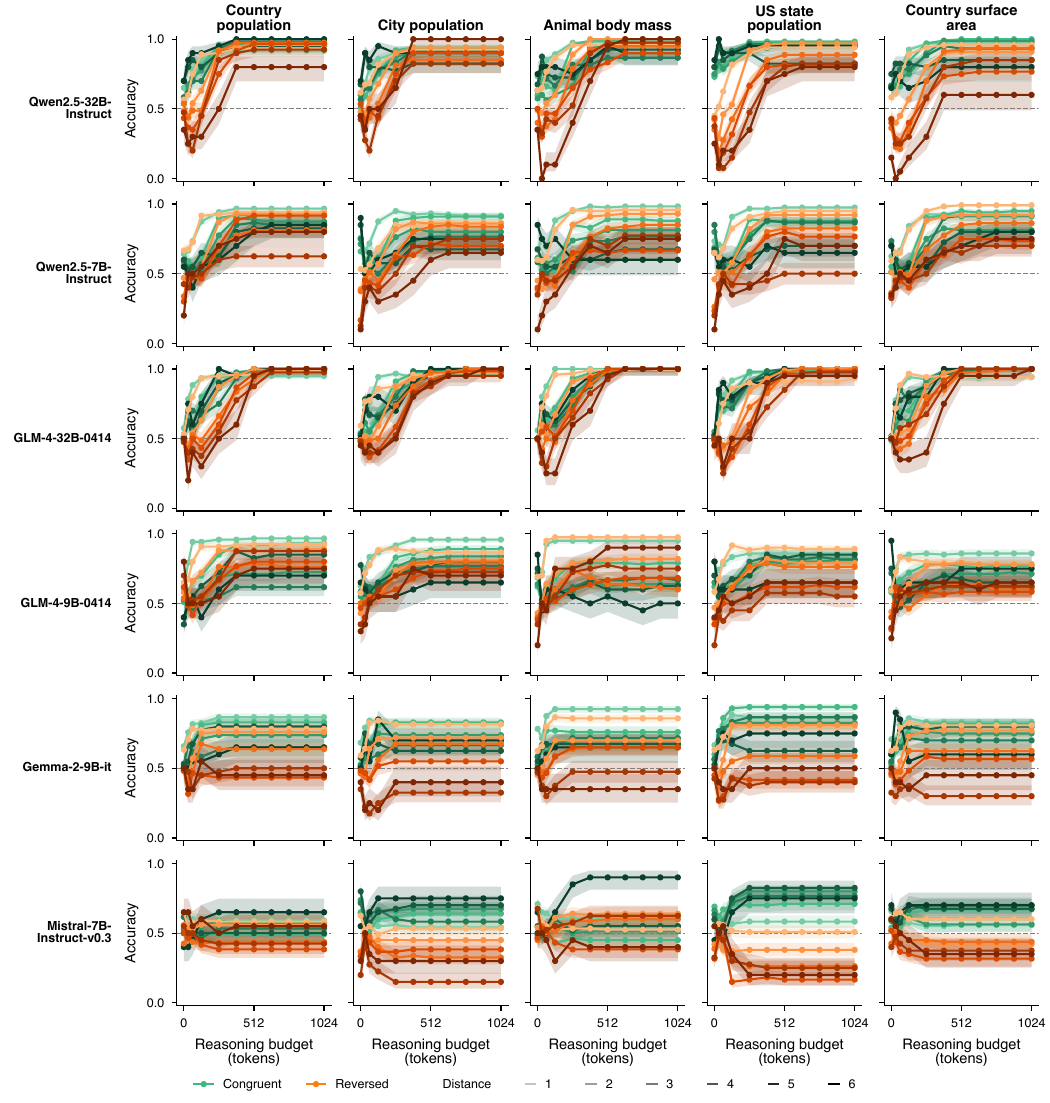}
  \caption{\textbf{Recovery with reasoning: Qwen2.5, GLM, Gemma, and Mistral models.}
  Congruent- and reversed-context accuracy by reasoning budget, formatted as in Fig.~\ref{fig:app-llm-recovery-qwen3}.
  Darker lines indicate larger distances.
  Bands show 68\% confidence intervals over trials; dashed lines indicate chance.}
  \label{fig:app-llm-recovery-qwen25-glm-gemma-mistral}
\end{figure}

\begin{figure}[p]
  \centering
  \includegraphics{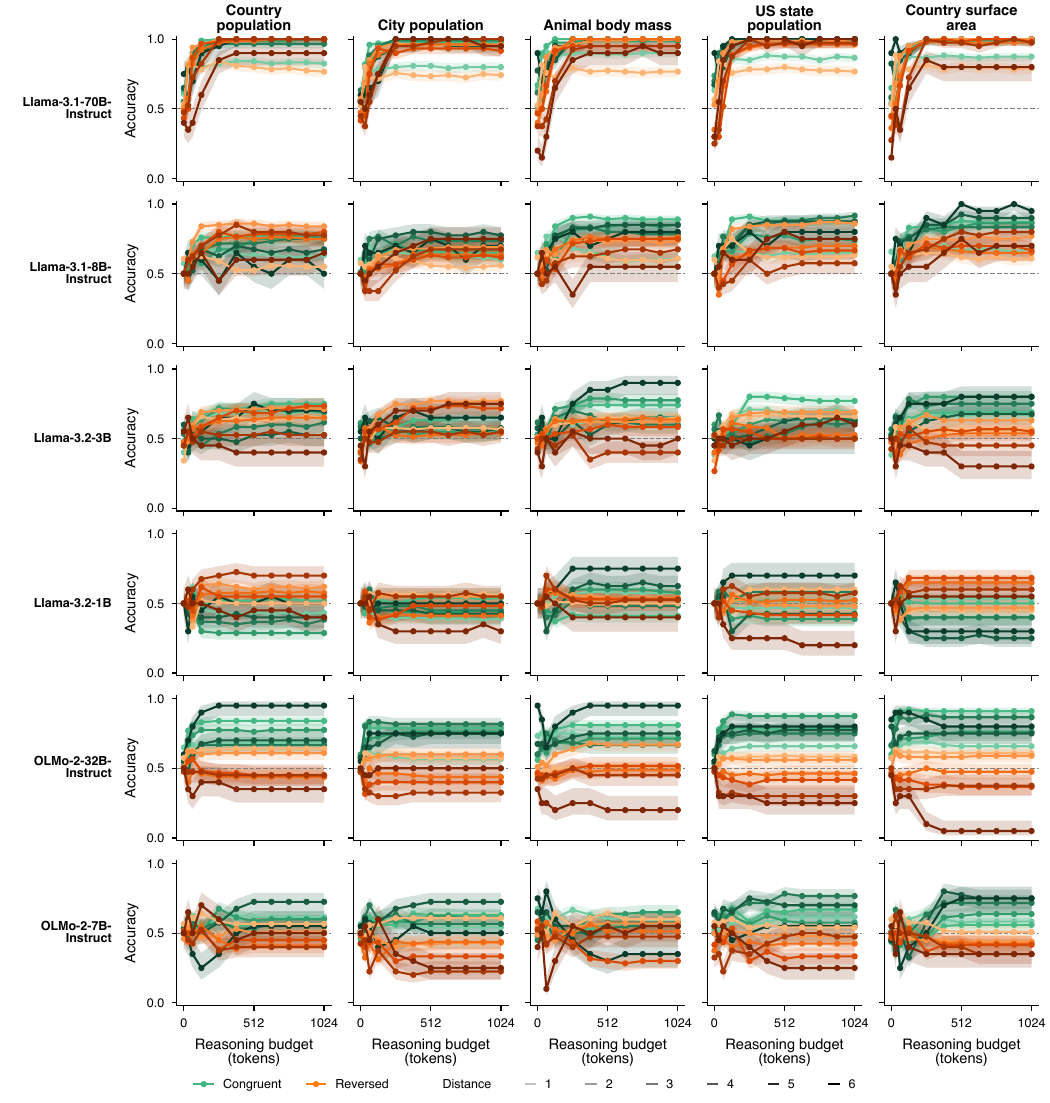}
  \caption{\textbf{Recovery with reasoning: Llama and OLMo models.}
  Congruent- and reversed-context accuracy by reasoning budget, formatted as in Fig.~\ref{fig:app-llm-recovery-qwen3}.
  Darker lines indicate larger distances.
  Bands show 68\% confidence intervals over trials; dashed lines indicate chance.}
  \label{fig:app-llm-recovery-llama-olmo}
\end{figure}

\paragraph{Recovery with reasoning.}
We examined whether reasoning could overcome interference on reversed-context trials across the same models and dimensions (Figs.~\ref{fig:app-llm-recovery-qwen3}--\ref{fig:app-llm-recovery-llama-olmo}).
At 1,024 tokens, accuracy on reversed-context trials exceeded direct accuracy by at least 10 percentage points in 89 of 110 model--dimension combinations.
At a budget of 256 tokens, accuracy was higher for adjacent than the most distant pairs in 98 of 110 combinations, consistent with more distant comparisons requiring more reasoning.

\FloatBarrier

\begin{figure}[H]
  \centering
  \includegraphics{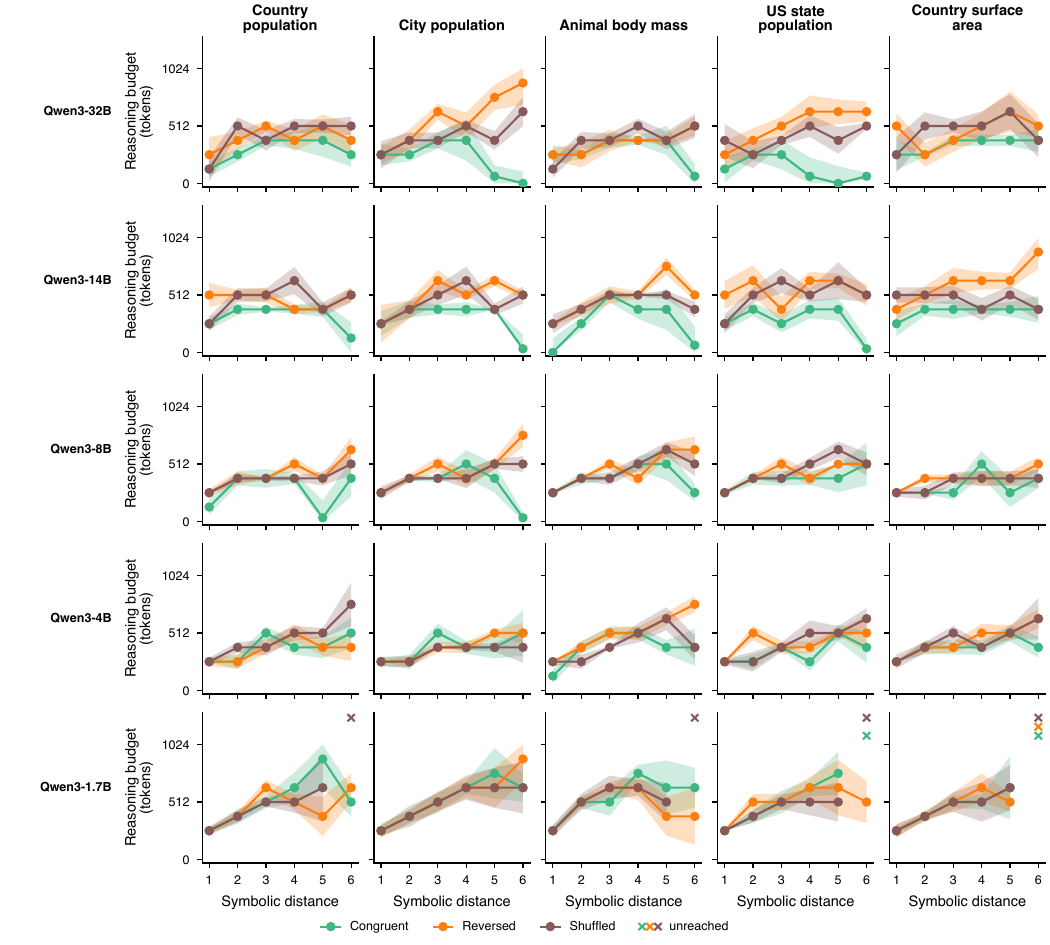}
  \caption{\textbf{LLM reasoning requirements under content conditions: Qwen3 models.}
  Smallest reasoning budget reaching 90\% of each series' maximum accuracy, by distance, under congruent, reversed, and shuffled contexts; zero denotes a direct answer.
  Shuffled distances are measured in the ordering shown.
  Each series contains 120 trials at distance one, decreasing to 20 at distance six.
  Crosses mark distances where accuracy was not reliably above chance at any tested budget (the 95\% Wilson lower bound on maximum accuracy did not exceed .5), so no budget is reported.
  Bands show $\pm1$ bootstrap SE over trials, bounded by zero and 1,024 tokens.}
  \label{fig:app-llm-criterion-qwen3}
\end{figure}

\begin{figure}[p]
  \centering
  \includegraphics{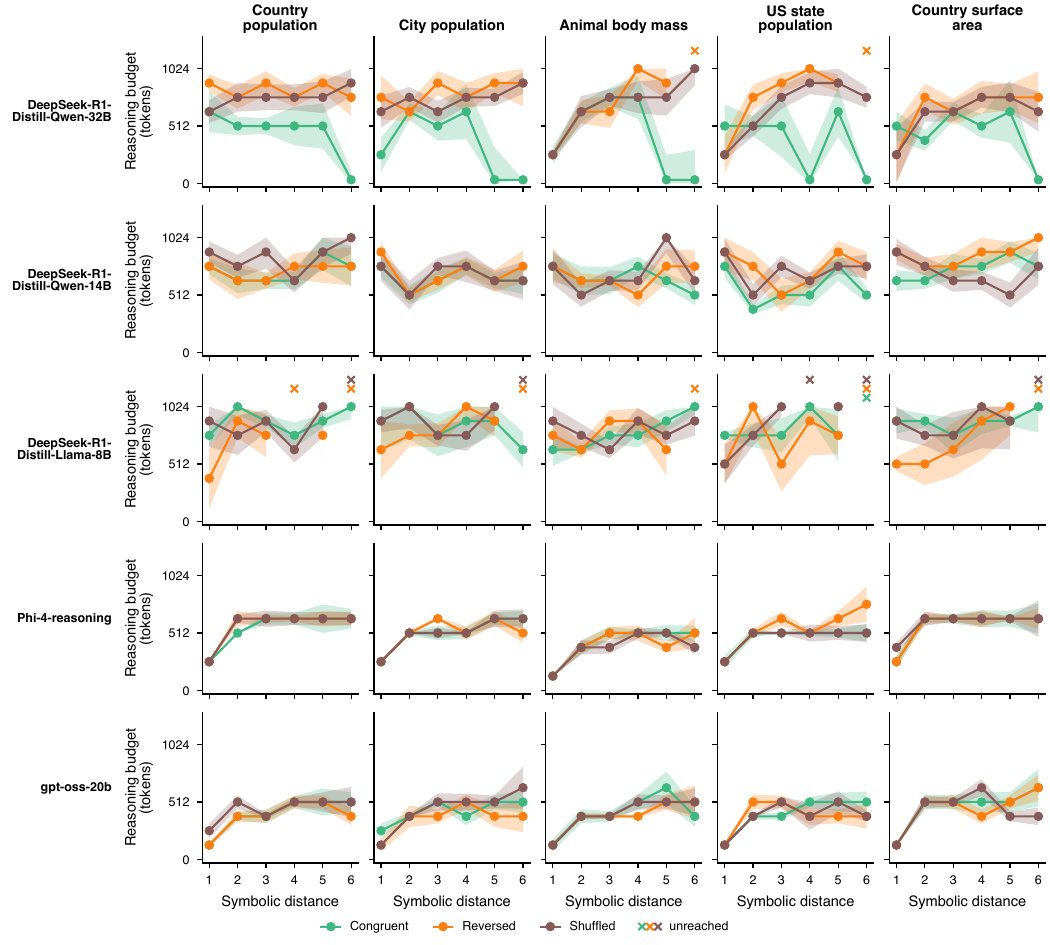}
  \caption{\textbf{LLM reasoning requirements under content conditions: DeepSeek, Phi, and gpt-oss models.}
  Budget estimates are defined as in Fig.~\ref{fig:app-llm-criterion-qwen3}.
  Bands show $\pm1$ bootstrap SE over trials.
  Crosses mark distances where accuracy was not reliably above chance at any tested budget, so no budget could be estimated.}
  \label{fig:app-llm-criterion-deepseek-phi-gptoss}
\end{figure}

\begin{figure}[p]
  \centering
  \includegraphics{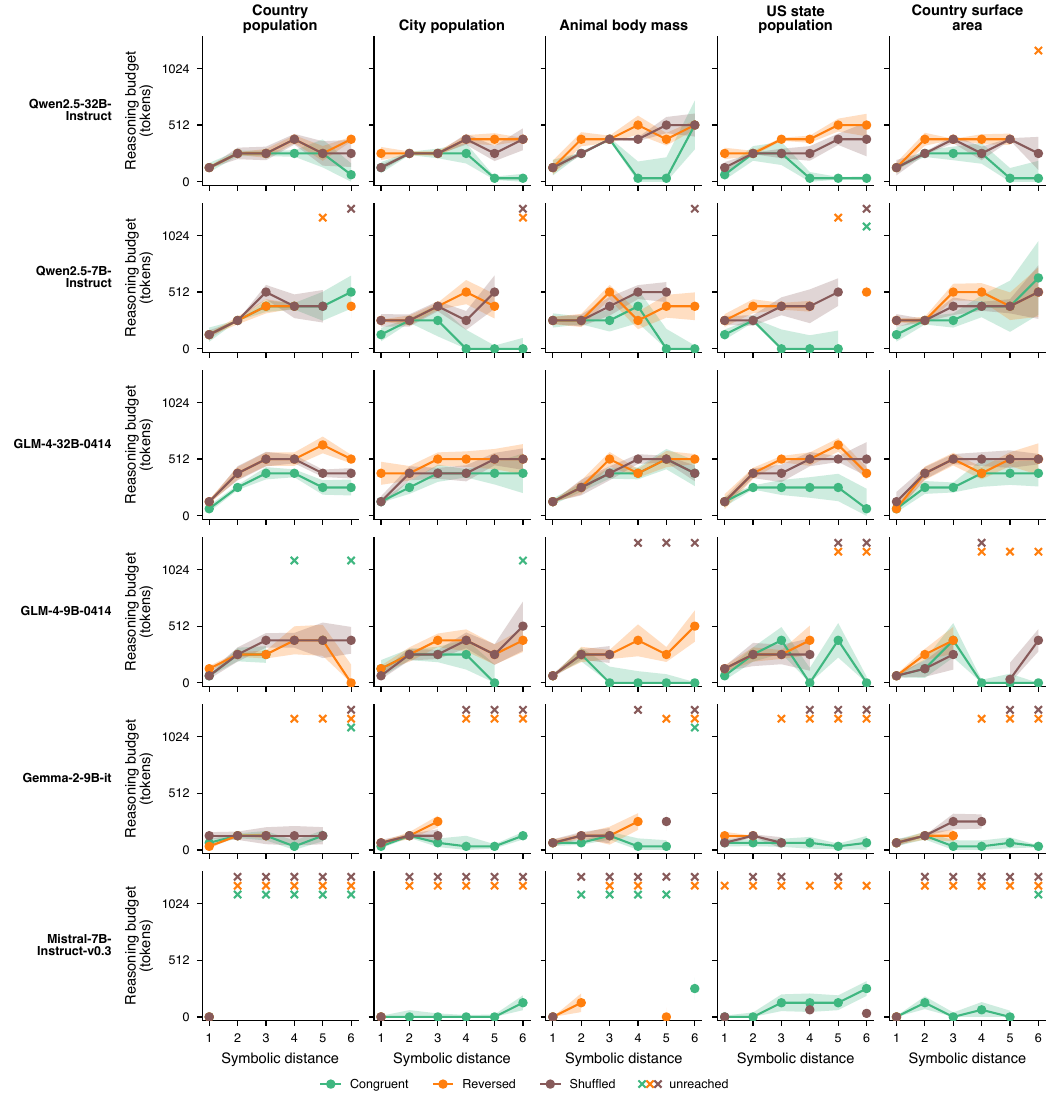}
  \caption{\textbf{LLM reasoning requirements under content conditions: Qwen2.5, GLM, Gemma, and Mistral models.}
  Budget estimates are defined as in Fig.~\ref{fig:app-llm-criterion-qwen3}.
  Bands show $\pm1$ bootstrap SE over trials.
  Crosses mark distances where accuracy was not reliably above chance at any tested budget, so no budget could be estimated.}
  \label{fig:app-llm-criterion-qwen25-glm-gemma-mistral}
\end{figure}

\begin{figure}[p]
  \centering
  \includegraphics{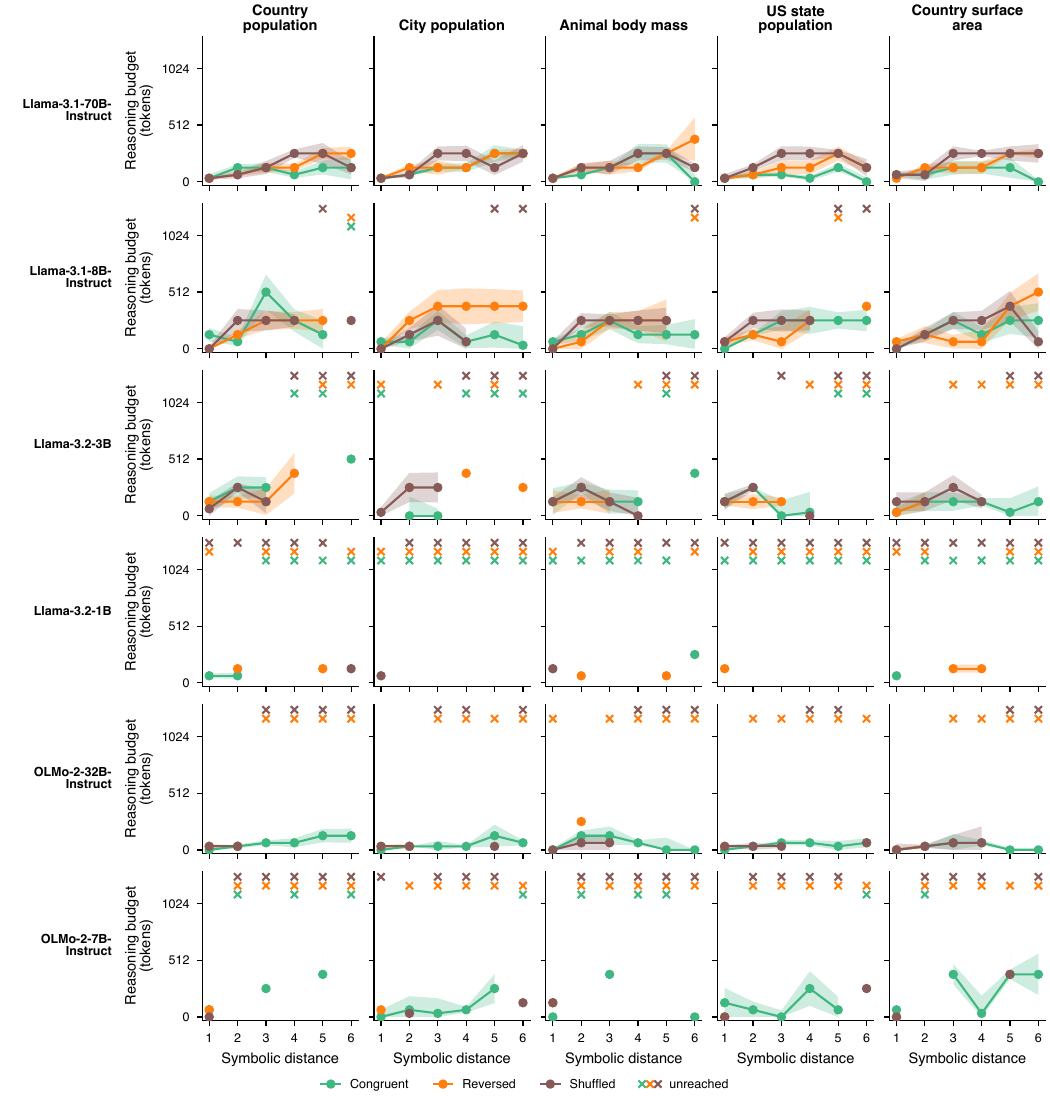}
  \caption{\textbf{LLM reasoning requirements under content conditions: Llama and OLMo models.}
  Budget estimates are defined as in Fig.~\ref{fig:app-llm-criterion-qwen3}.
  Bands show $\pm1$ bootstrap SE over trials.
  Crosses mark distances where accuracy was not reliably above chance at any tested budget, so no budget could be estimated.}
  \label{fig:app-llm-criterion-llama-olmo}
\end{figure}

\paragraph{Reasoning requirements under content conditions.}
Figures~\ref{fig:app-llm-criterion-qwen3}--\ref{fig:app-llm-criterion-llama-olmo} show the reasoning budget required to reach 90\% of maximum accuracy under congruent, reversed, and shuffled contexts for all models and dimensions.
Among model--dimension combinations with an estimate at both distances, the budget was larger at distance six than at distance one in 57 of 66 combinations under reversed contexts and 60 of 66 under shuffled contexts, and in 52 of 89 under congruent contexts.

\FloatBarrier

\paragraph{Additional mechanism results.}
To supplement the LLM results shown in Fig.~\ref{fig:mechanisms}C, we performed additional experiments investigating the mechanisms underlying each strategy in Qwen3-32B, and repeated the main interventions on two further dimensions (Fig.~\ref{fig:app-llm-representation}; procedures in Appendix~\ref{app:llm-methods}).
Figure~\ref{fig:app-llm-representation}A shows the held-out pairwise ordering accuracy of the rank direction (probe accuracy) at each layer for country, city, and US state population.
Ranks were decodable above chance from every layer for all three dimensions, and became more decodable over the first several layers before leveling off.

In the main text (Fig.~\ref{fig:mechanisms}C), we report results for US state population, averaged over distances.
Figure~\ref{fig:app-llm-representation}B shows IWL test accuracy at each symbolic distance, for both erasing and inverting the rank direction.
As in the transformers, erasing the rank direction reduced IWL performance and largely removed the SDE shown by the intact model (Fig.~\ref{fig:app-llm-representation}B, green).
Unlike in the transformers, however, inverting the rank direction did not drive responses below chance or reverse the distance effect (Fig.~\ref{fig:app-llm-representation}B, pink).
We also experimented with writing a random unrelated ordering along the rank direction (Fig.~\ref{fig:app-llm-representation}C).
This shifted direct answers only slightly toward the unrelated ordering, and agreement after reasoning remained near chance.
These results are consistent with the fact that incongruent accuracy rarely dipped far below chance in these models (Fig.~\ref{fig:app-llm-content-qwen3}), and suggest that the LLM's direct answers depend on the rank direction, but that, unlike in the transformers trained from scratch, the presence of interference (whether from incongruent examples in context or from inverted or arbitrary rank directions) drives behavior toward chance rather than toward specific prepotent responses.

Finally, Fig.~\ref{fig:app-llm-representation}D,E shows the interventions of Fig.~\ref{fig:mechanisms}C for country and city population.
In all dimensions, erasing the rank direction reduced IWL accuracy but left accuracy after reasoning intact, whereas ablating context-reading heads reduced accuracy after reasoning on reversed-context and ICL trials but had little effect on IWL accuracy.
Erasing a random direction or ablating random heads had little effect.

\begin{figure}[!htbp]
  \centering
  \includegraphics[width=\textwidth]{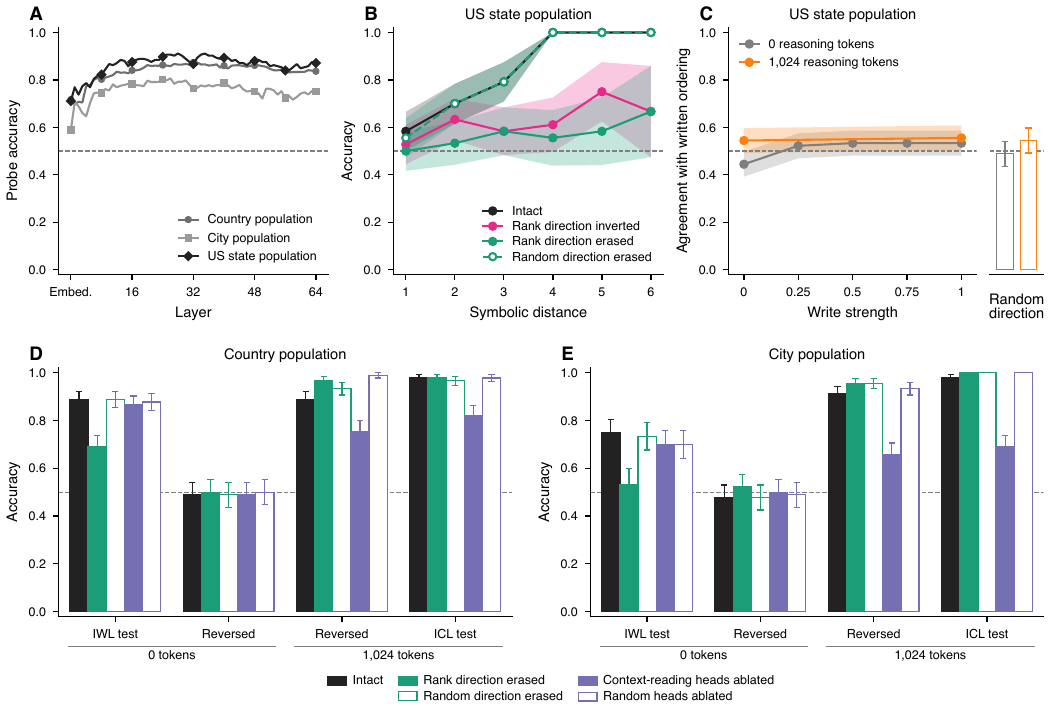}
  \caption{\textbf{LLM mechanism results: Qwen3-32B.}
  (\textbf{A})~Pairwise ordering accuracy of the rank direction on held-out items, by layer, for three different real-world dimensions.
  (\textbf{B})~Direct IWL test accuracy for US state population when the rank direction is intact, erased, or inverted, or a random direction is erased.
  (\textbf{C})~Agreement with another ordering written along the rank direction on reversed-context trials for US state population, answering directly or after 1,024 reasoning tokens; the control shows full-strength writes along a random direction.
  (\textbf{D},~\textbf{E})~The interventions of Fig.~\ref{fig:mechanisms}C for country population and city population.
  Bands and error bars show $\pm1$ binomial SE over trials; dashed lines indicate chance.}
  \label{fig:app-llm-representation}
\end{figure}
\FloatBarrier

\begin{figure}[!htbp]
  \centering
  \includegraphics{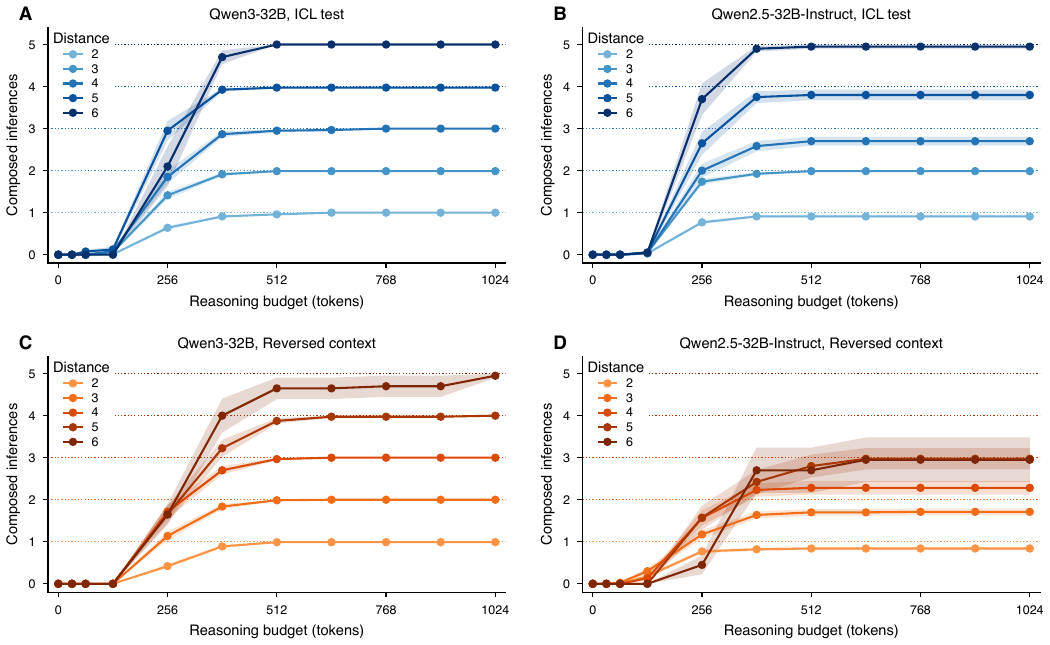}
  \caption{\textbf{Composed inferences in LLM reasoning chains.}
  Mean number of inferences that the reasoning chain had composed by each reasoning budget, on (\textbf{A},~\textbf{B})~the ICL test and (\textbf{C},~\textbf{D})~a reversed context (US state population), for Qwen3-32B (left) and Qwen2.5-32B-Instruct (right).
  Darker lines indicate larger symbolic distances (2--6; distance one requires no inference); dotted lines mark the number required at each distance.
  Bands show $\pm1$~SE over trials.}
  \label{fig:app-llm-chain-steps}
\end{figure}

\paragraph{Reasoning traces.}
To examine whether the larger reasoning budgets required at larger distances (Figs.~\ref{fig:sde}N and~\ref{fig:content}L) reflected additional inferences being made in context, we counted the inferences composed in the reasoning chains of Qwen3-32B and Qwen2.5-32B-Instruct on ICL and reversed-context trials (Fig.~\ref{fig:app-llm-chain-steps}).
An inference combines two relations that share an item (e.g., \texttt{A > B} and \texttt{B > C} give \texttt{A > C}), so a query at distance $d$ requires $d-1$ inferences.
At each budget, we counted how many of these inferences the relations in the chain had composed, counting only relations consistent with the ordering shown and ignoring restatements of the query.
On the ICL test, both models had composed at least 90\% of the required inferences at every distance by 512 tokens.
Under a reversed context, Qwen3-32B reached the same level by 512 tokens, whereas Qwen2.5-32B-Instruct fell short of it at every distance, increasingly so at larger distances.

Traces~\ref{trace:qwen3-correct}--\ref{trace:qwen25-incorrect} show one correct and one incorrect reversed-context trial per model at distance four.
In both of its traces, Qwen3-32B listed all 10 relations, combined them into longer chains until the queried items were connected, and then rechecked each step; in the incorrect trial, the chain reached the correct conclusion, but the answer read after the scoring cue did not follow it.
Qwen2.5-32B-Instruct generated shorter chains, and in the incorrect trial it joined two chains through a relation not given in the context (\texttt{Texas > Kansas}, the real-world relation).

\FloatBarrier

\clearpage
\paragraph{Example reasoning traces.}
Traces~\ref{trace:qwen3-correct}--\ref{trace:qwen25-incorrect} are the trials described above (reversed context, US state population, distance four, 1,024-token budget).
For each model and outcome, we chose the trial whose chain length was closest to the median among chains that ended within 1,024 tokens and, for correct trials, composed all three required inferences.
Only one of Qwen3-32B's seven incorrect trials ended within 1,024 tokens, so Trace~\ref{trace:qwen3-incorrect} is that trial.
Each prompt is shown verbatim, including the chat template, followed by the chain as generated (Appendix~\ref{app:llm-methods}).

\input{appendix/llm_trace_examples}

%% file: tables/experiment1_iwl_accuracy.tex
\begin{tabular}{lrrcrr}
\toprule
Effect & Estimate & \textit{SE} & 95\% CI & $z$ & $p$ \\
\midrule
\multicolumn{6}{l}{\textit{Fixed effects}} \\
\quad Intercept & 0.759 & 0.245 & [0.279, 1.240] & 3.10 & .002 \\
\quad Symbolic distance & 0.286 & 0.072 & [0.145, 0.427] & 3.97 & $<$.001 \\
\multicolumn{6}{l}{\textit{Random effects (SD)}} \\
\quad Intercept & 1.183 \\
\quad Symbolic distance & 0.287 \\
\multicolumn{6}{l}{\textit{Random-effect correlations}} \\
\quad Intercept, symbolic distance & $-$.28 \\
\midrule
Participants & 40 \\
Observations & 1,600 \\
\bottomrule
\end{tabular}

%% file: tables/experiment1_icl_accuracy.tex
\begin{tabular}{lrrcrr}
\toprule
Effect & Estimate & \textit{SE} & 95\% CI & $z$ & $p$ \\
\midrule
\multicolumn{6}{l}{\textit{Fixed effects}} \\
\quad Intercept & 3.183 & 0.427 & [2.345, 4.020] & 7.45 & $<$.001 \\
\quad Symbolic distance & $-$0.387 & 0.069 & [$-$0.521, $-$0.253] & $-$5.65 & $<$.001 \\
\multicolumn{6}{l}{\textit{Random effects (SD)}} \\
\quad Intercept & 2.130 \\
\quad Symbolic distance & 0.217 \\
\multicolumn{6}{l}{\textit{Random-effect correlations}} \\
\quad Intercept, symbolic distance & $-$.84 \\
\midrule
Participants & 40 \\
Observations & 1,600 \\
\bottomrule
\end{tabular}

%% file: tables/experiment1_accuracy_by_test.tex
\begin{tabular}{lrrcrr}
\toprule
Effect & Estimate & \textit{SE} & 95\% CI & $z$ & $p$ \\
\midrule
\multicolumn{6}{l}{\textit{Fixed effects}} \\
\quad Intercept & 3.250 & 0.447 & [2.374, 4.127] & 7.27 & $<$.001 \\
\quad Symbolic distance & $-$0.399 & 0.071 & [$-$0.538, $-$0.260] & $-$5.63 & $<$.001 \\
\quad Test (IWL) & $-$2.488 & 0.468 & [$-$3.405, $-$1.572] & $-$5.32 & $<$.001 \\
\quad Symbolic distance $\times$ test (IWL) & 0.685 & 0.107 & [0.475, 0.895] & 6.40 & $<$.001 \\
\multicolumn{6}{l}{\textit{Random effects (SD)}} \\
\quad Intercept & 2.248 \\
\quad Symbolic distance & 0.234 \\
\quad Test (IWL) & 2.190 \\
\quad Symbolic distance $\times$ test (IWL) & 0.430 \\
\multicolumn{6}{l}{\textit{Random-effect correlations}} \\
\quad Intercept, symbolic distance & $-$.87 \\
\quad Intercept, test (IWL) & $-$.86 \\
\quad Intercept, symbolic distance $\times$ test (IWL) & .73 \\
\quad Symbolic distance, test (IWL) & .66 \\
\quad Symbolic distance, symbolic distance $\times$ test (IWL) & $-$.79 \\
\quad Test (IWL), symbolic distance $\times$ test (IWL) & $-$.72 \\
\midrule
Participants & 40 \\
Observations & 3,200 \\
\bottomrule
\end{tabular}

%% file: tables/experiment1_iwl_rt.tex
\begin{tabular}{lrrcrrr}
\toprule
Effect & Estimate & \textit{SE} & 95\% CI & $t$ & \textit{df} & $p$ \\
\midrule
\multicolumn{7}{l}{\textit{Fixed effects}} \\
\quad Intercept & 7.491 & 0.064 & [7.361, 7.621] & 116.61 & 39.68 & $<$.001 \\
\quad Symbolic distance & $-$0.009 & 0.012 & [$-$0.033, 0.015] & $-$0.78 & 38.26 & .442 \\
\multicolumn{7}{l}{\textit{Random effects (SD)}} \\
\quad Intercept & 0.361 \\
\quad Symbolic distance & 0.052 \\
\quad Residual & 0.480 \\
\multicolumn{7}{l}{\textit{Random-effect correlations}} \\
\quad Intercept, symbolic distance & .69 \\
\midrule
Participants & 40 \\
Observations & 1,223 \\
\bottomrule
\end{tabular}

%% file: tables/experiment1_icl_rt.tex
\begin{tabular}{lrrcrrr}
\toprule
Effect & Estimate & \textit{SE} & 95\% CI & $t$ & \textit{df} & $p$ \\
\midrule
\multicolumn{7}{l}{\textit{Fixed effects}} \\
\quad Intercept & 8.187 & 0.112 & [7.960, 8.414] & 73.14 & 36.17 & $<$.001 \\
\quad Symbolic distance & 0.211 & 0.017 & [0.177, 0.245] & 12.53 & 34.28 & $<$.001 \\
\multicolumn{7}{l}{\textit{Random effects (SD)}} \\
\quad Intercept & 0.672 \\
\quad Symbolic distance & 0.083 \\
\quad Residual & 0.589 \\
\multicolumn{7}{l}{\textit{Random-effect correlations}} \\
\quad Intercept, symbolic distance & .05 \\
\midrule
Participants & 40 \\
Observations & 1,251 \\
\bottomrule
\end{tabular}

%% file: tables/experiment1_rt_by_test.tex
\begin{tabular}{lrrcrrr}
\toprule
Effect & Estimate & \textit{SE} & 95\% CI & $t$ & \textit{df} & $p$ \\
\midrule
\multicolumn{7}{l}{\textit{Fixed effects}} \\
\quad Intercept & 8.180 & 0.111 & [7.956, 8.404] & 73.72 & 40.28 & $<$.001 \\
\quad Symbolic distance & 0.214 & 0.013 & [0.187, 0.241] & 16.14 & 42.62 & $<$.001 \\
\quad Test (IWL) & $-$0.690 & 0.105 & [$-$0.901, $-$0.479] & $-$6.58 & 43.80 & $<$.001 \\
\quad Symbolic distance $\times$ test (IWL) & $-$0.222 & 0.018 & [$-$0.258, $-$0.187] & $-$12.58 & 43.86 & $<$.001 \\
\multicolumn{7}{l}{\textit{Random effects (SD)}} \\
\quad Intercept & 0.672 \\
\quad Symbolic distance & 0.058 \\
\quad Test (IWL) & 0.594 \\
\quad Symbolic distance $\times$ test (IWL) & 0.073 \\
\quad Residual & 0.540 \\
\multicolumn{7}{l}{\textit{Random-effect correlations}} \\
\quad None (uncorrelated random effects) \\
\midrule
Participants & 40 \\
Observations & 2,474 \\
\bottomrule
\end{tabular}

%% file: tables/experiment2_accuracy.tex
\begin{tabular}{lrrcrr}
\toprule
Effect & Estimate & \textit{SE} & 95\% CI & $z$ & $p$ \\
\midrule
\multicolumn{6}{l}{\textit{Fixed effects}} \\
\quad Intercept & 2.569 & 0.149 & [2.277, 2.860] & 17.29 & $<$.001 \\
\quad Condition (shuffled) & $-$2.442 & 0.287 & [$-$3.004, $-$1.880] & $-$8.52 & $<$.001 \\
\quad Log symbolic distance & 0.182 & 0.127 & [$-$0.067, 0.431] & 1.43 & .152 \\
\quad Trial number (per 10 questions) & 0.108 & 0.033 & [0.043, 0.172] & 3.26 & .001 \\
\quad Condition $\times$ log symbolic distance & $-$1.325 & 0.200 & [$-$1.718, $-$0.932] & $-$6.61 & $<$.001 \\
\multicolumn{6}{l}{\textit{Distance slope within condition}} \\
\quad Congruent & 0.845 & 0.199 & [0.454, 1.235] & 4.24 & $<$.001 \\
\quad Shuffled & $-$0.480 & 0.113 & [$-$0.702, $-$0.259] & $-$4.25 & $<$.001 \\
\multicolumn{6}{l}{\textit{Random effects (SD)}} \\
\quad Intercept & 1.176 \\
\quad Log symbolic distance & 0.252 \\
\multicolumn{6}{l}{\textit{Random-effect correlations}} \\
\quad Intercept, log symbolic distance & $-$.02 \\
\midrule
Participants & 99 \\
Observations & 4,752 \\
\bottomrule
\end{tabular}

%% file: tables/experiment2_rt.tex
\begin{tabular}{lrrcrrr}
\toprule
Effect & Estimate & \textit{SE} & 95\% CI & $t$ & \textit{df} & $p$ \\
\midrule
\multicolumn{7}{l}{\textit{Fixed effects}} \\
\quad Intercept & 8.429 & 0.080 & [8.270, 8.588] & 105.19 & 96.47 & $<$.001 \\
\quad Condition (shuffled) & 0.937 & 0.160 & [0.619, 1.256] & 5.85 & 96.47 & $<$.001 \\
\quad Log symbolic distance & 0.099 & 0.026 & [0.047, 0.152] & 3.75 & 95.75 & $<$.001 \\
\quad Trial number (per 10 questions) & $-$0.163 & 0.007 & [$-$0.176, $-$0.150] & $-$24.81 & 3808.04 & $<$.001 \\
\quad Condition $\times$ log symbolic distance & 0.345 & 0.053 & [0.240, 0.450] & 6.52 & 95.75 & $<$.001 \\
\multicolumn{7}{l}{\textit{Distance slope within condition}} \\
\quad Congruent & $-$0.073 & 0.037 & [$-$0.147, 0.000] & $-$1.98 & 90.49 & .050 \\
\quad Shuffled & 0.272 & 0.038 & [0.197, 0.347] & 7.19 & 101.25 & $<$.001 \\
\multicolumn{7}{l}{\textit{Random effects (SD)}} \\
\quad Intercept & 0.792 \\
\quad Log symbolic distance & 0.218 \\
\quad Residual & 0.567 \\
\multicolumn{7}{l}{\textit{Random-effect correlations}} \\
\quad Intercept, log symbolic distance & .76 \\
\midrule
Participants & 99 \\
Observations & 3,988 \\
\bottomrule
\end{tabular}

%% file: appendix/llm_trace_examples.tex
\tracehead{trace:qwen3-correct}{Qwen3-32B, correct answer.}
\begin{tracetext}[label={\textsf{Prompt}}]
<|im_start|>system
In the following question, > means "has a larger population" and < means "has a smaller population". Answer whether the statement is true or false. Base your answer only on the comparisons given — not on outside knowledge.<|im_end|>
<|im_start|>user
Alabama > Indiana
Florida > Texas
Kansas > Oklahoma
Illinois > New York
New Mexico > Kansas
Indiana > New Jersey
New York > Florida
Texas > California
New Jersey > Illinois
Oklahoma > Alabama

Is it true or false that Oklahoma > Illinois?<|im_end|>
<|im_start|>assistant
\end{tracetext}
\begin{tracetext}[label={\textsf{Reasoning chain (693 tokens)}}]
<think>
Okay, let's see. I need to figure out if Oklahoma has a larger population than Illinois based on the given comparisons. Let me start by listing all the provided statements to see if I can create a chain of comparisons that connects Oklahoma and Illinois.

First, the given statements are:

1. Alabama > Indiana
2. Florida > Texas
3. Kansas > Oklahoma
4. Illinois > New York
5. New Mexico > Kansas
6. Indiana > New Jersey
7. New York > Florida
8. Texas > California
9. New Jersey > Illinois
10. Oklahoma > Alabama

Hmm, let's try to map these out step by step. Maybe I can create a hierarchy or order the states based on these relations.

Starting with statement 1: Alabama > Indiana. So Alabama is larger than Indiana.

Statement 6: Indiana > New Jersey. So Indiana is larger than New Jersey. Combining these two, Alabama > Indiana > New Jersey.

Statement 10: Oklahoma > Alabama. So Oklahoma is larger than Alabama. Therefore, Oklahoma > Alabama > Indiana > New Jersey.

Now, statement 3: Kansas > Oklahoma. So Kansas is larger than Oklahoma. So Kansas > Oklahoma > Alabama > Indiana > New Jersey.

Statement 5: New Mexico > Kansas. So New Mexico > Kansas > Oklahoma > Alabama > Indiana > New Jersey.

Now, looking at statement 4: Illinois > New York.

Statement 7: New York > Florida. So Illinois > New York > Florida.

Statement 2: Florida > Texas. So Illinois > New York > Florida > Texas.

Statement 8: Texas > California. So Illinois > New York > Florida > Texas > California.

Statement 9: New Jersey > Illinois. Wait, this is interesting. New Jersey is larger than Illinois. But earlier, we had Indiana > New Jersey, and Alabama > Indiana. So combining these, Alabama > Indiana > New Jersey > Illinois. But statement 4 says Illinois > New York, and statement 7 says New York > Florida, etc. So putting that together: Alabama > Indiana > New Jersey > Illinois > New York > Florida > Texas > California.

But wait, statement 9 is New Jersey > Illinois. So that would mean New Jersey is larger than Illinois. But statement 4 says Illinois > New York. So New Jersey > Illinois > New York. Then New York > Florida (statement 7), and so on.

Now, the question is whether Oklahoma > Illinois. Let's see where Oklahoma stands. From earlier, we have Oklahoma > Alabama > Indiana > New Jersey > Illinois. Wait, no. Wait, Oklahoma is greater than Alabama (statement 10), and Alabama is greater than Indiana (statement 1), Indiana is greater than New Jersey (statement 6), and New Jersey is greater than Illinois (statement 9). So the chain would be Oklahoma > Alabama > Indiana > New Jersey > Illinois. Therefore, Oklahoma is greater than Illinois. So the answer should be true.

But let me double-check to make sure I didn't miss any conflicting statements. Let's verify each step:

- Oklahoma > Alabama (10)
- Alabama > Indiana (1)
- Indiana > New Jersey (6)
- New Jersey > Illinois (9)

So yes, Oklahoma > Alabama > Indiana > New Jersey > Illinois. Therefore, Oklahoma > Illinois. So the statement is true.
</think>
\end{tracetext}
\nopagebreak

\tracehead{trace:qwen3-incorrect}{Qwen3-32B, incorrect answer.}
\begin{tracetext}[label={\textsf{Prompt}}]
<|im_start|>system
In the following question, > means "has a larger population" and < means "has a smaller population". Answer whether the statement is true or false. Base your answer only on the comparisons given — not on outside knowledge.<|im_end|>
<|im_start|>user
New Mexico > Kansas
Florida > Texas
New Jersey > Illinois
Kansas > Oklahoma
Oklahoma > Alabama
Illinois > New York
Alabama > Indiana
Texas > California
New York > Florida
Indiana > New Jersey

Is it true or false that Indiana > Florida?<|im_end|>
<|im_start|>assistant
\end{tracetext}
\begin{tracetext}[label={\textsf{Reasoning chain (1,022 tokens)}}]
<think>
Okay, let's see. I need to figure out if Indiana has a larger population than Florida based on the given comparisons. Let me start by listing all the provided statements and see if I can create a chain of comparisons that connects Indiana to Florida.

First, the given statements are:

1. New Mexico > Kansas
2. Florida > Texas
3. New Jersey > Illinois
4. Kansas > Oklahoma
5. Oklahoma > Alabama
6. Illinois > New York
7. Alabama > Indiana
8. Texas > California
9. New York > Florida
10. Indiana > New Jersey

So, the question is whether Indiana > Florida. Let's try to connect these.

Starting from Indiana. From statement 10: Indiana > New Jersey. Then, from statement 3: New Jersey > Illinois. So combining these, Indiana > New Jersey > Illinois. Then, statement 6 says Illinois > New York. So now we have Indiana > New Jersey > Illinois > New York. Then, statement 9 says New York > Florida. So putting that together: Indiana > New Jersey > Illinois > New York > Florida. If that's the case, then Indiana would be greater than Florida. But wait, let me check if all these connections are valid.

Wait, let me map each step carefully.

From statement 10: Indiana > New Jersey.

From statement 3: New Jersey > Illinois. So combining these two, Indiana > New Jersey > Illinois. So Indiana > Illinois.

From statement 6: Illinois > New York. So Indiana > Illinois > New York.

From statement 9: New York > Florida. So Indiana > Illinois > New York > Florida. Therefore, Indiana > Florida. That would make the statement true. But wait, let me check if there are any contradictions or other paths that might affect this.

Alternatively, let's see if there's another path from Indiana to Florida. Let's see.

From statement 7: Alabama > Indiana. So Alabama is greater than Indiana. Then, from statement 5: Oklahoma > Alabama. So Oklahoma > Alabama > Indiana. Then, statement 4: Kansas > Oklahoma. So Kansas > Oklahoma > Alabama > Indiana. Then, statement 1: New Mexico > Kansas. So New Mexico > Kansas > Oklahoma > Alabama > Indiana. But how does this connect to Florida?

Looking at statement 2: Florida > Texas. And statement 8: Texas > California. So Florida > Texas > California. Then, statement 9: New York > Florida. So New York is greater than Florida. But earlier, we had Indiana > New York. So if Indiana > New York and New York > Florida, then by transitivity, Indiana > Florida. That seems to hold.

But wait, let me check if there's any conflicting information. For example, if there's a loop or contradiction. Let's see:

From the chain Indiana > New Jersey > Illinois > New York > Florida, that's a direct path. But also, from the other side, we have New Mexico > Kansas > Oklahoma > Alabama > Indiana. So Indiana is less than New Mexico, Kansas, Oklahoma, Alabama. But that doesn't conflict with Indiana being greater than Florida, as long as those other regions are not connected to Florida in a way that would create a contradiction.

Is there any other connection? Let's check all statements again.

Statement 2: Florida > Texas. Statement 8: Texas > California. So Florida > Texas > California. Statement 9: New York > Florida. So New York is above Florida. Then, from statement 6: Illinois > New York. So Illinois > New York > Florida. Then, from statement 3: New Jersey > Illinois. So New Jersey > Illinois > New York > Florida. Then, from statement 10: Indiana > New Jersey. So Indiana > New Jersey > Illinois > New York > Florida. Therefore, Indiana is indeed greater than Florida. So the answer should be true.

But wait, let me make sure there's no other path that would contradict this. For example, if there's a way to get from Indiana to Florida through another route that would show Indiana < Florida. Let's see.

From Indiana, we have statement 7: Alabama > Indiana. Then, statement 5: Oklahoma > Alabama. Then, statement 4: Kansas > Oklahoma. Then, statement 1: New Mexico > Kansas. But how does that connect to Florida? The only connection to Florida is through New York. So unless there's a connection between New Mexico or Kansas to Texas or something else, but I don't see that. The only other connection is statement 2: Florida > Texas, and statement 8: Texas > California. So Florida is above Texas, which is above California. But New York is above Florida (statement 9), and Indiana is above New York (from the chain). So no contradictions here. Therefore, the conclusion that Indiana > Florida is true based on the given statements.
</think>
\end{tracetext}
\nopagebreak

\tracehead{trace:qwen25-correct}{Qwen2.5-32B-Instruct, correct answer.}
\begin{tracetext}[label={\textsf{Prompt}}]
<|im_start|>system
In the following question, > means "has a larger population" and < means "has a smaller population". Answer whether the statement is true or false. Base your answer only on the comparisons given — not on outside knowledge.<|im_end|>
<|im_start|>user
Indiana > New Jersey
New York > Florida
New Jersey > Illinois
Oklahoma > Alabama
Kansas > Oklahoma
Texas > California
Florida > Texas
Alabama > Indiana
Illinois > New York
New Mexico > Kansas

Is it true or false that Florida > Indiana?
Let's think step by step.<|im_end|>
<|im_start|>assistant
\end{tracetext}
\begin{tracetext}[label={\textsf{Reasoning chain (267 tokens)}}]
To determine if the statement "Florida > Indiana" is true or false based on the given comparisons, we need to establish a chain of relationships between these states.

Here are the relevant comparisons from the provided information:

1. **Indiana > New Jersey**
2. **New Jersey > Illinois**
3. **Illinois > New York** (This is the reverse of what was given, so we know New York < Illinois)
4. **New York > Florida**
5. **Florida > Texas**
6. **Texas > California**
7. **Alabama > Indiana**

From the above, we can derive the following relationships:

- From 1 and 2: **Indiana > New Jersey > Illinois**
- From 3: **Illinois > New York**
- Combining the above: **Indiana > New Jersey > Illinois > New York**
- From 4: **New York > Florida**
- Combining with the previous: **Indiana > New Jersey > Illinois > New York > Florida**
- From 7: **Alabama > Indiana**

Now, let's check the relationship between Florida and Indiana:

- We have established that **Indiana > New Jersey > Illinois > New York > Florida**.
- This implies that **Indiana > Florida**.

Therefore, the statement "Florida > Indiana" is **false**.
\end{tracetext}
\nopagebreak

\tracehead{trace:qwen25-incorrect}{Qwen2.5-32B-Instruct, incorrect answer.}
\begin{tracetext}[label={\textsf{Prompt}}]
<|im_start|>system
In the following question, > means "has a larger population" and < means "has a smaller population". Answer whether the statement is true or false. Base your answer only on the comparisons given — not on outside knowledge.<|im_end|>
<|im_start|>user
New Mexico > Kansas
New Jersey > Illinois
Kansas > Oklahoma
Florida > Texas
New York > Florida
Texas > California
Illinois > New York
Alabama > Indiana
Oklahoma > Alabama
Indiana > New Jersey

Is it true or false that New York > Alabama?
Let's think step by step.<|im_end|>
<|im_start|>assistant
\end{tracetext}
\begin{tracetext}[label={\textsf{Reasoning chain (311 tokens)}}]
To determine if the statement "New York > Alabama" is true or false based on the given comparisons, we need to establish a chain of relationships between these states.

1. **Florida > Texas**
2. **Texas > California** (This doesn't directly help us with New York or Alabama)
3. **New York > Florida**
   - From 1 and 3: New York > Florida > Texas
4. **Illinois > New York**
   - From 3 and 4: Illinois > New York > Florida > Texas
5. **New Jersey > Illinois**
   - From 4 and 5: New Jersey > Illinois > New York > Florida > Texas
6. **Kansas > Oklahoma**
7. **Oklahoma > Alabama**
   - From 6 and 7: Kansas > Oklahoma > Alabama
8. **Alabama > Indiana**
   - From 7 and 8: Kansas > Oklahoma > Alabama > Indiana
9. **Indiana > New Jersey**
   - From 5, 8, and 9: New Jersey > Illinois > New York > Florida > Texas > Kansas > Oklahoma > Alabama > Indiana > New Jersey (This creates a loop, but we can still use the direct paths)

Now, let's focus on the relevant parts:
- **New York > Florida > Texas > Kansas > Oklahoma > Alabama**

From this chain, we can see that New York has a larger population than Alabama.

Therefore, the statement "New York > Alabama" is **true**.
\end{tracetext}
\nopagebreak